\documentclass[a4paper, 12pt]{article}
\usepackage{geometry}
\usepackage{amsmath,amsthm,amssymb, mathtools,mathrsfs}
\usepackage{physics}
\usepackage{amsfonts, latexsym, amssymb}
\usepackage{bbm, dsfont}
\usepackage{graphicx}
\usepackage{float}
\usepackage{url}
\usepackage{xcolor}
\usepackage[colorlinks, linkcolor=darkblue, citecolor=darkblue, urlcolor=darkblue, linktocpage]{hyperref} 
\usepackage{tikz}
\usetikzlibrary{decorations.pathmorphing,cd,decorations.markings,calc,fadings,math}
\hypersetup{
    colorlinks=true,
    linkcolor=BLUE,
    filecolor=magenta,      
    urlcolor=cyan,
}
\usepackage{enumerate}
\usepackage{bm}% bold math

\usepackage[export]{adjustbox}
\usepackage{cite}
\usepackage{cleveref}
\usepackage{appendix}
\usepackage{bbold}

\usepackage{subcaption}

\definecolor{BLUE}{cmyk}{1,0.5,0,.47}
\definecolor{PINK}{cmyk}{0,.55,.45,.27}
\definecolor{YELLOW}{cmyk}{0,.23,.77,133}

\definecolor{darkblue}{cmyk}{0.9,0.9,0,0}
\definecolor{darkgreen}{cmyk}{0.9,0,0.9,0}
\definecolor{blueblue}{cmyk}{0.73,0.28,0,0.5}
\definecolor{lightblue}{RGB}{55,171,200}
\definecolor{grey}{gray}{0.55}
\definecolor{pink}{cmyk}{0., 0.9859943977591037, 0.3571428571428571, 0.16000000000000003}
\definecolor{lightpink}{cmyk}{0., 0.5, 0.5, 0.}
\definecolor{lightgreen}{cmyk}{0.24175824175824182, 0., 0.9615384615384616, 0.28627450980392155}

\renewcommand{\d}{\mathrm{d}}

\newcommand{\ds}{\displaystyle}

\newcommand{\comment}[1]{}

\definecolor{darkgreen}{rgb}{0.1,0.6,0.1}

\newcommand{\rrangle}{\rangle\!\rangle}

\newcommand{\be}{\begin{equation}} \newcommand{\ee}{\end{equation}}
\newcommand{\ben}{\begin{equation*}} \newcommand{\een}{\end{equation*}}

\newcommand{\bea}{\begin{equation} \begin{aligned}} \newcommand{\eea}{\end{aligned} \end{equation}}

\newcommand{\cA}{\mathcal{A}}

\newcommand{\cC}{\mathcal{C}}

\newcommand{\cF}{\mathcal{F}}

\newcommand{\cH}{\mathcal{H}}

\newcommand{\cL}{\mathcal{L}}
\newcommand{\calL}{\mathscr{L}}
\newcommand{\cM}{\mathcal{M}}
\newcommand{\calM}{\mathscr{M}}

\newcommand{\cZ}{\mathcal{Z}}

\newcommand{\bZ}{\mathbb{Z}}

\newcommand{\unit}{\mathbbm{1}}

\def\repa{\raise4pt\hbox{$\square$}\mkern-14mu\raise-4pt\hbox{$\square$}}
\def\repab{\overline{\raise4pt\hbox{$\square$}\mkern-14mu\raise-4pt\hbox{$\square$}\mkern-1mu}}

\def\({\left(}
\def\){\right)}
\def\[{\left[}
\def\]{\right]}
\def\<{\langle}
\def\>{\rangle}

\begin{document}

\thispagestyle{empty}
\fontsize{12pt}{20pt}
\vspace{13mm}  %\vspace{10mm}
\begin{center}
	{\huge  S-Matrix Bootstrap and \\[.5em]  Non-Invertible Symmetries}
	\\[13mm]
    	{\large Christian Copetti$^{a} \,$, Luc\'ia C\'ordova$^{b} \,$ and Shota Komatsu$^{b} \,$} 
	%\\[6mm]
	
	\bigskip
	{\it
		$^a$ Mathematical Institute, University of Oxford, Woodstock Road,\\ Oxford, OX2 6GG, United Kingdom \\[.2em]
		$^b$ Department of Theoretical Physics, CERN, 1211 Meyrin, Switzerland \\[.2em]
	}
\end{center}

\begin{abstract}
We initiate the S-matrix bootstrap analysis of theories with non-invertible symmetries in (1+1) dimensions. Our previous work \cite{Copetti:2024rqj} showed that crossing symmetry of S-matrices in such theories is modified, with modification characterized by the fusion category data. By imposing unitarity, symmetry and the modified crossing, we constrain the space of consistent S-matrices, identifying integrable theories with non-invertible symmetries at the cusps of allowed regions. We also extend the modified crossing rules to cases where vacua transform in non-regular representations of fusion category, utilizing a connection to a dual category $\mathcal{C}^{\ast}_{\mathscr{M}}$ and Symmetry Topological Field Theory (SymTFT). This highlights the utility of SymTFT in the analysis of scattering amplitudes.
	
\end{abstract}
\newpage
\tableofcontents
\newpage

%%%%%%%%%%%%%%%%%%%%%%%%%%%%%%%%%%%%%%%%%%%%%%%%%%%%%%%%%%%%%%%%%%%%%
\section{Introduction and summary}
\subsection{Introduction}
The idea of constraining quantum field theories using fundamental principles, such as symmetry, unitarity and causality originated in the 1960s. At that time, researchers aimed to understand the physics of strong interaction without knowing the underlying Lagrangian. These early efforts were eventually set aside following the discovery and successful application of quantum chromodynamics (QCD). However this idea, known as the {\it S-matrix bootstrap} program, has recently been revived \cite{Paulos:2016fap,Paulos:2016but}. Today  it serves as a tool to explore the landscape of quantum field theories (QFTs), rather than focusing on a specific theory as was initially intended.

In parallel, a similar idea has been applied to conformal field theories (CFTs), which are quantum field theories with conformal symmetry. This approach, often called the {\it conformal bootstrap} program, achieved impressive success in $1+1$ dimensions in the 1980s. More recently, with the aid of judicious numerical implementation, it has made remarkably precise predictions for theories of physical interest in higher dimensions, such as the Ising model in $2+1$ dimensions \cite{Kos:2016ysd}.

Both bootstrap programs share a common philosophy: they use fundamental principles as inputs to delineate regions in theory space consistent with these principles. However, there is also a key difference. Most CFTs are ``isolated," meaning small deformations typically break conformal symmetry. Partly due to this property, in the best cases the conformal bootstrap identifies a tiny ``island" in parameter space, with theories of physical interest at the boundary of this region. In contrast, the landscape of quantum field theories is much broader. Most quantum field theories allow for continuous deformations by adjusting parameters like particle masses and interaction strengths. Thus, there is no guarantee that theories of our interest lie near the boundary\footnote{Important exceptions are theories in $(1+1)$ dimensions for which integrable theories were found to lie at the boundary of the allowed region.} of the allowed region determined by the S-matrix bootstrap.

To narrow down the landscape carved out by the S-matrix bootstrap, one needs to input theory-specific features. This can include data from observables, either computed from a theory or obtained from experiments. Alternatively, one can use detailed structures of the theory that distinguish it from others. In this paper, we take the latter approach; we discuss the interplay between the S-matrix bootstrap and a refined notion of symmetries intensively studied in the past few years, called {\it non-invertible symmetries}.

Non-invertible symmetries generalize the notion of symmetries in QFT. The key insight that led to this generalization is to identify symmetry operators with extended topological operators \cite{Gaiotto:2014kfa}. From this perspective, standard symmetries are special cases in which topological operators follow a group-like multiplication law, and more generally they obey {\it categorical} fusion algebras. Studies in recent years have shown that non-invertible symmetries are more common than initially thought \cite{Chang:2018iay,Choi:2021kmx,Choi:2022zal,Kaidi:2021xfk,Kaidi:2022uux,Bhardwaj:2022yxj,Damia:2022seq}. However they are (or at least some of them are) still less ubiquitous than standard symmetries: for example, $\mathcal{N}=4$ supersymmetric Yang-Mills theory has a non-invertible S-duality symmetry only when the coupling constant is tuned to a particular value \cite{Kaidi:2022uux} and requiring it to survive selects a subset of relevant deformations \cite {Damia:2023ses}. Thus incorporating such symmetries in the bootstrap program can potentially help narrow the search space and improve the bounds. 

In addition, we showed in the previous paper \cite{Copetti:2024rqj} that non-invertible symmetries have direct implications on the S-matrices. Specifically, in 1+1 dimensional theories with non-invertible symmetries, crossing symmetry is modified, with the modification characterized by the categorical data of these symmetries:
\begin{align}\label{eq:modified}
S^{ab}_{dc}(s)=\sqrt{\dfrac{\d_a \d_c}{\d_b \d_d}} \, S^{bc}_{ad}(t)\,.
\end{align}
Here $s$ and $t$ are Mandelstam variables and $\d_{a,b,c,d}$ are quantum dimensions of the relevant fusion category \cite{etingof2016tensor,Barkeshli:2014cna,Aasen:2020jwb}. See Section \ref{sec: symaction} for details.

Building on this finding, in this paper we initiate the S-matrix bootstrap for such theories by imposing unitarity and modified crossing relations.
 Below we outline several additional reasons why this analysis is of physical interest:
 \paragraph{Modified crossing disallows trivial scattering.} A simple yet important observation is that the modified crossing relation \eqref{eq:modified} cannot be satisfied by a trivial (non-interacting) S-matrix $S(s)=\pm {\bf 1}$; this means that the modified crossing alone ensures that the theory is interacting. This contrasts with the standard S-matrix bootstrap, where unitarity and crossing allow for a non-interacting S-matrix, necessitating additional steps to focus on physically interesting theories. This feature can be potentially useful in higher dimensions, where methods to isolate physically relevant theories are less developed compared to $1+1$ dimensions. We will come back to this point in Section \ref{sec: conclusions} (Conclusion). 
% the bounds obtained from the S-matrix bootstrap are often saturated by unphysical amplitudes without particle production, and an efficient way to exclude them has not been established.  
\paragraph{Bootstrap from IR to UV.} Theories with non-invertible symmetries provide a unique arena for the bootstrap program, offering the possibility to bootstrap the dynamics from infrared (IR) to ultraviolet (UV), defying the standard renormalization group paradigm. As we explain below, non-invertible symmetries impose nontrivial constraints at every energy scale and the data bootstrapped at lower energy scale can be used as inputs for the bootstrap at higher energy scale.
\begin{enumerate}
\item {\it TQFT bootstrap:} In the deep IR, non-invertible symmetries constrain the number of vacua, often disallowing a single vacuum. The number of vacua and their symmetry actions can be determined by bootstrapping topological quantum field theory (TQFT). See \cite{Thorngren:2019iar,Komargodski:2020mxz,Huang:2021zvu,Bhardwaj:2023idu} for recent analysis.
\item {\it S-matrix bootstrap:} The TQFT data bootstrapped in the IR determine the modified crossing relation, a fundamental input for the S-matrix bootstrap, as shown in our previous work \cite{Copetti:2024rqj} and this paper. In addition, non-invertible symmetries give constraints on the spectrum of particles as discussed in \cite{Cordova:2024vsq}. By bootstrapping the S-matrix using these data, one can constrain the dynamics of the theory along the renormalization group (RG) flow.
\item {\it Integrable bootstrap:} In the S-matrix bootstrap, integrable S-matrices often appear at the boundary of the allowed region. These S-matrices can be used to determine the finite volume spectrum of the theory through the Thermodynamic Bethe ansatz (TBA) \cite{Zamolodchikov:1989cf}. Alternatively, one can impose the Yang-Baxter equation and look for integrable S-matrices directly. 
\item {\it Conformal bootstrap:} By taking the UV limit of the finite volume spectrum, one can infer the spectrum of operators in the UV CFT. This data can then be used to constrain the UV CFT via the conformal bootstrap.
\item[4$^{\ast}$.] {\it Form factor bootstrap:} Even if integrable S-matrices are not found, progress can be made using the form factor bootstrap, see \cite{Karateev:2019ymz,Correia:2022dyp,Cordova:2023wjp,He:2024nwd} for recent discussions. The resulting UV data can then be used in the conformal bootstrap. (See also  \cite{Lin:2023uvm} for a different approach that combines non-invertible symmetries and the conformal bootstrap.)
\end{enumerate}
In this paper, we focus on the S-matrix bootstrap analysis. The interplay with integrability and TBA is a subject of the ongoing work \cite{inprogress}.

\subsection{Summary}
Let us now summarize the punchlines of our paper.
\paragraph{Vacua in general representations.} In our previous paper, we focused on the cases where vacua in the IR transform as the {\it regular representation} of the fusion category $\mathcal{C}$. Physically, they correspond to the complete spontaneous symmetry breaking of $\mathcal{C}$.
In Section \ref{sec: symaction}, we extend our analysis to vacua transforming in general representations of $\mathcal{C}$, which correspond to partial symmetry breaking\footnote{See \cite{Thorngren:2019iar,Huang:2021zvu,Bhardwaj:2023idu,Bhardwaj:2024kvy} for the classification of gapped phases with various symmetry breaking patterns.} of $\mathcal{C}$, by identifying irreducible representations with simple lines in a dual category $\mathcal{C}^{\ast}_{\mathscr{M}}$. Specifically, we derive Ward identities and modified crossing rules applicable to these broader cases.  We will encounter one such example in Section \ref{ssec: Fibboot} as an S-matrix saturating the bootstrap bound for Fibonacci fusion category.

\paragraph{Symmetry action on kinks and modified crossing from SymTFT.} In Section \ref{sec: symtft}, we develop a formalism to study symmetry actions on kinks interpolating between different vacua and their scattering amplitudes using Symmetry Topological Field Theory (SymTFT). SymTFT offers a universal and ``holographic" framework for studying symmetries and anomalies. Its key advantage is decoupling theory-specific dynamics from the categorical structures of symmetries and anomalies. Applied to our context, it leads to a clean derivation of the symmetry action on kinks and their modified crossing rules, including the extension to general representations mentioned above. Furthermore, SymTFT can naturally extend to higher dimensions. Thus we hope that the results in this paper lay the groundwork for higher-dimensional generalizations of modification of crossing rules due to categorical symmetries.

\paragraph{S-matrix bootstrap for $\cA_n$ and Fibonacci fusion categories.}
 By imposing the modified crossing rules together with unitarity and analyticity, we perform the S-matrix bootstrap analysis in Section \ref{sec: Bootstrap}. 
We focus on two well-known examples: the $\cA_n$ symmetry category, which is the symmetry preserved by e.g. the $\phi_{1,3}$ deformation of unitary minimal models\footnote{In the simplest non-invertible case, $\cA_4 = \bZ_2 \text{Tambara-Yamagami}$ is the symmetry of Ising CFT.} $\cM_n$, and the smallest non-invertible symmetry category, the Fibonacci fusion category ($\text{Fib}$).
In the former case we perform the bootstrap analysis for the regular representation and the minimum spectrum required by symmetry, that is the set of kinks interpolating between neighboring vacua and no other stable particles such as bound states, and find that the known integrable models studied by Zamolodchikov \cite{Zamolodchikov:1991vh} sit at vertices of the carved out parameter space. 
In the Fibonacci case, there are two vacua and the minimum spectrum consists of a kink, antikink and a breather which can be interpreted as a kink-antikink bound state with a cubic coupling $g$. Once more integrable theories are found sitting at vertices of the allowed space: the $\phi_{2,1}$ deformation of the tricritical Ising CFT \cite{Smirnov:1991uw} and the 3-state Potts CFT deformed by the relevant operator $Z+Z^{\ast}$. The latter model is located at a vertex of the $g=0$ slice of the allowed space, where we see a symmetry enhancement to $\text{Fib} \times \bZ_2$.  

\paragraph{} Several future directions are discussed in Section \ref{sec: conclusions}. A few appendices are included to explain technical details.

\paragraph{Note Added:} While this article was in preparation, the work \cite{Cordova:2024iti}, which has some overlap with Section \ref{sec: symaction}, appeared in arXiv. We also became aware of related upcoming works \cite{Choi:2024toappear,CC:boundaries}, which also have overlap with Section \ref{sec: symaction}. We are grateful to the authors of \cite{Choi:2024toappear,CC:boundaries} for communications prior to their publications.

%%%%%%%%%%%%%%%%%%%%%%%%%%%%%%%%%%%%%%%%%%%%%%%%%%%%%%%%%%%%%%%%
\section{Non-invertible symmetries and kink scattering}\label{sec: symaction}
The non-invertible symmetries discussed in this paper are spontaneously broken in the IR, at least partially, resulting in multiple degenerate vacua. This is due to anomalies, which forbid a unique gapped ground state. Stable particles around such degenerate vacua are typically kinks that interpolate between them.

In this section, we derive constraints on scattering amplitudes of these kinks imposed by non-invertible symmetries, which we will use in the bootstrap analysis in Section \ref{sec: Bootstrap}. Specifically, we extend our previous work \cite{Copetti:2024rqj} to cases where IR vacua transform in general representations of the fusion category $\mathcal{C}$.

Section \ref{subsec:symmetryreview} begins with a brief review of fusion categories to establish the necessary notation. Our conventions will follow those in \cite{Barkeshli:2014cna, Aasen:2020jwb}. For more comprehensive introductions to the subject, we recommend \cite{Chang:2018iay, Schafer-Nameki:2023jdn, Shao:2023gho, Bhardwaj:2023kri}. Section \ref{subsec:modulecat} summarizes the fundamental concepts required to discuss symmetry actions on gapped vacua in 1+1 dimensions following \cite{Huang:2021zvu}, specifically focusing on $\cC$-symmetric topological field theories (TFTs).
Next, in Section \ref{ssec: symkink}, we discuss kinks interpolating between different vacua and study how they organize into a multiplet of the fusion category. A clear physical formulation of this problem was recently given in  \cite{Cordova:2024vsq}, and we build upon their results by providing a way to identify irreducible multiplets with simple lines in a dual category $\mathcal{C}^{\ast}_{\mathcal{M}}$. Section \ref{sssec: fibz2} presents several concrete examples relevant to the subsequent analysis. 
Finally in Section \ref{sec: symtft}, we rephrase these concepts from the perspective of SymTFT, based on recent developments \cite{Copetti:2024onh,Choi:2024toappear,CC:boundaries}. Using this SymTFT framework, we derive Ward identities and modified crossing rules for general $\mathcal{C}$-symmetric S-matrices, setting the stage for the bootstrap analysis that follows.

%In this section, we elucidate the structure of symmetry multiplets within the space of massive kinks. Our setup follows \cite{Cordova:2024vsq}. For a related realization in lattice systems using the anyon chain formalism, see \cite{Aasen:2020jwb,Buican:2017rxc,Bhardwaj:2024kvy}.

%We begin with a brief review of fusion categories in 1+1 dimensions to establish the necessary notation. For more comprehensive introductions, we recommend \cite{Chang:2018iay, Schafer-Nameki:2023jdn, Shao:2023gho, Bhardwaj:2023kri}. Our conventions will follow those outlined in \cite{Barkeshli:2014cna, Aasen:2020jwb}. 

%Next, we introduce the fundamental concepts required to discuss the symmetry action on gapped vacua in 1+1 dimensions, specifically focusing on $\cC$-symmetric topological field theories (TFTs), the discussion here follows \cite{Huang:2021zvu}. 

%We then consider massive kinks that interpolate between different gapped vacua and examine their multiplet structure. A recent clear physical formulation of this problem was proposed in \cite{Cordova:2024vsq}, and we build upon their results by providing a straightforward method to identify irreducible multiplets through the SymTFT. This approach enables us to derive Ward identities for $\cC$-symmetric S-matrices. In Section \ref{sec: Bootstrap}, we use these Ward identities as constraints for the S-matrix bootstrap.

\subsection{\texorpdfstring{Categorical symmetry in $1+1$ dimensions}{Categorical Symmetry in 1+1d}}\label{subsec:symmetryreview}

In 1+1 dimensional QFTs, symmetry operations need not be group-like. Instead, they are described by the more general structure of a (unitary) fusion category $\cC$ \cite{etingof2016tensor}. A (simple) object $\cL$ in $\cC$ is an (indecomposable) topological line of the QFT, i.e., a line operator that commutes with the stress tensor $T_{\mu\nu}$.

\paragraph{Fusion of lines.}Two lines $\cL$ and $\cL'$ can be fused to form a new line $\cL \times \cL'$, with the fusion product also belonging to $\cC$:
\be
\begin{tikzpicture}
\draw[BLUE] (0,0) node[below] {$\cL$} -- (0,2); 
\draw[BLUE] (1,0) node[below] {$\cL'$} -- (1,2);
\node[right] at (1.5,1) {$\ds = \sum_{\cL''} N_{\cL \, \cL'}^{\cL''} $};
\draw[BLUE] (4,0) node[below] {$\cL''$} -- (4,2);
\end{tikzpicture}
\ee
The {\it fusion coefficients} $N_{\cL \cL'}^{\cL''}$ are non-negative integers representing the dimension of the vector space of trivalent topological junctions\footnote{This can be shown by considering radial quantization around the junction. See \cite{Chang:2018iay} for details.}:
\be
\begin{tikzpicture}
    \node[left] at (-1,0) {$x \, \in \, V_{\cL \cL'}^{\cL''}:$};
    \draw[BLUE]  (0,0) node[BLUE,circle,fill,scale=0.3] {} -- (210:1) node[below] {$\cL$};
    \draw[BLUE]  (0,0) -- (-30:1) node[below] {$\cL'$};
    \draw[BLUE]  (0,0) -- (90:1) node[above] {$\cL''$};
    \node[above left] at (0,0) {$x$};
\end{tikzpicture}
\ee
By a judicious choice of the basis of junctions, the orthogonality and completeness relations for these vector spaces take the form:
\be \label{eq:orthocompleteness}
\begin{tikzpicture}
\draw[BLUE] (0,0) node[below] {$\cL$} -- (0,0.5)  node[BLUE,circle,fill,scale=0.3] {}; \draw[BLUE] (0,1.5)  node[BLUE,circle,fill,scale=0.3] {} -- (0,2) node[above] {$\cL'$};
\draw[BLUE] (0,1) circle (0.5); \node[BLUE,left] at (-0.5,1) {$\cL_u$}; \node[BLUE,right] at (0.5,1) {$\cL_v$}; 
\node[above left] at (0,1.5) {$x$}; \node[below left] at (0,.5) {$y$}; 
\node[right] at (1,1) {$\ds = \sqrt{\frac{\d_{\cL_u} \d_{\cL_v}}{\d_{\cL}}} \delta_{\cL \, \cL'} \delta_{x \, y} $};
\draw[BLUE] (5,0) node[below] {$\cL$} -- (5,2);
\begin{scope}[shift={(9,0)}]
    \draw[BLUE] (0,0) node[below] {$\cL$} -- (0,2);    \draw[BLUE] (1,0) node[below] {$\cL'$} -- (1,2);
    \node[right] at (1.25,1) {$\ds = \sum_{\cL'', \, x} \sqrt{\frac{\d_{\cL''}}{\d_{\cL} \d_{\cL'}}}$};
    \begin{scope}[shift={(4.25,0)}]
     \draw[BLUE] (0,0) node[below] {$\cL$}   -- (0.5,0.5)  node[BLUE,circle,fill,scale=0.3] {} -- (0.5,1.5)  node[BLUE,circle,fill,scale=0.3] {} -- (0,2) node[above] {$\cL$};
     \draw[BLUE] (1,0) node[below] {$\cL'$} -- (0.5,0.5);  \draw[BLUE] (1,2) node[above] {$\cL'$} -- (0.5,1.5);
     \node[right,BLUE] at (0.5,1) {$\cL''$};
     \node[below left] at (0.5,1.5) {$x$}; \node[above left] at (0.5,0.5) {$x$};
    \end{scope}
\end{scope}
\end{tikzpicture}
\ee
Lines are typically oriented, with the orientation reversal $\cL^\vee$ being the unique line in $\cC$ such that 
$
\cL \times \cL^\vee = \unit + \ldots \,
$.

In what follows, we will make the following simplifications:
\begin{itemize}
\item We will omit the label $x$ of the basis vectors of junctions since $N_{\cL \cL'}^{\cL''}=0$ or $1$ in our examples, resulting in no loss of generality.
\item We will focus on cases where lines are self-dual (i.e. $\mathcal{L}=\mathcal{L}^{\vee}$). Thus we will dispense with orientation to simplify notation.
\end{itemize}

\paragraph{Quantum dimensions.}The expectation value of a loop gives the so-called {\it quantum dimension} $\d_\cL$ of the line $\cL$
\be
\begin{tikzpicture}
    \draw[BLUE] (0,0) circle (1); 
    \node[BLUE,left] at (-1,0) {$\cL$};
    \node[right] at (1,0) {$= \d_\cL$};
\end{tikzpicture}
\ee
It is possible to prove that  $\d_\cL \geq 1$, with the inequality saturated if and only if the $\cL$ line is \emph{invertible}, i.e. it is an element of some (possibly non-abelian) symmetry group.\footnote{In the present discussion we disregard the possibility of a non-trivial Frobenius-Shur indicator $\epsilon=-1$, which is possible for self-dual lines. Some physical implications of this quantity are discussed e.g. in \cite{Chang:2018iay,Hason:2020yqf,Cordova:2019wpi,Lin:2019kpn}.}

\paragraph{F-symbol and pentagon equation.} The fusion product is associative: $(\cL \times \cL') \times \cL'' = \cL \times (\cL' \times \cL'')$, which implies $\sum_{\cL_u} N_{\cL \cL'}^{\cL_u} N_{\cL_u \cL''}^{\cL'''} = \sum_{\cL_v} N_{\cL' \cL''}^{\cL_v} N_{\cL \cL_v}^{\cL'''}$. At the level of junction spaces, this requires the existence of an isomorphism $F$, called the {\it associator} or {\it F-symbol} of $\cC$:
\be
\begin{tikzpicture}
    \coordinate (g_hk) at (1.5, 1.2); \coordinate (h_k) at (2.25, 0.6); \coordinate (g_h) at (0.75, 0.6); \coordinate (gh_k) at (1.5, 1.2); \coordinate (g) at (0, 0); \coordinate (h) at (1.5, 0); \coordinate (k) at (3, 0); \coordinate (ghk) at (1.5, 1.8);
    \draw[color=BLUE!70!black, line width=1] (g) node[below] {$\cL$} to (g_h) to (gh_k) to (ghk) node[above] {$\cL'''$};
    \draw[color=BLUE!70!black, line width=1] (h) node[below] {$\cL'$} to (g_h);
    \draw[color=BLUE!70!black, line width=1] (k) node[below] {$\cL''$} to (gh_k);
    \node[BLUE,circle,fill,scale=0.3] at (g_h) {};  \node[BLUE,circle,fill,scale=0.3] at (gh_k) {};

    \node[right] at (3.5, 0.9) {$\ds = \sum_{\cL_v} \left[F^{\cL \cL' \cL''}_{\cL'''}\right]_{\cL_u \cL_v}$};
    \node[above left,BLUE] at (1.25,0.9) {$\cL_u$};
\begin{scope}[shift={(7,0)}]
    \coordinate (g_hk) at (1.5, 1.2); \coordinate (h_k) at (2.25, 0.6); \coordinate (g_h) at (0.75, 0.6); \coordinate (gh_k) at (1.5, 1.2); \coordinate (g) at (0, 0); \coordinate (h) at (1.5, 0); \coordinate (k) at (3, 0); \coordinate (ghk) at (1.5, 1.8);
    \draw[color=BLUE!70!black, line width=1] (k) node[below] {$\cL''$} to (h_k) to (gh_k) to (ghk) node[above] {$\cL'''$};
    \draw[color=BLUE!70!black, line width=1] (h) node[below] {$\cL'$} to (h_k);
    \draw[color=BLUE!70!black, line width=1] (g) node[below] {$\cL$} to (g_hk);
   \node[BLUE,circle,fill,scale=0.3] at (h_k) {};  \node[BLUE,circle,fill,scale=0.3] at (g_hk) {};
     \node[above right,BLUE] at (1.75,0.9) {$\cL_v$};
\end{scope}
\end{tikzpicture}
\ee
The $F$-symbols are strongly constrained by the pentagon equation \cite{Moore:1988qv}, which schematically reads 
\be
\resizebox{0.3\textwidth}{!}{
\begin{tikzpicture}[scale=0.5]
\coordinate (a) at (0,0); \coordinate (b) at (1,0); \coordinate (c) at (2,0); \coordinate(d) at (3,0); \coordinate (abcd) at (1.5,3); \coordinate (ab) at (0.5,1); \coordinate (bc) at (1.5,1); \coordinate (cd) at (2.5,1); \coordinate (abc) at (1,2) ; \coordinate (bcd) at (2,2); \coordinate (e) at (1.5, 3.5);
\draw[BLUE,thick] (a) -- (abcd); \draw[BLUE,thick] (b) -- (ab); \draw[BLUE,thick] (c) -- (abc); \draw[BLUE,thick] (d) -- (abcd) -- (e);
\node[rotate=30] at (-1.5,0) {$\Leftarrow$}; 
\node[rotate=-30] at (4.5,0) {$\Rightarrow$}; 

\begin{scope}[shift={(5,-3)}]
\coordinate (a) at (0,0); \coordinate (b) at (1,0); \coordinate (c) at (2,0); \coordinate(d) at (3,0); \coordinate (abcd) at (1.5,3); \coordinate (ab) at (0.5,1); \coordinate (bc) at (1.5,1); \coordinate (cd) at (2.5,1); \coordinate (abc) at (1,2) ; \coordinate (bcd) at (2,2); \coordinate (e) at (1.5, 3.5);
\draw[BLUE,thick] (a) -- (abcd); \draw[BLUE,thick] (b) -- (ab); \draw[BLUE,thick] (c) -- (cd); \draw[BLUE,thick] (d) -- (abcd) -- (e);        
\node[rotate=60] at (0,-1) {$\Leftarrow$};
\end{scope}

\begin{scope}[shift={(2,-6)}]
    \coordinate (a) at (0,0); \coordinate (b) at (1,0); \coordinate (c) at (2,0); \coordinate(d) at (3,0); \coordinate (abcd) at (1.5,3); \coordinate (ab) at (0.5,1); \coordinate (bc) at (1.5,1); \coordinate (cd) at (2.5,1); \coordinate (abc) at (1,2) ; \coordinate (bcd) at (2,2); \coordinate (e) at (1.5, 3.5);
  \draw[BLUE,thick] (a) -- (abcd) -- (e);   \draw[BLUE,thick] (b) -- (bcd); \draw[BLUE,thick] (c) -- (cd); \draw[BLUE,thick] (d) -- (abcd);
\end{scope}

\begin{scope}[shift={(-5,-3)}]
 \coordinate (a) at (0,0); \coordinate (b) at (1,0); \coordinate (c) at (2,0); \coordinate(d) at (3,0); \coordinate (abcd) at (1.5,3); \coordinate (ab) at (0.5,1); \coordinate (bc) at (1.5,1); \coordinate (cd) at (2.5,1); \coordinate (abc) at (1,2) ; \coordinate (bcd) at (2,2); \coordinate (e) at (1.5, 3.5);
     \draw[BLUE,thick] (a) -- (abcd) -- (e);   \draw[BLUE,thick] (b) -- (bc); \draw[BLUE,thick] (c) -- (abc); \draw[BLUE,thick] (d) -- (abcd);
     \node[rotate=-60] at (3,-1) {$\Rightarrow$};
\end{scope}

\begin{scope}[shift={(-2,-6)}]
 \coordinate (a) at (0,0); \coordinate (b) at (1,0); \coordinate (c) at (2,0); \coordinate(d) at (3,0); \coordinate (abcd) at (1.5,3); \coordinate (ab) at (0.5,1); \coordinate (bc) at (1.5,1); \coordinate (cd) at (2.5,1); \coordinate (abc) at (1,2) ; \coordinate (bcd) at (2,2); \coordinate (e) at (1.5, 3.5);
     \draw[BLUE,thick] (a) -- (abcd) -- (e);   \draw[BLUE,thick] (b) -- (bcd); \draw[BLUE,thick] (c) -- (cd); \draw[BLUE,thick] (d) -- (abcd);
     \node at (3.5,2) {$\Rightarrow$};
\end{scope}
\end{tikzpicture}
}
\ee
It is known that solutions to these equations modulo gauge transformations (i.e. unitary change of the basis for $V_{\cL \cL'}^{\cL''}$) do not admit smooth deformations, implying that fusion categories are `rigid' structures \cite{etingof2005fusion}.
Below, we often use the so-called {\it tetrahedral symbols}, defined by
\be
\begin{bmatrix}
    \cL_1 & \cL_2 & \cL_3 \\
    \cL_4 & \cL_5 & \cL_6
\end{bmatrix} = \frac{1}{\sqrt{\d_{\mathcal{L}_3} \d_{\mathcal{L}_6}}} \left[ F^{\cL_1 \cL_2 \cL_4}_{\cL_5} \right]_{\cL_3 \cL_6} \, .
\ee

\subsection{Symmetric gapped phases and module categories} \label{subsec:modulecat} 
To understand the interplay between solitons (kinks) and generalized symmetries, we need to examine the symmetry action on gapped vacua within the framework of $\cC$-symmetric TFT (see e.g. \cite{Huang:2021zvu} for in-depth discussions and \cite{Bhardwaj:2023idu}). This mathematical background, reviewed below, enables us to generalize the findings of \cite{Copetti:2024rqj} to situations where non-invertible symmetries are partially broken in the IR. Such scenarios occur, for example, in the RG flows starting from D-series minimal models, including a specific case of $Z+Z^{\ast}$ deformation of the 3-state Potts CFT, discussed in Section \ref{sec: Bootstrap}.

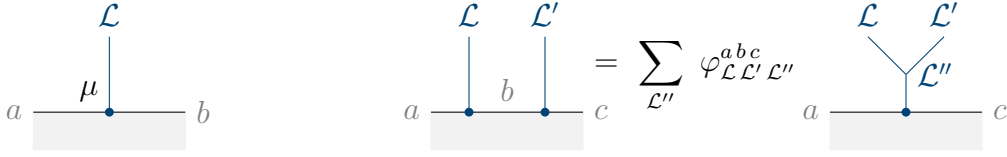
\begin{figure}
     \centering
   \begin{tikzpicture}
\draw[color=white!90!gray, fill=white!90!gray] (0,-0.5) -- (2,-0.5) -- (2,0) -- (0,0) -- cycle;
\draw (0,0) node[left,gray] {$a$} -- (2,0) node[right,gray] {$b$};
\draw[BLUE] (1,0) node[BLUE,circle,fill,scale=0.3] {} -- (1,1) node[above] {$\cL$};
\node[above left] at (1,0) {$\mu$};
\end{tikzpicture} \hspace{2cm} \begin{tikzpicture}
\draw[color=white!90!gray, fill=white!90!gray] (0,-0.5) -- (2,-0.5) -- (2,0) -- (0,0) -- cycle;
\draw (0,0) node[left,gray] {$a$} -- (2,0) node[right,gray] {$c$}; \node[gray,above] at (1,0) {$b$};
\draw[BLUE] (0.5,0) node[BLUE,circle,fill,scale=0.3] {} -- (0.5,1) node[above] {$\cL$};
\draw[BLUE] (1.5,0) node[BLUE,circle,fill,scale=0.3] {} -- (1.5,1) node[above] {$\cL'$}; \node[right] at (2,0.5) {$\ds = \ \sum_{\cL''} \ \varphi^{a \, b \, c}_{\cL \, \cL' \, \cL''}$}; 
\begin{scope}[shift={(5.25,0)}]
\draw[color=white!90!gray, fill=white!90!gray] (0,-0.5) -- (2,-0.5) -- (2,0) -- (0,0) -- cycle;
\draw (0,0) node[left,gray] {$a$} -- (2,0) node[right,gray] {$c$};    
\draw[BLUE] (1,0) node[BLUE,circle,fill,scale=0.3] {} -- (1,0.5) node[ right] {$\cL''$};
\draw[BLUE] (1,0.5) to (0.5,1) node[above] {$\cL$}; \draw[BLUE] (1,0.5) to (1.5,1) node[above] {$\cL'$};
\end{scope}
 \end{tikzpicture}
     \caption{Topological junction defining a module category $\calM$ over $\cC$ (Left). Boundary $F$-symbols implementing fusion of topological lines on the boundary (Right).}
     \label{fig: modulecat}
 \end{figure}

The input data, apart from the symmetry category $\cC$, is a set of boundary conditions $\vert a \rrangle$, which are in one-to-one correspondence with the Hilbert space of the theory on a circle $\cH^{S^1}$.
 The action of the topological lines $\cL$ of $\cC$ on these boundary conditions (b.c.) is encoded in a set of consistent topological junctions between boundaries and lines (see Figure \ref{fig: modulecat}), which endow them with the mathematical structure of a {\it module category} $\calM$ over $\cC$ \cite{etingof2016tensor}.\footnote{See e.g. \cite{Ostrik:2002ohv,Kitaev:2011dxc,Lan:2013wia,Bridgeman:2019wyu,Bullivant:2020xhy} for earlier studies of this type of representation.} These junction spaces are also vector spaces and we will denote their dimension by $(n_\cL)_a^b$. As before we will suppress the index for these junctions since $(n_{\cL})_a^b = 0, \, 1$ in all our examples. Multiplication over these indices is denoted by $\cdot$ in the formulas below. 
The fusion product of bulk lines attached to a boundary is given by a boundary $F$-symbol\footnote{Our convention in labelling the indices differs from other works, e.g. $\varphi^{abc}_{\cL\cL'\cL''}=(\tilde F^c_{\cL'\cL a})^{-1}_{\cL'' b}$ in \cite{Cordova:2024vsq}. In the regular representation $\varphi$ becomes the regular F-symbol $\varphi^{abc}_{\cL\cL'\cL''}={F^{a\cL\cL'}_{c}}_{b\cL''}$.} $\varphi^{a \, b \, c}_{\cL \, \cL' \, \cL''}.$ (Figure \ref{fig: modulecat}),
which is subject to the boundary pentagon equation\footnote{Notice that the boundary $F$-symbols $\varphi_{\cL \cL' \cL''}^{a b c}$ suffer from a large amount of redundancy stemming from a choice of basis of the bulk-boundary junction space. Denoting a unitary change of basis of boundary junctions by $u_{a b}^\cL$ and for bulk junctions by $U_{\cL \cL'}^{\cL''}$ we have that
\be
\varphi_{\cL \cL' \cL''}^{a b c} \simeq \varphi_{\cL \cL' \cL''}^{a b c} \, u_{a b}^\cL u_{b c}^{\cL'} (u^\dagger)_{a c}^{\cL''} (U^\dagger)_{\cL \cL'}^{\cL''} \, . 
\ee
}
\be
\sum_{\cL_u} \left[ F^{\cL \cL' \cL''}_{\cL_v} \right]_{\cL_u \cL'_u} \, \varphi^{ a b c}_{\cL \cL' \cL_u} \cdot \varphi^{a c d}_{\cL_u \cL'' \cL_v} = \varphi^{a b d}_{\cL \cL'_u \cL_v} \cdot  \varphi^{b c d}_{\cL' \cL'' \cL'_u } \, ,
\ee
that comes from the consistency of the following set of moves 
\be
\resizebox{0.3\textwidth}{!}{
\begin{tikzpicture}
    \filldraw[color=white!90!gray] (0,0) -- (2,0) -- (2,-0.5) -- (0,-0.5) -- cycle;
    \draw (0,0) -- (2,0);
    \draw[BLUE] (0.5,0) node[BLUE,circle,fill,scale=0.3] {} -- (0.5,1); \draw[BLUE] (1,0) node[BLUE,circle,fill,scale=0.3] {} -- (1,1); \draw[BLUE] (1.5,0) node[BLUE,circle,fill,scale=0.3] {} -- (1.5,1);
    \node[rotate=45] at (-1,-0.5) {\Large$\overset{\;\varphi}{\Leftarrow}$};
      \node[rotate=-45] at (3,-0.5) {\Large$\overset{\varphi}{\Rightarrow}$};
    \begin{scope}[shift={(-3,-2)}]
       \filldraw[color=white!90!gray] (0,0) -- (2,0) -- (2,-0.5) -- (0,-0.5) -- cycle;
    \draw (0,0) -- (2,0);
    \draw[BLUE] (0.75,0) node[BLUE,circle,fill,scale=0.3] {} -- (0.75,0.5); \draw[BLUE] (0.75,0.5) -- (0.5,1); \draw[BLUE] (0.75,0.5) -- (1,1);
    \draw[BLUE] (1.5,0) node[BLUE,circle,fill,scale=0.3] {} -- (1.5,1);
    \node[rotate=90] at (2,-1) {\Large$\overset{\;\;\varphi}{\Leftarrow}$};
    \end{scope}
      \begin{scope}[shift={(3,-2)}]
       \filldraw[color=white!90!gray] (0,0) -- (2,0) -- (2,-0.5) -- (0,-0.5) -- cycle;
    \draw (0,0) -- (2,0);
    \draw[BLUE] (1.25,0) node[BLUE,circle,fill,scale=0.3] {} -- (1.25,0.5); \draw[BLUE] (1.25,0.5) -- (1,1); \draw[BLUE] (1.25,0.5) -- (1.5,1);
    \draw[BLUE] (0.5,0) node[BLUE,circle,fill,scale=0.3] {} -- (0.5,1);
    \node[rotate=-90] at (0,-1) {\Large$\overset{\varphi}{\Rightarrow}$};
    \end{scope}
    \begin{scope}[shift={(-2,-5)}]
    \filldraw[color=white!90!gray] (0,0) -- (2,0) -- (2,-0.5) -- (0,-0.5) -- cycle;
    \draw (0,0) -- (2,0);
     \draw[BLUE] (1,0) node[BLUE,circle,fill,scale=0.3] {} -- (1,0.25);
     \draw[BLUE] (1,0.25) -- (0.5,1); 
     \draw[BLUE] (1,0.25) -- (1.5,1);
     \draw[BLUE] (0.825,0.5) -- (1,1);
     \node at (3,0) {\Large$\overset{F}{\Rightarrow}$};
    \end{scope}
      \begin{scope}[shift={(2,-5)}]
    \filldraw[color=white!90!gray] (0,0) -- (2,0) -- (2,-0.5) -- (0,-0.5) -- cycle;
    \draw (0,0) -- (2,0);
     \draw[BLUE] (1,0) node[BLUE,circle,fill,scale=0.3] {} -- (1,0.25);
     \draw[BLUE] (1,0.25) -- (0.5,1); 
     \draw[BLUE] (1,0.25) -- (1.5,1);
     \draw[BLUE] (1.175,0.5) -- (1,1);
    \end{scope}
\end{tikzpicture}
}
\ee
In addition, compatibility between the dimensions of the vector spaces on the two sides of the right figure of Figure \ref{fig: modulecat} implies the relation
\be
\sum_{b} (n_\cL)_a^b (n_{\cL'})_b^c = \sum_{\cL''} N_{\cL \cL'}^{\cL''} (n_{\cL''})_a^c \, .
\ee
Different solutions $(n_\cL)_a^b$ to these equations are known as  NIM (non-negative integer matrix) representations of the module category. 
The analogue of the bulk quantum dimension is given by the relative Euler terms $g_a$ of the boundary conditions. These can be defined as the partition function of the TFT on the disk with b.c. $|a\rrangle$:
\bea\label{eq:gdef}
\begin{tikzpicture}
    \node[left] at (-1.5,0) {$g_a =$};
    \filldraw[color=white!90!gray] (-1.5,-1.5) -- (1.5,-1.5) -- (1.5,1.5) -- (-1.5,1.5) -- cycle;
\filldraw[color=black, fill=white] (0,0) circle (1);
\node[gray,right] at (1,0) {$a$};
\end{tikzpicture}
\eea
Their overall scale is unphysical due to the finite Euler counterterm on the disk, but their ratios $g_a/g_b$ are physical observables.

\subsection{Symmetry action on the kink Hilbert space}\label{ssec: symkink}
Having reviewed the symmetry actions on gapped vacua, we now discuss the symmetry actions on kinks that interpolate between different vacua. 

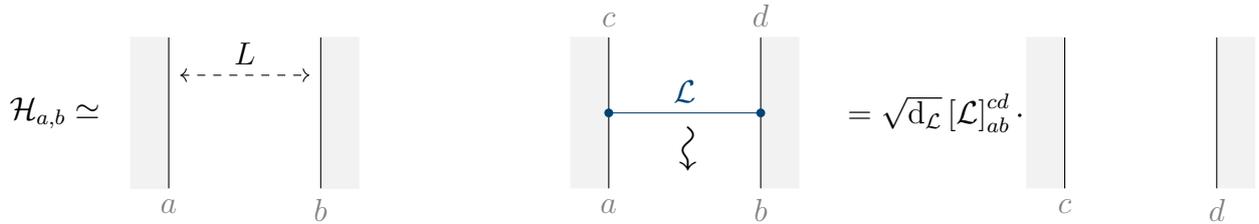
\begin{figure}[t]
    \centering
    \begin{tikzpicture}
    \node at (-1.5,1) {$\cH_{a,b} \simeq$};
    \filldraw[color=white!90!gray] (0,0) node[below,gray] {$a$} -- (0,2) -- (-0.5,2) -- (-0.5,0) -- cycle;
     \filldraw[color=white!90!gray] (2,0) node[below,gray] {$b$} -- (2,2) -- (2.5,2) -- (2.5,0) -- cycle;
     \draw (0,0) -- (0,2); \draw (2,0) -- (2,2);
     \draw[<->, dashed] (0.15,1.5) -- (1.85,1.5);
     \node[above] at (1,1.5) {$L$};
\end{tikzpicture} \hspace{2.5cm} \begin{tikzpicture}
\filldraw[color=white!90!gray, fill=white!90!gray] (-0.5,0) -- (-0.5,2) -- (0,2) -- (0,0);
\draw[black] (0,0) node[below,gray] {$a$} -- (0,2) node [above,gray] {$c$};
\filldraw[color=white!90!gray, fill=white!90!gray] (2.5,0) -- (2.5,2) -- (2,2) -- (2,0);
\draw[black] (2,0) node[below,gray] {$b$} -- (2,2) node[above,gray] {$d$};
\draw[BLUE] (0,1) node[BLUE,circle,fill,scale=0.3] {} -- (2,1) node[BLUE,circle,fill,scale=0.3] {};
\node[above,BLUE] at (1,1) {$\cL$}; 
\node[rotate=-90] at (1,0.5) {\Large $\leadsto$};
\node[right] at (3,1) {$= \sqrt{\d_\cL} \left[ \cL\right]_{a b}^{c d} \cdot $};
\begin{scope}[shift={(6,0)}]
\filldraw[color=white!90!gray, fill=white!90!gray] (-0.5,0) -- (-0.5,2) -- (0,2) -- (0,0);
\draw[black] (0,0) node[below,gray] {$c$} -- (0,2);
\filldraw[color=white!90!gray, fill=white!90!gray] (2.5,0) -- (2.5,2) -- (2,2) -- (2,0);
\draw[black] (2,0) node[below,gray] {$d$} -- (2,2) ;    
\end{scope}
\end{tikzpicture}
    \caption{The kink Hilbert space as an interval Hilbert space (Left). Symmetry action on $\cH_{ab}$ by pushing down the $\cL$ line. The dots denote the topological junctions of the module category $\calM$ of vacua. }
    \label{fig: hab}
\end{figure}

\paragraph{Review of \cite{Cordova:2024vsq}.} We start by briefly reviewing \cite{Cordova:2024vsq}. Following them, we will describe a state in the kink Hilbert space $\vert \psi_{a,b} \rangle \in \cH_{a b}$ as the $L \to \infty$ limit\footnote{More precisely the dimensionless parameter $ML \gg 1$, where $M$ is the mass of the lightest kink.} of the strip Hilbert space with boundary conditions $a,b$ at the two ends. These should be identified with vacua of the IR gapped phase, see Figure \ref{fig: hab}.
An excitation in this Hilbert space describes a solitonic field configuration interpolating between the vacua $a$ at $x \to -\infty$ and $b$ at $x \to + \infty$ --- that is--- a massive kink $K_{ab}$ of the theory. We therefore identify the strip Hilbert space with the Hilbert space of kinks. For a special case of $a=b$, the excitations do not change the vacuum, and they are often called \emph{breathers} $B_a$. 

A topological line $\cL$ stretching between the left and right boundaries defines a map $\left[ \cL \right]_{ab}^{cd}: \, \cH_{ab} \to \cH_{cd}$ by being pushed downwards, as shown in Figure \ref{fig: hab}.
Choosing the bases $|\psi_{ab}\rangle \in \cH_{ab}$ and $|\psi_{cs}\rangle \in \cH_{cd}$, this reads
\be\label{eq:symactionstrip}
\cL |\psi_{ab}\rangle = \sum_{\psi_{cd}} \sqrt{\d_{\cL}}\left[ \cL \right]_{ab}^{cd} |\psi_{cd}\rangle \, . 
\ee
Here we introduced the factor of $\sqrt{\d_\cL}$ for later convenience. The composition of $\cL$ and $\cL'$ actions can be evaluated by applying the definition twice or by using the boundary $F$-symbols:
\bea
\resizebox{.65\textwidth}{!}{
\begin{tikzpicture}[scale=0.8]
 \filldraw[color=white!90!gray, fill=white!90!gray] (-0.5,0) -- (-0.5,2) -- (0,2) -- (0,0);
\draw[black] (0,0) node[below,gray] {$a$} -- (0,2) node [above,gray] {$e$};
\filldraw[color=white!90!gray, fill=white!90!gray] (2.5,0) -- (2.5,2) -- (2,2) -- (2,0);
\draw[black] (2,0) node[below,gray] {$b$} -- (2,2) node[above,gray] {$f$};
\node[gray,left] at (0,1) {$c$}; \node[gray,right] at (2,1) {$d$}; 
\draw[BLUE] (0,1.5) node[BLUE,circle,fill,scale=0.3] {} -- (2,1.5) node[BLUE,circle,fill,scale=0.3] {};
\node[above,BLUE] at (1,1.5) {$\cL'$};    
\draw[BLUE] (0,0.5) node[BLUE,circle,fill,scale=0.3] {} -- (2,0.5) node[BLUE,circle,fill,scale=0.3] {};
\node[above,BLUE] at (1,0.5) {$\cL$};  
%{ \varphi^{\cL a e}_{\cL'}}_{c \cL''}\, {\varphi^{\cL b f}_{\cL'}}_{d \cL''}
\begin{scope}[shift={(-5,-3.5)}]H
\node[left] at (-0.5,1) {$\ds \sum_{\cL''} \varphi^{a c e}_{\cL \cL' \cL''} \varphi^{c d f}_{\cL \cL' \cL''} $};
 \filldraw[color=white!90!gray, fill=white!90!gray] (-0.5,0) -- (-0.5,2) -- (0,2) -- (0,0);
\draw[black] (0,0) node[below,gray] {$a$} -- (0,2) node [above,gray] {$e$};
\filldraw[color=white!90!gray, fill=white!90!gray] (2.5,0) -- (2.5,2) -- (2,2) -- (2,0);
\draw[black] (2,0) node[below,gray] {$b$} -- (2,2) node[above,gray] {$f$};
\draw[BLUE] (0,1) node[BLUE,circle,fill,scale=0.3] {} -- (0.5,1); 
\draw[BLUE] (2,1) node[BLUE,circle,fill,scale=0.3] {} -- (1.5,1);
\node[above right,BLUE] at (-0.1,1.1) {$\cL''$};
\draw[BLUE] (1,1) circle (0.5);
\node[above,BLUE] at (1,1.5) {$\cL'$};    
\node[below,BLUE] at (1,0.5) {$\cL$};      
\end{scope}
\begin{scope}[shift={(5,-3.5)}]
   \filldraw[color=white!90!gray, fill=white!90!gray] (-0.5,0) -- (-0.5,2) -- (0,2) -- (0,0);
\draw[black] (0,0) node[below,gray] {$c$} -- (0,2) node [above,gray] {$e$};
\filldraw[color=white!90!gray, fill=white!90!gray] (2.5,0) -- (2.5,2) -- (2,2) -- (2,0);
\draw[black] (2,0) node[below,gray] {$d$} -- (2,2) node[above,gray] {$f$};
\draw[BLUE] (0,1) node[BLUE,circle,fill,scale=0.3] {} -- (2,1) node[BLUE,circle,fill,scale=0.3] {};
\node[above,BLUE] at (1,1) {$\cL'$}; 
\node[left] at (-0.5,1) {$\sqrt{\d_\cL}\left[ \cL\right]_{a b}^{c d} \cdot $}; 
\end{scope}

\begin{scope}[shift={(5,-8)}]
   \filldraw[color=white!90!gray, fill=white!90!gray] (-0.5,0) -- (-0.5,2) -- (0,2) -- (0,0);
\draw[black] (0,0) node[below,gray] {$e$} -- (0,2) ;
\filldraw[color=white!90!gray, fill=white!90!gray] (2.5,0) -- (2.5,2) -- (2,2) -- (2,0);
\draw[black] (2,0) node[below,gray] {$f$} -- (2,2);
\node[left] at (-0.5,1) {$\sqrt{\d_{\cL} \d_{\cL'}}\left[ \cL' \right]_{cd}^{ef}\left[ \cL\right]_{a b}^{c d} \cdot $}; 
\end{scope}

\begin{scope}[shift={(-5,-8)}]
   \filldraw[color=white!90!gray, fill=white!90!gray] (-0.5,0) -- (-0.5,2) -- (0,2) -- (0,0);
\draw[black] (0,0) node[below,gray] {$a$} -- (0,2) node [above,gray] {$e$};
\filldraw[color=white!90!gray, fill=white!90!gray] (2.5,0) -- (2.5,2) -- (2,2) -- (2,0);
\draw[black] (2,0) node[below,gray] {$b$} -- (2,2) node[above,gray] {$f$};
\draw[BLUE] (0,1) node[BLUE,circle,fill,scale=0.3] {} -- (2,1) node[BLUE,circle,fill,scale=0.3] {};
\node[above,BLUE] at (1,1) {$\cL''$}; 
\node[left] at (-0.5,1) {$\ds \sum_{\cL''} \varphi^{a c e}_{\cL \cL' \cL''} \varphi^{c d f}_{\cL \cL' \cL''} \sqrt{\frac{\d_\cL \d_{\cL'}}{\d_{\cL''}}}$}; 
\end{scope}
\node[rotate=-135] at (-1.5,-1) {\Large$\Rightarrow$};
\node[rotate=-45] at (3.5,-1) {\Large$\Rightarrow$};
\node[rotate=-90] at (-4,-5) {\Large$\Rightarrow$};
\node[rotate=-90] at (6,-5) {\Large$\Rightarrow$};
\end{tikzpicture}
}
\eea
The consistency between the two leads to the identity,
\be
\left[ \cL' \right]_{ c d}^{e f} \cdot \left[ \cL \right]_{ab}^{c d} = \sum_{\cL ''} \varphi^{a c e}_{\cL \cL' \cL''} \varphi^{c d f}_{\cL \cL' \cL''} \left[ \cL'' \right]_{ab}^{ef}\,. \label{eq: symmaction}
\ee
Kinks $K^v_{a,b}$ and breathers $B_a^{v}$ form irreducible representations (irreps) $v$ of this algebra. Importantly, a multiplet can contain both kinks $K_{a,b}^v$ and breathers $B_a^v$, leading to degeneracy in mass \cite{Cordova:2024vsq}. 

\paragraph{Relation to dual category.}
We now build on \cite{Cordova:2024vsq} and give a simple characterization of such irreps using a relation to the dual category. The key idea is to interpret a module category $\cM$ as an interface between a QFT with symmetry $\cC$ and a QFT with a {\it dual category} symmetry $\cC^*_{\calM}$, obtained from $\cC$ by generalized orbifolding.\footnote{To be precise, generalized orbifolding corresponds to gauging a symmetric separable Frobenius algebra object $\cA$ of $\cC$ \cite{Fuchs:2002cm,Komargodski:2020mxz}. The dual symmetry $\cC^*_{\calM}$ is identified with the category of $\cA$-$\cA$ bimodules ${}_\cA \cC_{\cA}$. We will not need details of this construction for the purpose of this paper. However see \cite{Choi:2023xjw} for a physics motivated discussion about the relation between gauging and module categories and \cite{Diatlyk:2023fwf} for a recent rephrasing of the gauging prescription.}
Mathematically, the interface provides 
 a category of bimodules over $\cC_{\calM}^*$-$\cC$. In practice, it introduces ``dual" boundary $F$-symbols $\left(\varphi^*\right)_{v v' v''}^{abc}$ and an isomorphism $\left[\cL ; v \right]$ allowing us to commute junctions on the two sides:
\bea\label{eq:Lcommute}
\begin{tikzpicture}
\filldraw[color=white!90!gray] (0,0) -- (0,2) -- (-1,2) -- (-1,0) -- cycle; 
\draw[black] (0,0) node[below,gray] {$b$} -- (0,2) node[above,gray] {$c$};
\node[left,gray] at (0,1) {$a$};
\draw[BLUE] (1,1.5) node[right] {$\cL$} -- (0,1.5) node[BLUE,circle,fill,scale=0.3] {};
\draw[YELLOW] (-1,0.5) node[left] {$v$} -- (0,0.5) node[YELLOW,circle,fill,scale=0.3] {};
\node[right] at (1.5,1) {$\ds = \sum_{d} \left[\cL ; v \right]_{a b}^{c d}$}; 
\begin{scope}[shift={(6,0)}]
    \filldraw[color=white!90!gray] (0,0) -- (0,2) -- (-1,2) -- (-1,0) -- cycle; 
\draw[black] (0,0) node[below,gray] {$b$} -- (0,2) node[above,gray] {$c$};
\node[left,gray] at (0,1) {$d$};
\draw[BLUE] (1,0.5) node[right] {$\cL$} -- (0,0.5) node[BLUE,circle,fill,scale=0.3] {};
\draw[YELLOW] (-1,1.5) node[left] {$v$} -- (0,1.5) node[YELLOW,circle,fill,scale=0.3] {};
\end{scope}
\end{tikzpicture}
\eea
Similarly to how the pentagon equation is derived, one can show that $\left[ \cL ; v \right]$ satisfies \eqref{eq: symmaction}, which suggests a relationship between a dual category and a  representation of the algebra \eqref{eq: symmaction}.  The connection can be made more explicit by computing the action of $\mathcal{L}$ using the following diagram,
\bea
\begin{tikzpicture}
  \draw[color=white!90!gray, fill=white!90!gray] (0,-1) -- (2,-1) -- (2,0) -- (0,0) -- cycle;
\draw[gray] (0,0) node[left] {$c$} -- (2,0) node[right] {$d$};
\draw[BLUE] (0.25,0) node[above right,gray] {$a$} arc (180:0:0.75 and 0.75) node[above left,gray] {$b$}; \node[BLUE,above] at (1,0.75) {$\cL$};
\draw[YELLOW] (1,0) node[YELLOW,circle,fill,scale=0.3] {} -- (1,-1) node[below] {$v$};    
\node[right] at (2.5,0) {$= \sqrt{\dfrac{\d_\cL \,g_b}{g_d}} \left[ \cL ; v \right]_{ab}^{cd} $};
\begin{scope}[shift={(6.3,0)}]
 \draw[color=white!90!gray, fill=white!90!gray] (0,-1) -- (2,-1) -- (2,0) -- (0,0) -- cycle;
\draw[gray] (0,0) node[left] {$c$} -- (2,0) node[right] {$d$};
\draw[YELLOW] (1,0) node[YELLOW,circle,fill,scale=0.3] {} -- (1,-1) node[below] {$v$};    
\end{scope}
\end{tikzpicture}
\eea
where the extra $\sqrt{\frac{\d_{\mathcal{L}} g_b}{g_d}}$ factor comes from the expectation value of a half loop of $\mathcal{L}$ anchored on the interface: %\cite{Huang:2021zvu}
\be\label{eq:bdybubbleremoval}
\begin{tikzpicture}
  \draw[color=white!90!gray, fill=white!90!gray] (0,-1) -- (2,-1) -- (2,0) -- (0,0) -- cycle;
\draw[gray] (0,0) node[left] {$b$} -- (2,0) node[right] {$c$};
\draw[BLUE] (0.25,0)  arc (180:0:0.75 and 0.75); \node[BLUE,above] at (1,0.75) {$\cL$};
\node[gray , above] at (1,0) {$a$};
%\draw[YELLOW] (1,0) node[YELLOW,circle,fill,scale=0.3] {} -- (1,-1) node[below] {$v$};    
\node[right] at (2.5,0) {$= \sqrt{\dfrac{\d_\cL \,g_a}{g_b}} \, \delta_{bc}$};
\begin{scope}[shift={(5.5,0)}]
 \draw[color=white!90!gray, fill=white!90!gray] (0,-1) -- (2,-1) -- (2,0) -- (0,0) -- cycle;
\draw[gray] (0,0) node[left] {$b$} -- (2,0) node[right] {$b$};
%\draw[YELLOW] (1,0) node[YELLOW,circle,fill,scale=0.3] {} -- (1,-1) node[below] {$v$};    
\end{scope}
\end{tikzpicture}
\ee
This follows from a particular choice of normalization for junction vector spaces, explained in  Appendix \ref{app:junctions}.

By performing a radial quantization around a junction of $v$, we can interpret this as the action of $\mathcal{L}$ on the strip Hilbert space, i.e. the action on kinks\footnote{Note that $\sqrt{\d_{\cL}g_b/g_d}[\cL;v]$ cannot be directly identified with a matrix element of $\mathcal{L}$ in $\cH_{a,b}$ since kink states $K_{a,b}$ are not properly normalized here. See \cite{Copetti:2024rqj} for related discussions for the regular representation. The normalization of kink states will be taken into account when we derive the Ward identity in Section \ref{sec: symtft}.}: 
\be
\cL \cdot K_{a,b}= \sum_{K_{cd}}  \sqrt{\dfrac{\d_\cL \,g_b}{g_d}} \left[ \cL ; v \right]_{ab}^{cd} K_{c , d} \, .
\ee
Comparing this with \eqref{eq:symactionstrip}, we find that $\sqrt{g_b/g_d}[\cL; v]$ and $[\cL]$  can be identified. In addition, it is straightforward to check the extra factor $\sqrt{g_b/g_d}$ does not affect the relation \eqref{eq: symmaction}, and therefore $\sqrt{g_b/g_d}[\cL; v]$ provides a representation of the algebra \eqref{eq: symmaction}. In particular, the representation is irreducible if the line $v$ is a simple line of $\cC_{\calM}^*$.

This establishes that simple lines in the dual category $\cC^*_{\calM}$ correspond to irreps of the algebra \eqref{eq: symmaction}. In fact, the converse is also true \cite{ostrik2003module}: irreps of \eqref{eq: symmaction} are in one-to-one correspondence
%\footnote{A straightforward proof of this correspondence can be provided using the SymTFT framework, which we will discuss in Section \ref{sec: symtft}. } 
with simple lines in $\cC^*_{\mathcal{M}}$. Based on this correspondence,
we will represent the restriction of the kink Hilbert space $\cH_{ab}$ to a given representation $v$ by a path integral with a dual $v$ line insertion at the bottom\footnote{Here we are talking about the full QFT, not just IR TQFT. Thus the bottom junction should not be thought of as topological.}:
\bea
\resizebox{0.25\textwidth}{!}{
\begin{tikzpicture}
    \node at (-1.5,0.75) {$\cH_{a,b}^v \simeq$};
    \filldraw[color=white!90!gray] (0,0)  -- (0,2) node[above,gray] {$a$} -- (-0.5,2) -- (-0.5,0) -- cycle;
     \filldraw[color=white!90!gray] (2,0) -- (2,2)  node[above,gray] {$b$} -- (2.5,2) -- (2.5,0) -- cycle;
       \filldraw[color=white!90!gray] (-0.5,0) -- (0,0)  arc(-180:0:1 and 0.75) -- (2.5,0) -- (2.5,-1) -- (-0.5,-1) -- (-0.5,0);
       \draw (0,2) -- (0,0) arc(-180:0:1 and 0.75) -- (2,2);
       \draw[YELLOW] (1,-0.75) node[YELLOW,circle,fill,scale=0.3] {} -- (1,-1) node[below] {$v$};
       \node[above] at (1,-0.75) {$K^v_{a,b}$};
\end{tikzpicture}
}
\eea
 A SymTFT description of this picture will be given in Section \ref{sec: symtft}.

The relation \eqref{eq: symmaction} is one of the consistency conditions for the bimodule category. In addition, compatibility with the fusion structure of $v$ lines
\be \label{eq: vfusion}
v \times v' = \sum_{v''} {N^*}_{v v'}^{v''} \, v'' \, ,
\ee
leads to other consistency conditions like
\bea\label{eq:pentagonlikephistar}
 \left[\cL ; v \times v'\right]_{a b}^{c d} = \sum_{v''} {N^*}_{v v'}^{v''} \left[ \cL ; v'' \right]_{ab}^{cd} \, ,\\ 
 \sum_{f} \left( \varphi^* \right)_{v' v'' v }^{d f e} \, \left[ \cL ; v' \right]_{a b}^{d f} \, \left[\cL ; v'' \right]_{ c b}^{e f}\, \sqrt{\frac{g_b g_e}{g_f g_c}} &= \left( \varphi^* \right)_{v' v'' v }^{a b c} \, \left[ \cL ; v \right]_{a c}^{d e} \, .
\eea

\paragraph{Selection rules.}The structure we have explained gives strong constraints on the spectrum of bound states, as the dual category $\cC_{\calM}^*$ is also endowed with its own fusion ring structure \eqref{eq: vfusion}.
For example, we may wonder if, given a multiplet of kinks $K^v_{a,b}$, the breathers in this multiplet can be realized as bound states of kinks:
\be
B^v_a \sim K_{a,b}^v K_{b,a}^v \, .
\ee
A necessary condition for this to happen is ${N^*}_{v v}^{v} \neq 0$. If  ${N^*}_{v v}^{v} = 0$ instead, the cubic coupling between two kinks and the breather must vanish. We will encounter an example of this phenomenon related to the deformation of 3-state Potts CFT in the study of Fibonacci-symmetric S-matrices in Section \ref{sec: Bootstrap}.

\paragraph{Remarks.}
\begin{itemize}
\item
While our formulas are simple and general, computing the data $\varphi, \varphi^*$ and $[\cL;v]$ for a given $(\cC, \, \calM)$, can be challenging. For small categories $\cC$, this can be done semi-analytically, but systematic implementation, such as in computer-algebra programs, is hindered by large gauge redundancy, making it difficult to identify physically distinct solutions. Comprehensive results exist in the literature for specific symmetries like discrete groups \cite{davydov2010modular} and Tambara-Yamagami categories (generalizations of Ising symmetry) \cite{meir2012module}. However, for other categories, such as the Haagerup category $\cH_3$ mentioned in the introduction, the full structure is still undetermined \cite{Huang:2021ytb}.
Another useful tool for computing the boundary $F$-symbols $\varphi$ and $\varphi^*$ is the internal-Hom construction of Ostrik \cite{ostrik2003module}, which relates the boundary $F$-symbols to the data of the generalized gauging $\cA$ connecting $\cC$ with $\cC^*_\calM$. See also Appendix A of \cite{Choi:2023xjw} for explicit expressions.
\item The structures discussed above do not directly determine to which symmetry multiplet $K^v$  excitations of a given theory belong. This information needs to be manually input in the bootstrap analysis. Typically, it can be inferred by studying allowed field configurations between vacua, which encode the dual representation coefficients $(n_v)_a^b$. However, this method can be inadequate especially if some symmetry remains unbroken in the IR (i.e. if $\calM \neq Reg$, the regular module category).\footnote{In many such cases the model is related to one in which the SSB is maximal by a generalized orbifold. One can then study the spectrum of kinks in the orbifolded theory and map it back to the original system.}
An alternative approach that works in some cases is to identify the kink creation operator in the UV theory, which lives in the twisted Hilbert space of  the $v$ line \cite{Copetti:2024rqj}.
However, to the best of our knowledge, there is no universally applicable procedure to extract this information. \end{itemize}

\subsection{Examples}\label{sssec: fibz2}

\paragraph{Regular representation.} The simplest example is the regular module category $\calM = Reg$, which describes the complete spontaneous breaking of the non-invertible symmetry. In this case the boundary conditions are isomorphic to the space of lines, and there exists an ``identity" b.c. $\vert \unit \rrangle$ such that:
\be
\vert \cL \rrangle = \cL \vert \unit \rrangle \, .
\ee
The NIM coefficients are identified with the fusion coefficients of $\cC$: $n_{\cL \cL'}^{\cL''} = N_{\cL \cL'}^{\cL''}$ and the boundary $F$-symbols are just bulk $F$-symbols.  The regular module implements a trivial interface, which means $\cC^*_{\calM} = \cC$ and therefore from \eqref{eq:Lcommute} we see that $[\cL; v]$ is simply the bulk $F$-symbol: 
\be\label{eq:Fregular}
[\cL; v]_{ab}^{cd} = {F^{\cL c v}_b}_{ad} =   \sqrt{\d_a \d_d} \left[ \begin{array}{ccc}
   \cL  & c & a  \\
   v  & b & d
\end{array}  \right] \, .
\ee
The regular representation appears in the study of e.g. the deformation of the unitary minimal models $\cM_{n}$ by the relevant $\phi_{1,3}$ operator. The theory, with the negative sign of the deforming operator, is known to flow to a set of $n-1$ gapped vacua which form the regular representation of the $\cA_n$ category. The interested reader can find more detailed material about these flows and the preserved symmetry in Appendix \ref{app: flows}.

 \paragraph{\texorpdfstring{$\text{Fib}$ and $\text{Fib} \times\bZ_2$}{Fib and FibxZ2}.}
As a second example we consider theories with Fibonacci symmetry:
\be
1, \, W, \, \ \ \ W^2 = 1 + W \, ,
\ee
and its $\bZ_2$ extension $\text{Fib} \times \bZ_2$:
\be
1, \, W , \, \eta , \, W'=\eta W \, .
\ee
The Fibonacci symmetry admits a single module category (the regular one) with two vacua. Kinks between these vacua belong to the $W$ multiplet \cite{Cordova:2024vsq}:
\be \label{eq:WFibmultiplet}
K^W_{1, W} , \, K^W_{W, 1} , \, B^W_W \, ,
\ee
and the breather $B_W^W$ can be interpreted as a bound state of two kinks $K^W_{W,1} \times K^W_{1,W}$.
A more interesting situation appears if one studies RG flows from D-series minimal models. The first non-trivial example is a gapped RG flow from the Potts model triggered by the $Z + Z^*$ deformation \cite{Chang:2018iay}. This has been studied in the integrability literature by using the parafermion description \cite{Fateev:1991bv}. Also, it is the $\bZ_2$-orbifolded version of the $\cA_5$-preserving flow $\cM_{5,4} + \phi_{1,3}$.\footnote{To be precise, $\text{Fib} \times \bZ_2$ describes the fusion ring of the symmetry category. There are various choices of $F$-symbols given this Fusion ring, we will denote them by $\cA_5$ and $\cA_5/\bZ_2$. The symmetry category of Potts is $\cA_5/\bZ_2$.} 
The symmetry preserved in the Potts description is $\text{Fib} \times \bZ_2$ and the theory flows to two gapped vacua ${1, \, W}$. This is not the Regular TFT for the total symmetry, as the $\bZ_2$ is unbroken.
The kink multiplets are labelled by lines in the $\cA_5$ category:
\be
1, \, \widetilde{W}, \, \widetilde{\eta}, \, \widetilde{W}' \, ,
\ee
and the kink is identified with the $\widetilde{W}'$ line, corresponding to the first node of the $A_5$ Dynkin diagram.
The kink multiplet is now:
\be
K^{\widetilde{W}'}_{1 , \, W} , \, K^{\widetilde{W}'}_{W , \, 1} , \, B_W^{\widetilde{W}'} \, .
\ee
However the fusion rule
\be\label{eq:fusionW'}
\widetilde{W}' \times \widetilde{W}' = 1 + \widetilde{W} \, .
\ee
now implies that the breather $B_W^{\widetilde{W}'}$ is \emph{not} a kink-antikink bound state (although, being part of the same multiplet, they are still degenerate in mass). We will see later that this has nontrivial consequence on the structure of the S-matrix for this flow.

\subsection{SymTFT description of kink scattering}\label{sec: symtft}

We now explore kink scattering using the SymTFT framework for boundary conditions, developed in recent works\cite{Choi:2024toappear,CC:boundaries} (see also \cite{Copetti:2024onh} and \cite{Huang:2023pyk}).

The primary advantage of this approach is its ability to decouple the dynamics of the theory from the structure of symmetries, disentangling in our case the symmetry action from the dynamical data of a scattering process. 
This perspective allows us to discuss various symmetry aspects, such as representations and 't Hooft anomalies, independently of the specific QFT involved. Though the formalism is relatively new \cite{Freed:2022qnc}\footnote{Early examples however date back to e.g. \cite{Witten:1998wy,Fuchs:2002cm}.}, it has already been applied to a range of areas, including the analysis of representations \cite{Bhardwaj:2023ayw,Bartsch:2023wvv} and 't Hooft anomalies \cite{Antinucci:2023ezl,Cordova:2023bja} for non-invertible symmetries in higher dimensions, as well as the description of symmetric gapped phases and their transitions in (1+1) dimensions \cite{Chatterjee:2022tyg,Huang:2023pyk,Bhardwaj:2023idu,Bhardwaj:2024qrf}. 
 In theories with a holographic dual, the SymTFT can be derived by reduction over the compact dimensions, as pioneered in \cite{Apruzzi:2021nmk}. See also e.g. \cite{DelZotto:2022ras,Apruzzi:2022rei,Antinucci:2022vyk}.
While primarily applied to discrete symmetries, extensions to continuous symmetries have been discussed recently\cite{Antinucci:2024zjp,Brennan:2024fgj,Bonetti:2024cjk,Apruzzi:2024htg,Antinucci:2024bcm}.

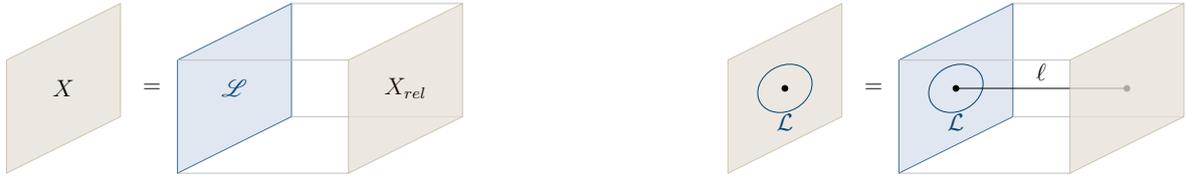
\begin{figure}[t]
    \centering
    \resizebox{.75\width}{!}{
   \begin{tikzpicture}
 \draw[color=white!70!YELLOW, fill=white!90!YELLOW, opacity=0.75] (0,0) -- (2,1) -- (2,3) -- (0,2) -- cycle;   \node at (1,1.5) {$X$};
\node[right] at (2.25,1.5) {$=$};
\begin{scope}[shift={(6,0)}]
     \draw[color=BLUE, fill=white!90!BLUE, opacity=0.75] (-3,0) -- (-1,1) -- (-1,3) -- (-3,2) -- cycle; \node[BLUE] at (-2,1.5) {$\calL$};
   \draw[white!75!black] (0,0) -- (-3,0);    \draw[white!75!black] (2,1) -- (-1,1);   \draw[white!75!black] (2,3) -- (-1,3);   \draw[white!75!black] (0,2) -- (-3,2);        \draw[color=white!70!YELLOW, fill=white!90!YELLOW, opacity=0.75] (0,0) -- (2,1) -- (2,3) -- (0,2) -- cycle;   \node[YELLOW] at (1,1.5) {$X_{rel}$};
\end{scope}
\end{tikzpicture} }
\hspace{3cm}
\resizebox{.75\width}{!}{
\begin{tikzpicture}
     \draw[color=white!70!YELLOW, fill=white!90!YELLOW, opacity=0.75] (0,0) -- (2,1) -- (2,3) -- (0,2) -- cycle;  \draw[rotate around={30:(1,1.5)},BLUE] (1,1.5) ellipse (0.5 and 0.4);
       \node[BLUE] at (1,0.9) {$\cL$};
       \draw[fill=black] (1,1.5) circle (0.05);
\node[right] at (2.25,1.5) {$=$};
\begin{scope}[shift={(6,0)}]
       \draw[color=BLUE, fill=white!90!BLUE, opacity=0.75] (-3,0) -- (-1,1) -- (-1,3) -- (-3,2) -- cycle; 
       \draw[rotate around={30:(-2,1.5)},BLUE] (-2,1.5) ellipse (0.5 and 0.4);
       \node[BLUE] at (-2,0.9) {$\cL$};
       \draw[white!75!black] (0,0) -- (-3,0);    \draw[white!75!black] (2,1) -- (-1,1);   \draw[white!75!black] (2,3) -- (-1,3);  
       \draw (1,1.5) -- (-2,1.5); \draw[fill=black] (1,1.5) circle (0.05); \draw[fill=black] (-2,1.5) circle (0.05);
       \node[above] at (-0.5,1.5) {$\ell$};
       \draw[white!75!black] (0,2) -- (-3,2);        \draw[color=white!70!YELLOW, fill=white!90!YELLOW, opacity=0.75] (0,0) -- (2,1) -- (2,3) -- (0,2) -- cycle;
----       \end{scope}
\end{tikzpicture} }
    \caption{The SymTFT sandwich construction of a theory $X$ (Left) and its representation of symmetries and charged operators (Right).}
    \label{fig: SymTFTbasics}
\end{figure}

Below we will not provide a comprehensive review of the SymTFT formalism. The interested reader can refer to \cite{Freed:2022qnc,Apruzzi:2021nmk} for a more complete exposition.\footnote{Also, the interested reader can consult some of the recent papers on the subject, e.g. \cite{Kaidi:2022cpf,Antinucci:2022vyk,Antinucci:2023ezl,Cordova:2023bja,Argurio:2024oym,Lu:2024lzf,Franco:2024mxa,Nardoni:2024sos,Bhardwaj:2024qrf,DelZotto:2024tae,Brennan:2024fgj,Antinucci:2024zjp,Apruzzi:2023uma,Kaidi:2023maf,Antinucci:2024bcm}.} More detailed analysis of the interplay between SymTFT and boundary conditions will be presented elsewhere \cite{CC:boundaries}.

\subsubsection{SymTFT and boundaries} 
\paragraph{Basics of SymTFT.} Let us briefly introduce the basic concepts. The SymTFT framework separates the dynamical QFT data from the rigid symmetry action by introducing a $(d+1)$-dimensional bulk. This approach is particularly powerful and explicit in $(1+1)$ dimensions, while it is less explored in higher dimensions\cite{Antinucci:2023ezl,Cordova:2023bja}. Below we focus on $(1+1)$ dimensions. We associate to a QFT $X$ with symmetry category $\cC$ a triplet 
\be
\left( \cZ(\cC) , \, \calL  , \, X_{rel}\right),
\ee
where $\cZ(\cC)$ is a 3d TQFT whose spectrum of lines forms the {\it Drinfeld center} of $\cC$ \cite{etingof2016tensor},  $\calL$ is a topological (gapped) boundary condition for $\cZ(\cC)$ and $X_{rel}$ is a free boundary condition coupling the bulk to the dynamical d.o.f. of $X$. The topological boundary condition is technically described by a Lagrangian algebra object $\calL$ in $\cZ(\cC)$.\footnote{A Lagrangian algebra is a line algebra $\calL= \sum_\lambda Z_\lambda \, \lambda$ which is associative, commutative and whose dimension is maximal $\d_\calL^2 = \sum_\lambda \d_\lambda^2$. For a precise description see \cite{Kapustin:2010if,Kapustin:2010hk,Kong:2013aya,Kaidi:2021gbs,Benini:2022hzx}.} This setup provides a ``sandwich" realization of the theory $X$, as illustrated in Figure \ref{fig: SymTFTbasics}. The symmetry $\cC$ is realized by topological line operators confined to the gapped boundary condition $\calL$, while local charged operators (representations of the symmetry) are implemented by lines $\ell \in \calL$ terminating on the gapped surface. The symmetry action then follows from the bulk braiding (see Figure \ref{fig: SymTFTbasics}).

The ``decoupling" between symmetries and dynamical data can be achieved by inserting a resolution of identity of 3d TQFT at an intermediate location along the interval; the procedure known as {\it bulk surgery}. Specifically, for a bulk geometry $\Sigma \times I$, we select a basis $\vert \psi \rangle$ of the $\cZ(\cC)$ Hilbert space $\cH_\Sigma$ on $\Sigma$ (possibly with punctures corresponding to insertions of lines along $I$) and insert a resolution of identity in their path integral: 
\bea
\begin{tikzpicture}[baseline={(0,1.25)},scale=0.75]
 \draw[color=BLUE, fill=white!90!BLUE, opacity=0.75] (-3,0) -- (-1,1) -- (-1,3) -- (-3,2) -- cycle;
   \draw[white!75!black] (0,0) -- (-3,0);    \draw[white!75!black] (2,1) -- (-1,1);   \draw[white!75!black] (2,3) -- (-1,3);   \draw[white!75!black] (0,2) -- (-3,2);        \draw[color=white!70!YELLOW, fill=white!90!YELLOW, opacity=0.75] (0,0) -- (2,1) -- (2,3) -- (0,2) -- cycle; 
   \end{tikzpicture} =\sum_\psi \;
   \begin{tikzpicture}[baseline={(0,1.25)},scale=0.75]
 \draw[color=BLUE, fill=white!90!BLUE, opacity=0.75]  (-3,0) -- (-1,1) -- (-1,3) -- (-3,2) -- cycle;
   \draw[white!75!black] (-1.5,0) -- (-3,0);    \draw[white!75!black] (.5,1) -- (-1,1);   \draw[white!75!black] (.5,3) -- (-1,3);   \draw[white!75!black] (-1.5,2) -- (-3,2);        \draw[color=white!70!black] (-1.5,0) -- (.5,1) -- (.5,3) -- (-1.5,2) -- cycle; 
   \end{tikzpicture}
   \  |\psi\rangle\, \langle\psi|\
   \begin{tikzpicture}[baseline={(0,1.25)},scale=0.75]
 \draw[color=white!70!black]  (-3,0) -- (-1,1) -- (-1,3) -- (-3,2) -- cycle;
   \draw[white!75!black] (-1.5,0) -- (-3,0);    \draw[white!75!black] (.5,1) -- (-1,1);   \draw[white!75!black] (.5,3) -- (-1,3);   \draw[white!75!black] (-1.5,2) -- (-3,2);        \draw[color=white!70!YELLOW, fill=white!90!YELLOW, opacity=0.75] (-1.5,0) -- (.5,1) -- (.5,3) -- (-1.5,2) -- cycle; 
   \end{tikzpicture}
   % \begin{tikzpicture}[baseline={(0,1.25)},scale=0.75]  
   % \draw[color=BLUE, fill=white!90!BLUE, opacity=0.75] (-3,0) -- (-1,1) -- (-1,3) -- (-3,2) -- cycle;
   %  \draw[white!75!black] plot[smooth, tension=1.5] coordinates {(-3,0) (-2,1) (-3,2)}; 
   %     \draw[white!75!black] plot[smooth, tension=1.5] coordinates {(-1,1) (0,2) (-1,3)};
   %     \draw[black] (-2,1) -- (0,2); \node[above] at (-1.5,1.25) {$\psi$};
   %     \begin{scope}[shift={(1.25,0)}]
   %     \draw[white!75!black] plot[smooth, tension=1.5] coordinates {(0,0) (-1,1) (0,2)}; 
   %     \draw[white!75!black] plot[smooth, tension=1.5] coordinates {(2,1) (1,2) (2,3)};
   %     \draw[black] (1,2) -- (-1,1); \node[below] at (-0.5,1.25) {$\psi^\dagger$};
   %       \draw[color=white!70!YELLOW, fill=white!90!YELLOW, opacity=0.75] (0,0) -- (2,1) -- (2,3) -- (0,2) -- cycle;
   %       \end{scope}
   % \end{tikzpicture}
\eea
This separates the initial geometry into two pieces with the left piece encoding all the information on the $\cC$-symmetry, while the right piece carries the dynamical information.

\begin{figure}[h]
    \centering
    \resizebox{.75\width}{!}{
  \begin{tikzpicture}
  \draw[color=white!70!YELLOW, fill=white!90!YELLOW, opacity=0.75] (1,0.5) -- (2,1) -- (2,3) -- (1,2.5) -- cycle;   
\draw[gray, line width =2] (2,1) -- (2,3) node[above] {$a$};
\node at (1.5,1.75) {$X$};
\node[right] at (3,2) {$=$};
\begin{scope}[shift={(7,0)}]
    \draw[color=BLUE, fill=white!90!BLUE, opacity=0.75] (-2,0.5) -- (-1,1) -- (-1,3) -- (-2,2.5) -- cycle;
   \node[BLUE] at (-1.5,1.75) {$\calL$};
     \draw[color=gray, fill=white!90!gray, opacity=0.75] (-1,1) -- (-1,3) node[above] {$a$} -- (2,3) -- (2,1) -- cycle;
   
     \node[gray,above] at (0.5,3) {$\calL_{\calM}$};
       \draw[color=white!70!YELLOW, fill=white!90!YELLOW, opacity=0.75] (1,0.5) -- (2,1) -- (2,3) -- (1,2.5) -- cycle;   
      \node[YELLOW] at (1.5,1.75) {$X_{rel}$};
      \draw[gray, line width =2] (2,1) -- (2,3) node[above,black] {$\partial X_{rel}$};
      \end{scope}
\end{tikzpicture} } \hspace{3cm} \resizebox{.75\width}{!}{\begin{tikzpicture}
    \draw[color=white!70!YELLOW, fill=white!90!YELLOW, opacity=0.75] (1,0.5) -- (2,1) -- (2,3) -- (1,2.5) -- cycle;   
\draw[gray, line width =2] (2,1) node[below] {$a$} -- (2,3) node[above] {$b$};
\draw[fill=black] (2,2) node[right] {$K^v_{a b}$} circle (0.05);
\node[right] at (3,2) {$=$};
\begin{scope}[shift={(6,0)}]
     \draw[color=BLUE, fill=white!90!BLUE, opacity=0.75] (-2,0.5) -- (-1,1) -- (-1,3) -- (-2,2.5) -- cycle;
     \draw[color=gray, fill=white!90!gray, opacity=0.75] (-1,1) node[below] {$a$} -- (-1,3) node[above] {$b$} -- (2,3) -- (2,1) -- cycle;
     \draw[black,thick] (-1,2) -- (2,2); \node[black,above] at (0.5,2) {$v$}; \draw[fill=black] (-1,2) circle (0.05); \draw[fill=black] (2,2) circle (0.05);
      \draw[color=white!70!YELLOW, fill=white!90!YELLOW, opacity=0.75] (1,0.5) -- (2,1) -- (2,3) -- (1,2.5) -- cycle;
      \draw[gray, line width =2,opacity=0.75] (2,1) -- (2,3) node[above,black] {$\partial X_{rel}$};
      \end{scope}
\end{tikzpicture} }
    \caption{SymTFT representation of a boundary condition for $X$ (Left) and its boundary operators (Right).}
    \label{fig: bdysymtft}
\end{figure}
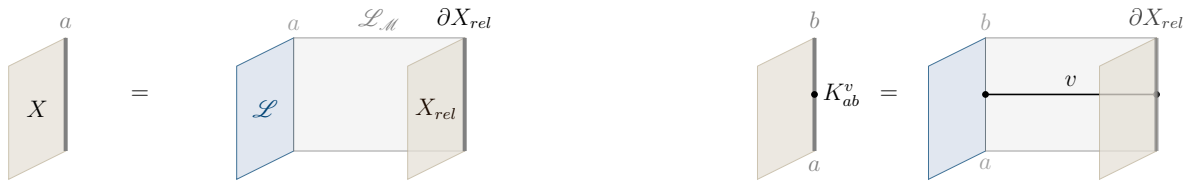

\paragraph{Boundaries and kink multiplet.} The framework above can be extended to include boundaries \cite{Choi:2024toappear,CC:boundaries,Copetti:2024onh,Huang:2023pyk}. A boundary condition corresponding to an irreducible object $a \in \calM$ (where $\calM$ is a module category over $\cC$) for the QFT $X$ is represented by a second topological boundary $\calL_{\calM}$ associated with the dual symmetry $\cC^*_{\calM}$ , which interpolates between $X_{rel}$ and $\calL$, together with the choice of a distinct junction $a$ between $\calL$ and $\calL_\calM$. See Figure \ref{fig: bdysymtft}.

Symmetry lines $\cL$ on the left boundary $\calL$ can terminate topologically on the interfaces $a,b,c ...$ endowing them with the structure of a module category over $\cC$. On the other hand, lines $v \in \cC^*_\calM$ along $\calL_{\calM}$ describe a boundary-changing operator, whose multiplet is labelled by $v$.
This is exactly the kink multiplet $K_{a b}^v$ we have previously described. The symmetry $\cC$ acts on it by half-braiding on the topological boundary $\calL$ \cite{Copetti:2024onh,Choi:2024toappear,CC:boundaries}:
\be
\begin{tikzpicture}[baseline={(0,1)}]
    \draw[color=white!90!BLUE, fill=white!90!BLUE, opacity=0.75] (0,0) -- (1,0) -- (1,2) -- (0,2) --cycle;
    \draw[color=white!90!gray, fill=white!90!gray, opacity=0.75] (1,0) -- (2,0) -- (2,2) -- (1,2) --cycle;
    \draw[color=gray, thick] (1,0) -- (1,2);\node[black,above] at (1.5,1) {$v$};
    \draw[black, thick] (1,1) -- (2,1); 
    \draw[fill=BLUE] (1,1) circle (0.05);
    \draw[BLUE,thick] (1,0.5) arc (-90:-270:0.5 and 0.5); \node[left,BLUE] at (0.5,1) {$\cL$};
    \node[gray,left] at (1,0.75) {\footnotesize$a$};   \node[gray,left] at (1,1.25) {\footnotesize$b$};   \node[gray,left] at (1,0.25) {\footnotesize$c$};   \node[gray,left] at (1,1.75) {\footnotesize$d$};
    \node[above,BLUE] at (0.5,2.05) {$\calL$};
     \node[above,gray] at (1.5,2) {$\calL_\calM$};
\end{tikzpicture} = \sqrt{\dfrac{\d_\cL\, g_b}{g_d}} \left[ \cL ; v \right]_{a b}^{c d}
\begin{tikzpicture}[baseline={(0,1)}]
    \draw[color=white!90!BLUE, fill=white!90!BLUE, opacity=0.75] (0,0) -- (1,0) -- (1,2) -- (0,2) --cycle;
    \draw[color=white!90!gray, fill=white!90!gray, opacity=0.75] (1,0) -- (2,0) -- (2,2) -- (1,2) --cycle;
    \draw[color=gray, thick] (1,0) -- (1,2);\node[black,above] at (1.5,1) {$v$}; \draw[fill=BLUE] (1,1) circle (0.05);
    \draw[black, thick] (1,1) -- (2,1); 
       \node[gray,left] at (1,0.25) {\footnotesize$c$};   \node[gray,left] at (1,1.75) {\footnotesize$d$};
        \node[above,BLUE] at (0.5,2.05) {$\calL$};
     \node[above,gray] at (1.5,2) {$\calL_\calM$};
\end{tikzpicture} 
\ee
To make contact with our previous description of the symmetry action on the interval Hilbert space $\cH_{a b}$, we consider a bulk geometry with two topological boundaries $\calL_\calM$ encasing $\calL$ on both ends. This describes the Hilbert space $\cH_{a, b}$ for the theory $X$. As discussed in the previous subsection, states in this Hilbert space are in irreps of the relevant symmetry algebra \eqref{eq: symmaction}, specified by a line $v\in \mathcal{C}^{\ast}_{\cM}$. In the SymTFT description, this corresponds to considering the following configuration, where the $v$ line stretches along $\calL_{\calM}$ connecting $\calL$ and $X_{rel}$ while the symmetry lines $\mathcal{L}$ sit on the left topological boundary $\calL$:
\bea
\begin{tikzpicture}
    \draw[color=BLUE, fill=white!90!BLUE, opacity=0.75,smooth] (-2,0.5) to[bend right=30, out=-45] (-1.5,0.25) to[bend right=30, out=-45] (-1,1) -- (-1,3) node[above,gray] {$d$} -- (-2,2.5) node[above,gray] {$c$} -- cycle;
    \draw[BLUE, line width=1.5] (-2,1.5) -- (-1,2);
    \draw[fill=BLUE] (-2,1.5) circle (0.05); \draw[fill=BLUE] (-1,2) circle (0.05);
    \node[below,BLUE] at (-1.5, 1.75) {$\cL$};
     \draw[color=gray, fill=white!90!gray, opacity=0.5] (-1,3) -- (-1,1) to[bend left=30, in= 135] (-1.5,0.25) -- (1.5,0.25) to[bend right=30, out=-45] (2,1) -- (2,3) -- cycle;
     \node[gray] at (-0.85,0.75) {$b$};
      \node[below,gray,opacity=0.75] at (-2,0.4) {$a$};
     %\draw[gray,opacity=0.5] (-1,1) node[at={(-0.85,0.75)},gray] {$b$} -- (2,1);
      \draw[color=gray, fill=white!90!gray, opacity=0.5] (-2,2.5) -- (-2,0.5)  to[bend right=30, out=-45] (-1.5,0.25) -- (1.5,0.25) to[bend left=30, out=45] (1,0.5) -- (1,2.5) -- cycle;
       %\draw[gray,opacity=0.5] (-2,0.5) node[below,gray] {$a$} -- (1,0.5);
      \draw[color=white!70!YELLOW, fill=white!90!YELLOW, opacity=0.75] (1,2.5) --  (1,0.5) to[bend right=30, out=-45] (1.5,0.25) to[bend right=30, out=-45] (2,1) -- (2,3) --cycle;
      \draw[gray, line width =1.5,opacity=0.75] (1,2.5) --  (1,0.5) to[bend right=30, out=-45] (1.5,0.25) to[bend right=30, out=-45] (2,1) -- (2,3)  ;
      \draw[black,thick] (-1.5,0.25) -- (1.5,0.25); \node[black,below] at (0,0.25) {$v$}; \draw[fill=black] (-1.5,0.25) circle (0.05); \draw[fill=black] (1.5,0.25) circle (0.05);
\end{tikzpicture}
\eea
In what follows, we will use this description to study implications of categorical symmetries on the scattering of kinks. 
%%%%%%%%%%%%%%%%%%%%%%%%%%%%%%%%%%%%%%%%%%%%%%%%
\subsubsection{S-matrix of kinks and modified crossing}\label{sssec:modifiedcross}
We will now generalize the results of \cite{Copetti:2024rqj}  using the SymTFT framework. 

To define the S-matrix of kinks, we follow \cite{Copetti:2024rqj} and consider a correlation function
on a (Euclidean) disk with insertions of boundary-changing operators that create kinks. This correlation function describes the scattering process, after the analytic continuation to Lorentzian kinematics and performing appropriate Fourier transformations on Cauchy slices (see e.g.~\cite[Ch.~5]{Itzykson:1980rh} for details). 
As a result, we obtain a schematic relation
\bea\label{eq:Sroughly}
S^{ab}_{dc}(s)\propto \quad\begin{tikzpicture}[baseline={(0,-0.1)}]
\filldraw[color=white!90!gray] (0,0) circle (1.5);
\filldraw[line width =1.0, color=gray, fill=white!90!YELLOW, opacity=0.65] (0,0) node {}  circle (1);
\filldraw[line width =.7,  color=white!30!YELLOW, fill=white!85!YELLOW, opacity=1] (0,0) node {$S$}  circle (.45);
\draw[line width =.7, color=white!30!YELLOW] (45:.45) -- (45:1.);
\draw[line width =.7, color=white!30!YELLOW] (-45:.45) -- (-45:1.);
\draw[line width =.7, color=white!30!YELLOW] (135:.45) -- (135:1.);
\draw[line width =.7, color=white!30!YELLOW] (-135:.45) -- (-135:1.);
 \draw[black,thick] (45:1) node[black,circle,fill,scale=0.3] {} -- (45:1.5);
     \draw[black,thick] (-45:1) node[black,circle,fill,scale=0.3] {} -- (-45:1.5);
      \draw[black,thick] (135:1) node[black,circle,fill,scale=0.3] {} -- (135:1.5);
       \draw[black,thick] (-135:1) node[black,circle,fill,scale=0.3] {} -- (-135:1.5);
       \node[gray] at (0:1.2) {$b$};    \node[gray] at (90:1.2) {$a$};    \node[gray] at (180:1.2) {$d$};    \node[gray] at (270:1.2) {$c$}; 
    \end{tikzpicture}\,.
\eea
Here the right hand side represents the analytically-continued disk correlation function, $a$-$d$ label the vacua, $s$ is the Mandelstam variable. In the SymTFT framework discussed above, the right hand side of \eqref{eq:Sroughly} admits the following three-dimensional description:
\bea\label{eq:SmatSymTFT}
S^{ab}_{dc}(s)\propto \quad\begin{tikzpicture}[baseline={(0,-0.1)}]
\filldraw[color=white!90!gray] (0,0) circle (1.5);
\filldraw[line width =1.0, color=gray, fill=white!90!YELLOW, opacity=0.65] (0,0) node {}  circle (1);
\filldraw[line width =.7,  color=white!30!YELLOW, fill=white!85!YELLOW, opacity=1] (0,0) node {$S$}  circle (.45);
\draw[line width =.7, color=white!30!YELLOW] (45:.45) -- (45:1.);
\draw[line width =.7, color=white!30!YELLOW] (-45:.45) -- (-45:1.);
\draw[line width =.7, color=white!30!YELLOW] (135:.45) -- (135:1.);
\draw[line width =.7, color=white!30!YELLOW] (-135:.45) -- (-135:1.);
 \draw[black,thick] (45:1) node[black,circle,fill,scale=0.3] {} -- (45:1.5);
     \draw[black,thick] (-45:1) node[black,circle,fill,scale=0.3] {} -- (-45:1.5);
      \draw[black,thick] (135:1) node[black,circle,fill,scale=0.3] {} -- (135:1.5);
       \draw[black,thick] (-135:1) node[black,circle,fill,scale=0.3] {} -- (-135:1.5);
       \node[gray] at (0:1.2) {$b$};    \node[gray] at (90:1.2) {$a$};    \node[gray] at (180:1.2) {$d$};    \node[gray] at (270:1.2) {$c$};
    \node at (2,0) {$=$};
    \begin{scope}[shift={(6.5,0)}]
       \draw[thick,color=BLUE, fill=white!90!BLUE] (-3,0) node {} ellipse (0.75 and 1);
           %Back, up
    \pgfmathsetmacro{\xA}{-.05+3/4 * cos(30)}
    \pgfmathsetmacro{\yA}{5/4 *sin(30)}
    \pgfmathsetmacro{\xB}{-3.05 + 3/4 * cos(30)}
    \pgfmathsetmacro{\yB}{5/4 *sin(30)}
\draw[black,thick] (\xA,\yA) -- (\xB,\yB);
\draw[fill=black]   (\xA,\yA) circle (0.05);
\draw[fill=BLUE]   (\xB,\yB) circle (0.05);
       %Back, down
       \pgfmathsetmacro{\xAA}{3/4 * cos(-30)}
\pgfmathsetmacro{\yAA}{sin(-30)}
\pgfmathsetmacro{\xBB}{-3 + 3/4 * cos(-30)}
\pgfmathsetmacro{\yBB}{sin(-30)}
\draw[black,thick] (\xAA,\yAA) -- (\xBB,\yBB);
\draw[fill=black]   (\xAA,\yAA) circle (0.05);
\draw[fill=BLUE]   (\xBB,\yBB) circle (0.05);
    \draw[color=white!70!gray, fill=white!90!gray, opacity=0.65] (0,1) arc (90:270: 0.75 and 1) -- (-3,-1) arc (270:90: 0.75 and 1) -- (0,1) -- cycle; 
        \draw[thick,color=gray, fill=white!90!YELLOW, opacity=0.65] (0,0) node[black] {} ellipse (0.75 and 1);
             %front, up
\pgfmathsetmacro{\xxA}{3/4 * cos(150)}
\pgfmathsetmacro{\yyA}{sin(150)}
\pgfmathsetmacro{\xxB}{-3 + 3/4 * cos(150)}
\pgfmathsetmacro{\yyB}{sin(150)}
\draw[black,thick] (\xxA,\yyA) -- (\xxB,\yyB);
\draw[fill=black]   (\xxA,\yyA) circle (0.05);
\draw[fill=BLUE]   (\xxB,\yyB) circle (0.05);
    %front, down
\pgfmathsetmacro{\xxAA}{3/4 * cos(140)}
\pgfmathsetmacro{\yyAA}{sin(-140)}
\pgfmathsetmacro{\xxBB}{-3 + 3/4 * cos(-140)}
\pgfmathsetmacro{\yyBB}{sin(-140)}
\draw[black,thick] (\xxAA,\yyAA) -- (\xxBB,\yyBB);
\draw[fill=black]   (\xxAA,\yyAA) circle (0.05);
\draw[fill=BLUE]   (\xxBB,\yyBB) circle (0.05);
        \draw[line width =.7, color=white!30!YELLOW] (\xA,\yA) -- (\xxAA,\yyAA);
        \draw[line width =.7, color=white!30!YELLOW] (\xAA,\yAA) -- (\xxA,\yyA);
        \draw[line width =.7, color=white!30!YELLOW, fill=white!85!YELLOW, opacity=1] (0,0) node[white!30!YELLOW] {$S$} ellipse (.45*.75 and .45);
        \node[gray,above] at (-3,1) {$a$}; \node[gray,below] at (-3,-1) {$c$}; \node[gray,right] at (-2.25,0) {$b$}; \node[gray,left] at (-3.75,0) {$d$}; 
    \end{scope}
    \end{tikzpicture}\,.
\eea

As pointed out in our previous work \cite{Copetti:2024rqj}, to ensure unitarity of the S-matrix, we also need to take into account corrections to the norms of in- and out-states due to the TQFT dynamics in the IR. To compute such corrections, let us first consider the  path-integral representations of in- and out-states\footnote{Here the analytic continuation to Lorentzian kinematics are made implicit on the right hand sides.}:
\bea \label{eq:inoutstates}
 \vert \text{In} \rangle =
 \begin{tikzpicture}[baseline={(0,-0.5)}]
 \filldraw[color=white!90!gray] (1.5,0) arc (0:-180: 1.5 and 1.5) -- (-1,0) arc (-180:0:1 and 1) -- (1.5,0)  --cycle;
 \filldraw[color=white!90!YELLOW,opacity=0.65] (1.0,0) arc (0:-180:1 and 1) -- (-1,0) ;
 \draw[line width =1.0,gray] (-1,0) arc (-180:0:1 and 1);
 \node[above,gray] at (-1.25,0) {$d$}; \node[above,gray] at (1.25,0) {$b$}; \node[below,gray] at (0,-1.0) {$c$};
 \draw[black,thick] (-135:1.5) node[below left] {$v$} -- (-0.7,-0.7) node[black,circle,fill,scale=0.3] {};
 \draw[black,thick] (-45:1.5) node[below right] {$v$} -- (0.7,-0.7) node[black,circle,fill,scale=0.3] {};
 \draw[line width =.7,color=white!30!YELLOW] (-0.7,-0.7) arc (-45:0:1 and 1) ;
 \draw[line width =.7,color=white!30!YELLOW] (0.7,-0.7) arc (-45:0:-1 and 1) ;
\end{tikzpicture} \,, \ \ \ \ \ \ 
\langle \text{Out} \vert =
\begin{tikzpicture}[baseline={(0,0.5)}]
 \filldraw[color=white!90!gray]  (1.5,0) arc (0:180: 1.5 and 1.5) -- (-1,0) arc (180:0:1 and 1) -- (1.5,0)  --cycle;
 \filldraw[color=white!90!YELLOW,opacity=0.65] (-1,0) arc (180:0:1 and 1)-- (-1,0);
 \draw[line width =1.0,gray] (-1,0) arc (180:0:1 and 1);
 \node[below,gray] at (-1.25,0) {$d$}; \node[below,gray] at (1.25,0) {$b$}; \node[above,gray] at (0,1) {$a$};
 \draw[black,thick] (135:1.5) node[above left] {$v$} -- (-0.7,0.7) node[black,circle,fill,scale=0.3] {};
 \draw[black,thick] (45:1.5) node[above right] {$v$} -- (0.7,0.7) node[black,circle,fill,scale=0.3] {};   
 \draw[line width =.7,color=white!30!YELLOW] (-0.7,0.7) arc (45:0:1 and 1) ;
 \draw[line width =.7,color=white!30!YELLOW] (0.7,0.7) arc (45:0:-1 and 1) ;
 \end{tikzpicture}\;.
 \eea
In principle, the norms $\langle \text{In}|\text{In}\rangle$ and $\langle \text{Out}|\text{Out}\rangle$ can be computed by gluing \eqref{eq:inoutstates} to their upside-down images and performing the path integral. This however does not separate the IR TQFT from theory-specific dynamics, making the analysis challenging.

To make progress, we use the SymTFT description. We propose that the norm of in-states can be computed by the following SymTFT configuration (a similar expression holds for out-states):
\be \label{eq:innorm}
 \langle {\rm In} \vert {\rm In} \rangle =
 \begin{tikzpicture}
[baseline= {(0,0)}]
  \pgfmathsetmacro{\yA}{0.75*sin(45)}
\pgfmathsetmacro{\xA}{cos(45)}
 \begin{scope}[shift={(6.5,0)}]
       \draw[thick,color=BLUE, fill=white!90!BLUE] (-2,0)  ellipse (0.75 and 1);
       \pgfmathsetmacro{\xA}{3/4 * cos(30)}
\pgfmathsetmacro{\yA}{sin(30)}
\pgfmathsetmacro{\xB}{-2 + 3/4 * cos(30)}
\pgfmathsetmacro{\yB}{sin(30)}
\draw[black,thick] (\xA,\yA) -- (\xB,\yB);
\draw[fill=BLUE]   (\xB,\yB) circle (0.05);

       \pgfmathsetmacro{\xAA}{3/4 * cos(-30)}
\pgfmathsetmacro{\yAA}{sin(-30)}
\pgfmathsetmacro{\xBB}{-2 + 3/4 * cos(-30)}
\pgfmathsetmacro{\yBB}{sin(-30)}
\draw[black,thick] (\xAA,\yAA) -- (\xBB,\yBB);
\draw[fill=BLUE]   (\xBB,\yBB) circle (0.05);

    \draw[color=white!70!gray, fill=white!90!gray, opacity=0.65] (0,1) arc (90:270: 0.75 and 1) -- (-2,-1) arc (270:90: 0.75 and 1) -- (0,1) -- cycle; 
\pgfmathsetmacro{\xxA}{3/4 * cos(140)}
\pgfmathsetmacro{\yyA}{sin(140)}
\pgfmathsetmacro{\xxB}{-2 + 3/4 * cos(140)}
\pgfmathsetmacro{\yyB}{sin(140)}
\draw[black,thick] (\xxA,\yyA) -- (\xxB,\yyB);
\draw[fill=BLUE]   (\xxB,\yyB) circle (0.05);

\pgfmathsetmacro{\xxAA}{3/4 * cos(-140)}
\pgfmathsetmacro{\yyAA}{sin(-140)}
\pgfmathsetmacro{\xxBB}{-2 + 3/4 * cos(-140)}
\pgfmathsetmacro{\yyBB}{sin(-140)}
\draw[black,thick] (\xxAA,\yyAA) -- (\xxBB,\yyBB);
\draw[fill=BLUE]   (\xxBB,\yyBB) circle (0.05);

        \draw[thick,color=gray, fill=white!90!gray, opacity=0.65] (0,0) node[black] {} ellipse (0.75 and 1);
        %\node[gray,above] at (-1,1) {$\calL_\calM$};
        \node[gray,above] at (-2,1) {$c$}; \node[gray,below] at (-2,-1) {$c$}; \node[gray,right] at (-1.25,0) {$b$}; \node[gray,left] at (-2.75,0) {$d$}; 
        \draw[black,thick] (135:1) (\xxA,\yyA) arc (60:-60:.9 and .75) ;
        \draw[black,thick] (0:.9) (\xA,\yA) arc (45:-45:-1 and .71) ;
    \end{scope}
%%%%%%%%%%%%%%%%%%%%%%%%%%%%%%%
    \begin{scope}[shift={(8.5,0)}]
       
       \pgfmathsetmacro{\xA}{3/4 * cos(30)}
\pgfmathsetmacro{\yA}{sin(30)}
\pgfmathsetmacro{\xB}{+2 + 3/4 * cos(30)}
\pgfmathsetmacro{\yB}{sin(30)}
\draw[black,thick] (\xA,\yA) -- (\xB,\yB);
\draw[fill=BLUE]   (\xB,\yB) circle (0.05);

       \pgfmathsetmacro{\xAA}{3/4 * cos(-30)}
\pgfmathsetmacro{\yAA}{sin(-30)}
\pgfmathsetmacro{\xBB}{+2 + 3/4 * cos(-30)}
\pgfmathsetmacro{\yBB}{sin(-30)}
\draw[black,thick] (\xAA,\yAA) -- (\xBB,\yBB);
\draw[fill=BLUE]   (\xBB,\yBB) circle (0.05);

    \draw[color=white!70!gray, fill=white!90!gray, opacity=0.65] (0,1) arc (90:270: 0.75 and 1) -- (2,-1) arc (270:90: 0.75 and 1) -- (0,1) -- cycle; 
\pgfmathsetmacro{\xxA}{3/4 * cos(140)}
\pgfmathsetmacro{\yyA}{sin(140)}
\pgfmathsetmacro{\xxB}{+2 + 3/4 * cos(140)}
\pgfmathsetmacro{\yyB}{sin(140)}

\pgfmathsetmacro{\xxAA}{3/4 * cos(-140)}
\pgfmathsetmacro{\yyAA}{sin(-140)}
\pgfmathsetmacro{\xxBB}{+2 + 3/4 * cos(-140)}
\pgfmathsetmacro{\yyBB}{sin(-140)}

        \draw[black,thick] (135:1) (\xxA,\yyA) arc (60:-60:.9 and .75) ;
        \draw[black,thick] (0:.9) (\xA,\yA) arc (45:-45:-1 and .71) ;
        \draw[thick,color=gray, fill=white!90!gray, opacity=0.65] (0,0) node[black] {} ellipse (0.75 and 1);
%yellow face right
\draw[line width =1.0, color=gray, fill=white!90!YELLOW, opacity=0.65] (2,0)  ellipse (0.75 and 1);
\draw[line width =.7,color=white!30!YELLOW] (135:1) (\xxA+2,\yyA) arc (60:-60:.9 and .75) ;
\draw[line width =.7,color=white!30!YELLOW] (0:.9) (\xA+2,\yA) arc (45:-45:-1 and .71) ;
    %front, up
    \draw[black,thick] (\xxA,\yyA) -- (\xxB,\yyB);
    \draw[fill=black]   (\xxB,\yyB) circle (0.05);
     %front, down
     \draw[black,thick] (\xxAA,\yyAA) -- (\xxBB,\yyBB);
    \draw[fill=black]   (\xxBB,\yyBB) circle (0.05);
    \end{scope}
    
\end{tikzpicture}\,.
\ee
Here we performed the bulk surgery and projected to a specific state depicted above, which is described by two $v$ lines going upward without any junctions in between. This is to ensure that the state at $t=0$ slice is a two-kink state; without it, the path integral would include all possible field configurations including multi-particle states. 

The key advantage of \eqref{eq:innorm} is that it cleanly separates the TQFT dynamics from everything else: the right half of the figure gives the standard QFT norm of in-states containing the momentum-conserving delta function\footnote{Written explicitly, it reads $\langle p_1',p_2'|p_1,p_2\rangle=(2\pi)^2\, 2\sqrt s \sqrt{s-4m^2}\,\delta^2(p_1+p_2-p_1'-p_2')\,.$} while the left half describes the correction to the norm due to the IR TQFT, which depends purely on the fusion category data. Evaluating the left half in the IR TQFT, we obtain
 \be \label{eq:innormTQFT}
 \langle {\rm In} \vert {\rm In} \rangle\big\vert_\text{TQFT} =
 \begin{tikzpicture}
[baseline= {(0,0)}]
  \pgfmathsetmacro{\yA}{0.75*sin(45)}
\pgfmathsetmacro{\xA}{cos(45)}
 \begin{scope}[shift={(6.5,0)}]
       \draw[thick,color=BLUE, fill=white!90!BLUE] (-2,0) %node {$\calL$}
       ellipse (0.75 and 1);
       \pgfmathsetmacro{\xA}{3/4 * cos(30)}
\pgfmathsetmacro{\yA}{sin(30)}
\pgfmathsetmacro{\xB}{-2 + 3/4 * cos(30)}
\pgfmathsetmacro{\yB}{sin(30)}
\draw[black,thick] (\xA,\yA) -- (\xB,\yB);
\draw[fill=BLUE]   (\xB,\yB) circle (0.05);

       \pgfmathsetmacro{\xAA}{3/4 * cos(-30)}
\pgfmathsetmacro{\yAA}{sin(-30)}
\pgfmathsetmacro{\xBB}{-2 + 3/4 * cos(-30)}
\pgfmathsetmacro{\yBB}{sin(-30)}
\draw[black,thick] (\xAA,\yAA) -- (\xBB,\yBB);
\draw[fill=BLUE]   (\xBB,\yBB) circle (0.05);

    \draw[color=white!70!gray, fill=white!90!gray, opacity=0.65] (0,1) arc (90:270: 0.75 and 1) -- (-2,-1) arc (270:90: 0.75 and 1) -- (0,1) -- cycle; 
\pgfmathsetmacro{\xxA}{3/4 * cos(140)}
\pgfmathsetmacro{\yyA}{sin(140)}
\pgfmathsetmacro{\xxB}{-2 + 3/4 * cos(140)}
\pgfmathsetmacro{\yyB}{sin(140)}
\draw[black,thick] (\xxA,\yyA) -- (\xxB,\yyB);
\draw[fill=BLUE]   (\xxB,\yyB) circle (0.05);

\pgfmathsetmacro{\xxAA}{3/4 * cos(-140)}
\pgfmathsetmacro{\yyAA}{sin(-140)}
\pgfmathsetmacro{\xxBB}{-2 + 3/4 * cos(-140)}
\pgfmathsetmacro{\yyBB}{sin(-140)}
\draw[black,thick] (\xxAA,\yyAA) -- (\xxBB,\yyBB);
\draw[fill=BLUE]   (\xxBB,\yyBB) circle (0.05);

        \draw[thick,color=gray, fill=white!90!gray, opacity=0.65] (0,0) node[black] {} ellipse (0.75 and 1);
        %\node[gray,above] at (-1,1) {$\calL_\calM$};
        \node[gray,above] at (-2,1) {$c$}; \node[gray,below] at (-2,-1) {$c$}; \node[gray,right] at (-1.25,0) {$b$}; \node[gray,left] at (-2.75,0) {$d$}; 
        \draw[black,thick] (135:1) (\xxA,\yyA) arc (60:-60:.9 and .75) ;
        \draw[black,thick] (0:.9) (\xA,\yA) arc (45:-45:-1 and .71) ;
    \end{scope}
\end{tikzpicture}
=
\begin{tikzpicture}[baseline={(0,0)},scale=.9]
    \pgfmathsetmacro{\yA}{1*sin(45)}
\pgfmathsetmacro{\xA}{cos(45)}
 \draw[color=BLUE, thick, fill=white!90!gray] (0,0) ellipse (1 and 1); 
 \node[black,circle,fill,scale=0.3]  at (\xA,\yA) {};
 \node[black,circle,fill,scale=0.3]  at (-\xA,\yA)  {};
 \node[black,circle,fill,scale=0.3]  at (\xA,-\yA)  {};
 \node[black,circle,fill,scale=0.3]  at (-\xA,-\yA)  {};
 \draw[black,thick] plot[smooth, tension=1.5] coordinates{(\xA,\yA) (0.5,0) (\xA,-\yA) }; 
 \draw[black,thick] plot[smooth, tension=1.5] coordinates{(-\xA,\yA) (-0.5,0) (-\xA,-\yA) }; 
 \node[gray,below] at (0,-1) {$c$}; \node[gray,above] at (0,1) {$c$}; \node[gray,right] at (1,0) {$b$}; \node[gray,left] at (-1,0) {$d$};
 %\node[gray] at (1.35,1.1) {$\cT_\calM^{\cC^*_\calM}$};
\end{tikzpicture}
= 
\d_v \sqrt{g_b g_d}\,,
 \ee
 where we have used basic identities for (1+1)d symmetric TFTs, \eqref{eq:gdef} and \eqref{eq:bdybubbleremoval}, to evaluate the diagram.

We thus conclude that the S-matrix between properly normalized in- and out-states is given by
\begin{equation}\label{eq:Snormalized}
S^{ab}_{dc}(s) = 
\frac{
\begin{tikzpicture}[baseline={(0,0)},scale=.75]
    \begin{scope}[shift={(-1.5,0)}]
\filldraw[color=white!90!gray] (0,0) circle (1.5);
\filldraw[line width =1.0, color=gray, fill=white!90!YELLOW, opacity=0.65] (0,0) node {}  circle (1);
\filldraw[line width =.7,  color=white!30!YELLOW, fill=white!85!YELLOW, opacity=1] (0,0) node {$S$}  circle (.45);
\draw[line width =.7, color=white!30!YELLOW] (45:.45) -- (45:1.);
\draw[line width =.7, color=white!30!YELLOW] (-45:.45) -- (-45:1.);
\draw[line width =.7, color=white!30!YELLOW] (135:.45) -- (135:1.);
\draw[line width =.7, color=white!30!YELLOW] (-135:.45) -- (-135:1.);
 \draw[black,thick] (45:1) node[black,circle,fill,scale=0.3] {} -- (45:1.5);
     \draw[black,thick] (-45:1) node[black,circle,fill,scale=0.3] {} -- (-45:1.5);
      \draw[black,thick] (135:1) node[black,circle,fill,scale=0.3] {} -- (135:1.5);
       \draw[black,thick] (-135:1) node[black,circle,fill,scale=0.3] {} -- (-135:1.5);
       \node[gray] at (0:1.2) {$b$};    \node[gray] at (90:1.2) {$a$};    \node[gray] at (180:1.2) {$d$};    \node[gray] at (270:1.2) {$c$};
     \end{scope}
\end{tikzpicture}}
{
\resizebox{0.7\width}{!}{$
\sqrt{
\begin{tikzpicture}[scale=0.75]
    \pgfmathsetmacro{\yA}{1*sin(45)}
\pgfmathsetmacro{\xA}{cos(45)}
 \draw[color=BLUE, thick, fill=white!90!gray] (0,0) ellipse (1 and 1); 
 \node[black,circle,fill,scale=0.3]  at (\xA,\yA) {};
 \node[black,circle,fill,scale=0.3]  at (-\xA,\yA)  {};
 \node[black,circle,fill,scale=0.3]  at (\xA,-\yA)  {};
 \node[black,circle,fill,scale=0.3]  at (-\xA,-\yA)  {};
 \draw[black,thick] plot[smooth, tension=1.5] coordinates{(\xA,\yA) (0.5,0) (\xA,-\yA) }; 
 \draw[black,thick] plot[smooth, tension=1.5] coordinates{(-\xA,\yA) (-0.5,0) (-\xA,-\yA) }; 
 \node[gray,below] at (0,-1) {$a$}; \node[gray,above] at (0,1) {$a$}; \node[gray,right] at (1,0) {$b$}; \node[gray,left] at (-1,0) {$d$};
 %\node[gray] at (1.35,1.1) {$\cT_\calM^{\cC^*_\calM}$};
\end{tikzpicture} \begin{tikzpicture}[scale=0.75]
    \pgfmathsetmacro{\yA}{1*sin(45)}
\pgfmathsetmacro{\xA}{cos(45)}
 \draw[color=BLUE, thick, fill=white!90!gray] (0,0) ellipse (1 and 1); 
 \node[black,circle,fill,scale=0.3]  at (\xA,\yA) {};
 \node[black,circle,fill,scale=0.3]  at (-\xA,\yA)  {};
 \node[black,circle,fill,scale=0.3]  at (\xA,-\yA)  {};
 \node[black,circle,fill,scale=0.3]  at (-\xA,-\yA)  {};
 \draw[black,thick] plot[smooth, tension=1.5] coordinates{(\xA,\yA) (0.5,0) (\xA,-\yA) }; 
 \draw[black,thick] plot[smooth, tension=1.5] coordinates{(-\xA,\yA) (-0.5,0) (-\xA,-\yA) }; 
 \node[gray,below] at (0,-1) {$c$}; \node[gray,above] at (0,1) {$c$}; \node[gray,right] at (1,0) {$b$}; \node[gray,left] at (-1,0) {$d$};
 %\node[gray] at (1.35,1.1) {$\cT_\calM^{\cC^*_\calM}$};
\end{tikzpicture}
}
$}
}\,.
\end{equation}
As in our previous work, the numerator is expected to be crossing symmetric while the denominator depends on the channels ($s$-, $t$-channels) we consider, leading to the modified crossing rule,
 \be\label{eq:modifcross}
S^{ab}_{dc}(s) = \resizebox{0.75\width}{!}{$\sqrt{\frac{\begin{tikzpicture}[scale=0.75]
    \pgfmathsetmacro{\yA}{1*sin(45)}
\pgfmathsetmacro{\xA}{cos(45)}
 \draw[color=BLUE, thick, fill=white!90!gray] (0,0) ellipse (1 and 1); 
 \node[black,circle,fill,scale=0.3]  at (\xA,\yA) {};
 \node[black,circle,fill,scale=0.3]  at (-\xA,\yA)  {};
 \node[black,circle,fill,scale=0.3]  at (\xA,-\yA)  {};
 \node[black,circle,fill,scale=0.3]  at (-\xA,-\yA)  {};
 \draw[black,thick] plot[smooth, tension=1.5] coordinates{(\xA,\yA) (0,0.3) (-\xA,\yA) }; 
 \draw[black,thick] plot[smooth, tension=1.5] coordinates{(\xA,-\yA) (0,-0.3) (-\xA,-\yA) };
 \node[gray,below] at (0,-1) {$c$}; \node[gray,above] at (0,1) {$a$}; \node[gray,right] at (1,0) {$b$}; \node[gray,left] at (-1,0) {$b$};
\end{tikzpicture} \begin{tikzpicture}[scale=0.75]
    \pgfmathsetmacro{\yA}{1*sin(45)}
\pgfmathsetmacro{\xA}{cos(45)}
 \draw[color=BLUE, thick, fill=white!90!gray] (0,0) ellipse (1 and 1); 
 \node[black,circle,fill,scale=0.3]  at (\xA,\yA) {};
 \node[black,circle,fill,scale=0.3]  at (-\xA,\yA)  {};
 \node[black,circle,fill,scale=0.3]  at (\xA,-\yA)  {};
 \node[black,circle,fill,scale=0.3]  at (-\xA,-\yA)  {};
 \draw[black,thick] plot[smooth, tension=1.5] coordinates{(\xA,\yA) (0,0.3) (-\xA,\yA) }; 
 \draw[black,thick] plot[smooth, tension=1.5] coordinates{(\xA,-\yA) (0,-0.3) (-\xA,-\yA) };
 \node[gray,below] at (0,-1) {$c$}; \node[gray,above] at (0,1) {$a$}; \node[gray,right] at (1,0) {$d$}; \node[gray,left] at (-1,0) {$d$};
\end{tikzpicture} }{ \begin{tikzpicture}[scale=0.75]
    \pgfmathsetmacro{\yA}{1*sin(45)}
\pgfmathsetmacro{\xA}{cos(45)}
 \draw[color=BLUE, thick, fill=white!90!gray] (0,0) ellipse (1 and 1); 
 \node[black,circle,fill,scale=0.3]  at (\xA,\yA) {};
 \node[black,circle,fill,scale=0.3]  at (-\xA,\yA)  {};
 \node[black,circle,fill,scale=0.3]  at (\xA,-\yA)  {};
 \node[black,circle,fill,scale=0.3]  at (-\xA,-\yA)  {};
 \draw[black,thick] plot[smooth, tension=1.5] coordinates{(\xA,\yA) (0.5,0) (\xA,-\yA) }; 
 \draw[black,thick] plot[smooth, tension=1.5] coordinates{(-\xA,\yA) (-0.5,0) (-\xA,-\yA) }; 
 \node[gray,below] at (0,-1) {$a$}; \node[gray,above] at (0,1) {$a$}; \node[gray,right] at (1,0) {$b$}; \node[gray,left] at (-1,0) {$d$};
\end{tikzpicture} \begin{tikzpicture}[scale=0.75]
    \pgfmathsetmacro{\yA}{1*sin(45)}
\pgfmathsetmacro{\xA}{cos(45)}
 \draw[color=BLUE, thick, fill=white!90!gray] (0,0) ellipse (1 and 1); 
 \node[black,circle,fill,scale=0.3]  at (\xA,\yA) {};
 \node[black,circle,fill,scale=0.3]  at (-\xA,\yA)  {};
 \node[black,circle,fill,scale=0.3]  at (\xA,-\yA)  {};
 \node[black,circle,fill,scale=0.3]  at (-\xA,-\yA)  {};
 \draw[black,thick] plot[smooth, tension=1.5] coordinates{(\xA,\yA) (0.5,0) (\xA,-\yA) }; 
 \draw[black,thick] plot[smooth, tension=1.5] coordinates{(-\xA,\yA) (-0.5,0) (-\xA,-\yA) }; 
 \node[gray,below] at (0,-1) {$c$}; \node[gray,above] at (0,1) {$c$}; \node[gray,right] at (1,0) {$b$}; \node[gray,left] at (-1,0) {$d$};
\end{tikzpicture} }} $ } \, S^{bc}_{ad}(t) = \sqrt{\frac{g_a g_c}{g_b g_d}} \, S^{bc}_{ad}(t) \, .
\ee
This provides a generalization of the modified crossing rule in \cite{Copetti:2024rqj}, applicable to vacua transforming in general representations of $\mathcal{C}$.
\subsubsection{Ward identities and projector basis}\label{ssubsec:smatrixward}
\paragraph{Ward identities for S-matrix.} Another benefit of SymTFT is that it provides a conceptually clean derivation of the Ward identities obeyed by the S-matrix. 

SymTFT geometrically separates the symmetry actions from the QFT dynamics, and allows us to derive the Ward identities purely from the topological boundary $\calL$ of SymTFT, where symmetry lines live. Concretely, we consider a symmetry line $\mathcal{L}$ extending between the in- and out-states and deforming it past the $v$ lines using $[\mathcal{L};v]$ matrices. Focusing on the (left) topological boundary $\calL$, this gives the following identity for the disk correlation functions
\bea\resizebox{.8\width}{!}{
\begin{tikzpicture}
 \filldraw[color=white!90!gray] (0,0) circle (1.5);
\filldraw[color=BLUE, fill=white!90!BLUE, opacity=0.75] (0,0) circle (1);
 \draw[black] (45:1) node[black,circle,fill,scale=0.3] {} -- (45:1.5);
     \draw[black] (-45:1) node[black,circle,fill,scale=0.3] {} -- (-45:1.5);
      \draw[black] (135:1) node[black,circle,fill,scale=0.3] {} -- (135:1.5);
       \draw[black] (-135:1) node[black,circle,fill,scale=0.3] {} -- (-135:1.5);
       \draw[BLUE] (-1,0)  node[BLUE,circle,fill,scale=0.3] {} -- (1,0)  node[BLUE,circle,fill,scale=0.3] {};
       \node[above,BLUE] at (0,0) {$\cL$};
       \node[gray] at (22.5:1.2) {$\, b$};   \node[gray] at (-22.5:1.2) {$b'$};    \node[gray] at (90:1.2) {$a$};    \node[gray] at (202.5:1.2) {$d'$};  \node[gray] at (157.5:1.2) {$d$};    \node[gray] at (270:1.2) {$c'$};   
       \node[rotate=45,scale=1.3] at (-2.5,-2) {$\underset{\text{deform down}}{=}$}; 
        \node[rotate=-45,scale=1.3] at (2.5,-2) {$\underset{\text{deform up}}{=}$};
        \begin{scope}[shift={(-3,-5)}]
            \node[left] at (-1.5,0) {$ \ds \sum_{e} \left[\cL;v\right]_{d' c'}^{d e} \left[\cL;v\right]_{b' c'}^{b e } \sqrt{\frac{\d_\cL\,g_{c'}}{g_{e}}}$};
            \filldraw[color=white!90!gray] (0,0) circle (1.5);
\filldraw[color=BLUE, fill=white!90!BLUE, opacity=0.75] (0,0) circle (1);
 \draw[black] (45:1) node[black,circle,fill,scale=0.3] {} -- (45:1.5);
     \draw[black] (-45:1) node[black,circle,fill,scale=0.3] {} -- (-45:1.5);
      \draw[black] (135:1) node[black,circle,fill,scale=0.3] {} -- (135:1.5);
       \draw[black] (-135:1) node[black,circle,fill,scale=0.3] {} -- (-135:1.5);
       \node[gray] at (0:1.2) {$b$};    \node[gray] at (90:1.2) {$a$};    \node[gray] at (180:1.2) {$d$};    \node[gray] at (270:1.2) {$e$};
        \end{scope}
 \begin{scope}[shift={(3,-5)}]
            \node[right] at (1.5,0) {$ \ds \sum_{e'} \left[\cL;v\right]_{ d a}^{d' e'} \left[\cL;v\right]_{b a}^{b' e'} \sqrt{\frac{\d_\cL\, g_{a}}{g_{e'}}}$};
       \filldraw[color=white!90!gray] (0,0) circle (1.5);
\filldraw[color=BLUE, fill=white!90!BLUE, opacity=0.75] (0,0) circle (1);
 \draw[black] (45:1) node[black,circle,fill,scale=0.3] {} -- (45:1.5);
     \draw[black] (-45:1) node[black,circle,fill,scale=0.3] {} -- (-45:1.5);
      \draw[black] (135:1) node[black,circle,fill,scale=0.3] {} -- (135:1.5);
       \draw[black] (-135:1) node[black,circle,fill,scale=0.3] {} -- (-135:1.5);
       \node[gray] at (0:1.2) {$b'$};    \node[gray] at (90:1.2) {$e'$};    \node[gray] at (180:1.2) {$d'$};    \node[gray] at (270:1.2) {$c'$};
        \end{scope}        
\end{tikzpicture}
}
\eea
Reinstating the normalization factors needed to define the S-matrix, this leads to the Ward identity for the S-matrix,
\be \label{eq:symWI}
\ds \sum_{e} \left[\cL;v\right]_{d' c'}^{d e} \left[\cL;v\right]_{b' c'}^{b e }  \sqrt{\frac{g_{c'}g_{b}g_{d}}{g_{e}}} \, S^{ab}_{ d e}(s)= \sum_{e'} \left[\cL;v\right]_{ d a}^{d' e'} \left[\cL;v\right]_{b a}^{b' e'} \sqrt{\frac{g_{a}g_{b'}g_{d'}}{g_{e'}}} \, S^{e' b'}_{d' c'} (s)  \, .
\ee
For the regular representations, this reduces to the Ward identity used in our previous work \cite{Copetti:2024rqj}.
\paragraph{Projector basis.}
In \cite{Copetti:2024rqj}, we showed that $2\to2$ S-matrix of kinks can be decomposed into representations of the fusion category by using projectors $P_{\chi}$:
\be\label{eq:Sdecompreview}
S^{a b}_{d c}(s) = \sum_{\chi \in v \times v} \left(P_\chi\right)^{a b}_{d c} \,  A_\chi (s)\, ,
\ee
where, for the regular representation, $P_\chi$ takes the explicit form \cite{Aasen:2020jwb}: 
\bea\label{eq:reviewprojector}
\left(P_\chi\right)^{a b}_{d c} = \frac{\sqrt{\d_\chi}}{\d_v^2 \sqrt{\d_b \, \d_d}}
\begin{tikzpicture}[baseline={(0,0)},scale=.95]
\draw[BLUE,thick] (0,0) circle (1);    
 \pgfmathsetmacro{\yA}{1*sin(45)}
\pgfmathsetmacro{\xA}{cos(45)}
  \node[BLUE] at (\xA-.1,\yA-.4) {$v$};
  \node[BLUE] at (-\xA+.1,\yA-.4) {$v$};
  \node[BLUE] at (\xA-.1,-\yA+.4) {$v$};
  \node[BLUE] at (-\xA+.1,-\yA+.4) {$v$};
  \draw[BLUE,thick] plot[smooth, tension=1.5] coordinates{(\xA,\yA) (0,0.35) (-\xA,\yA) }; 
 \draw[BLUE,thick] plot[smooth, tension=1.5] coordinates{(\xA,-\yA) (0,-0.35) (-\xA,-\yA) }; 
  \draw[BLUE,thick,double] (0,0.35) -- (0,-0.35);
  \node[BLUE,right] at (-0.05,0) {$\chi$};
 \node[BLUE,below] at (0,-1) {$c$}; \node[BLUE,above] at (0,1) {$a$}; \node[BLUE,right] at (1,0) {$b$}; \node[BLUE,left] at (-1,0) {$d$};
\end{tikzpicture}
= \sqrt{\d_a \d_c}\, \d_\chi 
    \begin{bmatrix}
     v & v & \chi\\
    d & b & a
    \end{bmatrix}
    \begin{bmatrix}
    v & v & \chi\\
    d & b & c
    \end{bmatrix} \, ,
\eea
This decomposition ensures that the $S$-matrix satisfies the Ward identities.

We now generalize this result to general module categories by applying the idea of separating the symmetry action of $\cC$ from the dynamical data using SymTFT and bulk surgery. This can be achieved simply by inserting a resolution of identity in the SymTFT description of the S-matrix, leading to the following schematic picture: (Here various normalization factors are made implicit for simplicity.)
\be\label{eq:Sproj}
\begin{tikzpicture}[baseline={(0,0)},scale=1]
\filldraw[color=white!90!gray] (0,0) circle (1.5);
\filldraw[line width =1.0, color=gray, fill=white!90!YELLOW, opacity=0.65] (0,0) node {}  circle (1);
\filldraw[line width =.7,  color=white!30!YELLOW, fill=white!85!YELLOW, opacity=1] (0,0) node {$S$}  circle (.45);
\draw[line width =.7, color=white!30!YELLOW] (45:.45) -- (45:1.);
\draw[line width =.7, color=white!30!YELLOW] (-45:.45) -- (-45:1.);
\draw[line width =.7, color=white!30!YELLOW] (135:.45) -- (135:1.);
\draw[line width =.7, color=white!30!YELLOW] (-135:.45) -- (-135:1.);
 \draw[black,thick] (45:1) node[black,circle,fill,scale=0.3] {} -- (45:1.5);
     \draw[black,thick] (-45:1) node[black,circle,fill,scale=0.3] {} -- (-45:1.5);
      \draw[black,thick] (135:1) node[black,circle,fill,scale=0.3] {} -- (135:1.5);
       \draw[black,thick] (-135:1) node[black,circle,fill,scale=0.3] {} -- (-135:1.5);
       \node[gray] at (0:1.2){$b$};  \node[gray] at (90:1.2){$a$};  \node[gray] at (180:1.2){$d$};  \node[gray] at (270:1.2){$c$};
\end{tikzpicture}
= \sum_\chi
\begin{tikzpicture}
[baseline= {(0,0)}]
  \pgfmathsetmacro{\yA}{0.75*sin(45)}
\pgfmathsetmacro{\xA}{cos(45)}
 \begin{scope}[shift={(6.5,0)}]
       \draw[thick,color=BLUE, fill=white!90!BLUE] (-2,0)  ellipse (0.75 and 1);
       \pgfmathsetmacro{\xA}{3/4 * cos(30)}
\pgfmathsetmacro{\yA}{sin(30)}
\pgfmathsetmacro{\xB}{-2 + 3/4 * cos(30)}
\pgfmathsetmacro{\yB}{sin(30)}
\draw[black,thick] (\xA,\yA) -- (\xB,\yB);
\draw[fill=BLUE]   (\xB,\yB) circle (0.05);

       \pgfmathsetmacro{\xAA}{3/4 * cos(-30)}
\pgfmathsetmacro{\yAA}{sin(-30)}
\pgfmathsetmacro{\xBB}{-2 + 3/4 * cos(-30)}
\pgfmathsetmacro{\yBB}{sin(-30)}
\draw[black,thick] (\xAA,\yAA) -- (\xBB,\yBB);
\draw[fill=BLUE]   (\xBB,\yBB) circle (0.05);

    \draw[color=white!70!gray, fill=white!90!gray, opacity=0.65] (0,1) arc (90:270: 0.75 and 1) -- (-2,-1) arc (270:90: 0.75 and 1) -- (0,1) -- cycle; 
\pgfmathsetmacro{\xxA}{3/4 * cos(140)}
\pgfmathsetmacro{\yyA}{sin(140)}
\pgfmathsetmacro{\xxB}{-2 + 3/4 * cos(140)}
\pgfmathsetmacro{\yyB}{sin(140)}
\draw[black,thick] (\xxA,\yyA) -- (\xxB,\yyB);
\draw[fill=BLUE]   (\xxB,\yyB) circle (0.05);

\pgfmathsetmacro{\xxAA}{3/4 * cos(-140)}
\pgfmathsetmacro{\yyAA}{sin(-140)}
\pgfmathsetmacro{\xxBB}{-2 + 3/4 * cos(-140)}
\pgfmathsetmacro{\yyBB}{sin(-140)}
\draw[black,thick] (\xxAA,\yyAA) -- (\xxBB,\yyBB);
\draw[fill=BLUE]   (\xxBB,\yyBB) circle (0.05);

        \draw[thick,color=gray, fill=white!90!gray, opacity=0.65] (0,0) node[black] {} ellipse (0.75 and 1);
        \node[gray,above] at (-2,1) {$a$}; \node[gray,below] at (-2,-1) {$c$}; \node[gray,right] at (-1.25,0) {$b$}; \node[gray,left] at (-2.75,0) {$d$}; 

        \draw[black,thick]  (\xAA,\yAA) arc (45:150:.78 and .7) ;
        \draw[black,thick]  (\xA,\yA) arc (-45:-150:.78 and .7) ;
        \draw[black,thick,double] (0,0.3) -- (0,-0.3);
        \node[black] at (0.3,0) {$\chi$};
    \end{scope}
%%%%%%%%%%%%%%%%%%%%%%%%%%%%%%%
    \begin{scope}[shift={(8.5,0)}]
       
       \pgfmathsetmacro{\xA}{3/4 * cos(30)}
\pgfmathsetmacro{\yA}{sin(30)}
\pgfmathsetmacro{\xB}{+2 + 3/4 * cos(30)}
\pgfmathsetmacro{\yB}{sin(30)}
\draw[black,thick] (\xA,\yA) -- (\xB,\yB);
\draw[fill=BLUE]   (\xB,\yB) circle (0.05);

       \pgfmathsetmacro{\xAA}{3/4 * cos(-30)}
\pgfmathsetmacro{\yAA}{sin(-30)}
\pgfmathsetmacro{\xBB}{+2 + 3/4 * cos(-30)}
\pgfmathsetmacro{\yBB}{sin(-30)}
\draw[black,thick] (\xAA,\yAA) -- (\xBB,\yBB);
\draw[fill=BLUE]   (\xBB,\yBB) circle (0.05);

    \draw[color=white!70!gray, fill=white!90!gray, opacity=0.65] (0,1) arc (90:270: 0.75 and 1) -- (2,-1) arc (270:90: 0.75 and 1) -- (0,1) -- cycle; 
\pgfmathsetmacro{\xxA}{3/4 * cos(140)}
\pgfmathsetmacro{\yyA}{sin(140)}
\pgfmathsetmacro{\xxB}{+2 + 3/4 * cos(140)}
\pgfmathsetmacro{\yyB}{sin(140)}

\pgfmathsetmacro{\xxAA}{3/4 * cos(-140)}
\pgfmathsetmacro{\yyAA}{sin(-140)}
\pgfmathsetmacro{\xxBB}{+2 + 3/4 * cos(-140)}
\pgfmathsetmacro{\yyBB}{sin(-140)}

        \draw[black,thick]  (\xAA,\yAA) arc (45:150:.78 and .7) ;
        \draw[black,thick]  (\xA,\yA) arc (-45:-150:.78 and .7) ;
        \draw[black,thick,double] (0,0.3) -- (0,-0.3);
        \node[black] at (0.3,0) {$\chi$};
        \draw[thick,color=gray, fill=white!90!gray, opacity=0.65] (0,0) node[black] {} ellipse (0.75 and 1);
%yellow face right
\draw[line width =1.0, color=gray, fill=white!90!YELLOW, opacity=0.65] (2,0)  ellipse (0.75 and 1);
%\draw[line width =.7,color=white!30!YELLOW] (135:1) (\xxA+2,\yyA) arc (60:-60:.9 and .75) ;
%\draw[line width =.7,color=white!30!YELLOW] (0:.9) (\xA+2,\yA) arc (45:-45:-1 and .71) ;
\draw[line width =.7, color=white!30!YELLOW] (\xA+2,\yA) -- (\xxAA+2,\yyAA);
\draw[line width =.7, color=white!30!YELLOW] (\xAA+2,\yAA) -- (\xxA+2,\yyA);
\draw[line width =.7, color=white!30!YELLOW, fill=white!85!YELLOW, opacity=1] (2,0) node[white!30!YELLOW] {$S$} ellipse (.45*.75 and .45);
        
    %front, up
    \draw[black,thick] (\xxA,\yyA) -- (\xxB,\yyB);
    \draw[fill=black]   (\xxB,\yyB) circle (0.05);
     %front, down
     \draw[black,thick] (\xxAA,\yyAA) -- (\xxBB,\yyBB);
    \draw[fill=black]   (\xxBB,\yyBB) circle (0.05);

    \end{scope}
    
\end{tikzpicture}\,.
\ee
 More precisely, we inserted the following orthogonal states of the bulk TQFT\footnote{The prefactor $\sqrt{\d_{\chi}}/\d_v$ was chosen to match the conventions of the projector basis. It does not affect the Ward identities.}
\be
\vert \chi \rangle = \frac{\sqrt{\d_\chi}}{\d_v} \
\begin{tikzpicture}[baseline= {(0,0)},scale=.9]
  \pgfmathsetmacro{\yA}{1*sin(45)}
\pgfmathsetmacro{\xA}{cos(45)}
 \draw[color=gray,thick, fill=white!90!gray, opacity=0.5] (0,0) ellipse (1 and 1); 
 % \node[black,circle,fill,scale=0.3]  at (\xA,\yA) {};
 % \node[black,circle,fill,scale=0.3]  at (-\xA,\yA)  {};
 % \node[black,circle,fill,scale=0.3]  at (\xA,-\yA)  {};
 % \node[black,circle,fill,scale=0.3]  at (-\xA,-\yA)  {};
 \node[black,right] at (0,0) {$\chi$};
 \draw[black,thick] plot[smooth, tension=1.5] coordinates{(\xA,\yA) (0,0.35) (-\xA,\yA) }; 
 \draw[black,thick] plot[smooth, tension=1.5] coordinates{(\xA,-\yA) (0,-0.35) (-\xA,-\yA) }; 
 \draw[black,thick,double] (0,0.35) -- (0,-0.35);
\end{tikzpicture} \,  .
\ee

The SymTFT picture \eqref{eq:Sproj} provides a geometric realization of the S-matrix decomposition \eqref{eq:Sdecompreview}: the left half corresponds to the projector $P_{\chi}$, which enforces the Ward identities, while the right half represents the partial wave amplitude $A_{\chi}$, capturing the dynamical information. Taking into account the normalizations of in- and out-states discussed above, $\langle {\rm In}|{\rm In}\rangle\big\vert_\text{TQFT}=\langle {\rm Out}|{\rm Out}\rangle\big\vert_\text{TQFT}=\d_{v}\sqrt{g_{b}g_{d}}$, the projector basis can be computed explicitly as follows:
\be\label{eq:projector}
\left( P_\chi \right)^{a b}_{d c} = \frac{\sqrt{\d_\chi}}{\d_v^2 \sqrt{g_b g_d}} 
\begin{tikzpicture}
[baseline= {(0,0)}]
  \pgfmathsetmacro{\yA}{0.75*sin(45)}
\pgfmathsetmacro{\xA}{cos(45)}
 \begin{scope}[shift={(6.5,0)}]
       \draw[thick,color=BLUE, fill=white!90!BLUE] (-2,0) %node {$\calL$} 
       ellipse (0.75 and 1);
       \pgfmathsetmacro{\xA}{3/4 * cos(30)}
\pgfmathsetmacro{\yA}{sin(30)}
\pgfmathsetmacro{\xB}{-2 + 3/4 * cos(30)}
\pgfmathsetmacro{\yB}{sin(30)}
\draw[black,thick] (\xA,\yA) -- (\xB,\yB);
\draw[fill=black]   (\xB,\yB) circle (0.05);

       \pgfmathsetmacro{\xAA}{3/4 * cos(-30)}
\pgfmathsetmacro{\yAA}{sin(-30)}
\pgfmathsetmacro{\xBB}{-2 + 3/4 * cos(-30)}
\pgfmathsetmacro{\yBB}{sin(-30)}
\draw[black,thick] (\xAA,\yAA) -- (\xBB,\yBB);
\draw[fill=black]   (\xBB,\yBB) circle (0.05);

    \draw[color=white!70!gray, fill=white!90!gray, opacity=0.65] (0,1) arc (90:270: 0.75 and 1) -- (-2,-1) arc (270:90: 0.75 and 1) -- (0,1) -- cycle; 
\pgfmathsetmacro{\xxA}{3/4 * cos(140)}
\pgfmathsetmacro{\yyA}{sin(140)}
\pgfmathsetmacro{\xxB}{-2 + 3/4 * cos(140)}
\pgfmathsetmacro{\yyB}{sin(140)}

\pgfmathsetmacro{\xxAA}{3/4 * cos(-140)}
\pgfmathsetmacro{\yyAA}{sin(-140)}
\pgfmathsetmacro{\xxBB}{-2 + 3/4 * cos(-140)}
\pgfmathsetmacro{\yyBB}{sin(-140)}

        \draw[thick,color=gray, fill=white!90!gray, opacity=0.65] (0,0) node[black] {} ellipse (0.75 and 1);
        %\node[gray,above] at (-1,1) {$\calL_\calM$};
        \draw[black,thick] (\xxAA+.008,\yyAA) -- (\xxBB,\yyBB);
        \draw[fill=black]   (\xxBB,\yyBB) circle (0.05);
        \draw[black,thick] (\xxA+.008,\yyA) -- (\xxB,\yyB);
        -\draw[fill=black]   (\xxB,\yyB) circle (0.05);
        \node[gray,above] at (-2,1) {$a$}; \node[gray,below] at (-2,-1) {$c$}; \node[gray,right] at (-1.25,0) {$b$}; \node[gray,left] at (-2.75,0) {$d$}; 
        \draw[black,thick]  (\xAA,\yAA) arc (45:150:.78 and .7) ;
        \draw[black,thick]  (\xA,\yA) arc (-45:-150:.78 and .7) ;
         \draw[black,thick,double] (0,0.3) -- (0,-0.3);
        \node[black] at (0.3,0) {$\chi$};
    \end{scope}
\end{tikzpicture}
= \frac{\sqrt{\d_\chi}}{\d_v^2 \sqrt{g_b g_d}}
\begin{tikzpicture}[baseline={(0,0)},scale=.9]
    \pgfmathsetmacro{\yA}{1*sin(45)}
\pgfmathsetmacro{\xA}{cos(45)}
 \draw[color=BLUE, thick, fill=white!90!gray] (0,0) ellipse (1 and 1); 
 \node[black,circle,fill,scale=0.3]  at (\xA,\yA) {};
 \node[black,circle,fill,scale=0.3]  at (-\xA,\yA)  {};
 \node[black,circle,fill,scale=0.3]  at (\xA,-\yA)  {};
 \node[black,circle,fill,scale=0.3]  at (-\xA,-\yA)  {};
 \node[black,right] at (0,0) {$\chi$};
 \draw[black,thick] plot[smooth, tension=1.5] coordinates{(\xA,\yA) (0,0.35) (-\xA,\yA) }; 
 \draw[black,thick] plot[smooth, tension=1.5] coordinates{(\xA,-\yA) (0,-0.35) (-\xA,-\yA) }; 
 \draw[black,thick,double] (0,0.35) -- (0,-0.35);
 \node[gray,below] at (0,-1) {$c$}; \node[gray,above] at (0,1) {$a$}; \node[gray,right] at (1,0) {$b$}; \node[gray,left] at (-1,0) {$d$};
 %\node[gray] at (1.25,1) {$\cT_{\calM}^{\cC^*_{\calM}}$};
\end{tikzpicture} =  \left(\varphi^*\right)_{v v \chi}^{d c b} \left( \varphi^* \right)_{v v \chi}^{b a d} \, .
\ee
Here we reduced the 3d picture to 2d TQFT in the second equality, and evaluated the diagram using the dual boundary $F$-symbols and the results from \cite{Huang:2021zvu} in the third equality. When applied to the regular representation, this correctly reproduces \eqref{eq:reviewprojector}. 

It is straightforward to show, following the arguments in \cite{Copetti:2024rqj, Aasen:2020jwb}, that the projector defined this way satisfies the Ward identities. Pictorially, they correspond to the following manipulation,
\bea
\begin{tikzpicture}[baseline= {(0,0)}]
\pgfmathsetmacro{\yA}{0.75*sin(45)}
\pgfmathsetmacro{\xA}{cos(45)}
\begin{scope}[shift={(0,0)}]
       \draw[thick,color=BLUE, fill=white!90!BLUE] (-2,0) ellipse (0.75 and 1);
\draw[BLUE,thick] (-2.75,0) node[BLUE,circle,fill,scale=0.3] {} -- (-1.25,0) node[BLUE,circle,fill,scale=0.3] {};
\node[BLUE,below] at (-2,0) {$\cL$};

\pgfmathsetmacro{\xA}{3/4 * cos(30)}
\pgfmathsetmacro{\yA}{sin(30)}
\pgfmathsetmacro{\xB}{-2 + 3/4 * cos(30)}
\pgfmathsetmacro{\yB}{sin(30)}
\draw[black,thick] (\xA,\yA) -- (\xB,\yB);
\draw[fill=black]   (\xB,\yB) circle (0.05);

\pgfmathsetmacro{\xAA}{3/4 * cos(-30)}
\pgfmathsetmacro{\yAA}{sin(-30)}
\pgfmathsetmacro{\xBB}{-2 + 3/4 * cos(-30)}
\pgfmathsetmacro{\yBB}{sin(-30)}
\draw[black,thick] (\xAA,\yAA) -- (\xBB,\yBB);
\draw[fill=black]   (\xBB,\yBB) circle (0.05);

\draw[color=white!70!gray, fill=white!90!gray, opacity=0.55] (0,1) arc (90:270: 0.75 and 1) -- (-2,-1) arc (270:90: 0.75 and 1) -- (0,1) -- cycle; 
\pgfmathsetmacro{\xxA}{3/4 * cos(140)}
\pgfmathsetmacro{\yyA}{sin(140)}
\pgfmathsetmacro{\xxB}{-2 + 3/4 * cos(140)}
\pgfmathsetmacro{\yyB}{sin(140)}

\pgfmathsetmacro{\xxAA}{3/4 * cos(-140)}
\pgfmathsetmacro{\yyAA}{sin(-140)}
\pgfmathsetmacro{\xxBB}{-2 + 3/4 * cos(-140)}
\pgfmathsetmacro{\yyBB}{sin(-140)}

        \draw[thick,color=gray, fill=white!90!gray, opacity=0.65] (0,0) node[black] {} ellipse (0.75 and 1);
        \draw[black,thick] (\xxAA+.008,\yyAA) -- (\xxBB,\yyBB);
        \draw[fill=black]   (\xxBB,\yyBB) circle (0.05);
        \draw[black,thick] (\xxA+.008,\yyA) -- (\xxB,\yyB);
        -\draw[fill=black]   (\xxB,\yyB) circle (0.05);
       \node[gray,above] at (-2,1) {$a$}; \node[gray,below] at (-2,-1) {$c'$}; 
        \node[gray,right] at (-1.3,-.27) {$b'$}; \node[gray,right] at (-3.25,-.29) {$d'$}; 
        \node[gray,right] at (-1.3,0.27) {$b$}; \node[gray,right] at (-3.25,0.29) {$d$}; 
        \draw[black,thick]  (\xAA,\yAA) arc (45:150:.78 and .7) ;
        \draw[black,thick]  (\xA,\yA) arc (-45:-150:.78 and .7) ;
         \draw[black,thick,double] (0,0.3) -- (0,-0.3);
        \node[black] at (0.3,0) {$\chi$};
    \end{scope}
\node[rotate=45,scale=1.3] at (-3.5,-2.5) {$\underset{\text{deform down}}{=}$}; 
\node[rotate=-45,scale=1.3] at (1.5,-2.5) {$\underset{\text{deform up}}{=}$};
%%%%%%%%%%%%%%%%%%%%%%%%%%%%%RIGHT
\begin{scope}[shift={(6,-5)},scale=.9]
\node[right,scale=.95] at (-7.2,0) {$ \ds \sum_{e'} \left[\cL;v\right]^{ d' e'}_{d a} \left[\cL;v\right]^{b' e'}_{b a} %\sqrt{\frac{\d_\cL\, g_{a}}{g_{e'}}}
$};
\pgfmathsetmacro{\yA}{0.75*sin(45)}
\pgfmathsetmacro{\xA}{cos(45)}
       \draw[thick,color=BLUE, fill=white!90!BLUE] (-2,0) ellipse (0.75 and 1);
\draw[BLUE,thick] plot[smooth, tension=1.5] coordinates{(-2+.6*\xA,.32+\yA) (-2,0.4) (-2-.6*\xA,.32+\yA) };
\node[BLUE,circle,fill,scale=0.3] at (-2-.6*\xA,.32+\yA) {} ;
\node[BLUE,circle,fill,scale=0.3] at (-2+.6*\xA,.32+\yA) {} ;
\node[BLUE,below] at (-2,0.4) {$\cL$};

\pgfmathsetmacro{\xA}{3/4 * cos(30)}
\pgfmathsetmacro{\yA}{sin(30)}
\pgfmathsetmacro{\xB}{-2 + 3/4 * cos(30)}
\pgfmathsetmacro{\yB}{sin(30)}
\draw[black,thick] (\xA,\yA) -- (\xB,\yB);
\draw[fill=black]   (\xB,\yB) circle (0.05);

\pgfmathsetmacro{\xAA}{3/4 * cos(-30)}
\pgfmathsetmacro{\yAA}{sin(-30)}
\pgfmathsetmacro{\xBB}{-2 + 3/4 * cos(-30)}
\pgfmathsetmacro{\yBB}{sin(-30)}
\draw[black,thick] (\xAA,\yAA) -- (\xBB,\yBB);
\draw[fill=black]   (\xBB,\yBB) circle (0.05);

\draw[color=white!70!gray, fill=white!90!gray, opacity=0.55] (0,1) arc (90:270: 0.75 and 1) -- (-2,-1) arc (270:90: 0.75 and 1) -- (0,1) -- cycle; 
\pgfmathsetmacro{\xxA}{3/4 * cos(140)}
\pgfmathsetmacro{\yyA}{sin(140)}
\pgfmathsetmacro{\xxB}{-2 + 3/4 * cos(140)}
\pgfmathsetmacro{\yyB}{sin(140)}

\pgfmathsetmacro{\xxAA}{3/4 * cos(-140)}
\pgfmathsetmacro{\yyAA}{sin(-140)}
\pgfmathsetmacro{\xxBB}{-2 + 3/4 * cos(-140)}
\pgfmathsetmacro{\yyBB}{sin(-140)}

        \draw[thick,color=gray, fill=white!90!gray, opacity=0.65] (0,0) node[black] {} ellipse (0.75 and 1);
        \draw[black,thick] (\xxAA+.008,\yyAA) -- (\xxBB,\yyBB);
        \draw[fill=black]   (\xxBB,\yyBB) circle (0.05);
        \draw[black,thick] (\xxA+.008,\yyA) -- (\xxB,\yyB);
        -\draw[fill=black]   (\xxB,\yyB) circle (0.05);
        \node[gray,above] at (-2,1) {$a$}; \node[gray,below] at (-2,-1) {$c'$}; 
        \node[gray,right] at (-1.5,1) {$e'$}; \node[gray,left] at (-2.5,1) {$e'$}; 
        \node[gray,right] at (-1.35,-0.15) {$b'$}; \node[gray,left] at (-2.65,-0.15) {$d'$}; 
        \draw[black,thick]  (\xAA,\yAA) arc (45:150:.78 and .7) ;
        \draw[black,thick]  (\xA,\yA) arc (-45:-150:.78 and .7) ;
         \draw[black,thick,double] (0,0.3) -- (0,-0.3);
        \node[black] at (0.3,0) {$\chi$};
    \end{scope}
\pgfmathsetmacro{\yA}{0.75*sin(45)}
\pgfmathsetmacro{\xA}{cos(45)}
%%%%%%%%%%%%%%%%%%%%%%%%LEFT
\begin{scope}[shift={(-2.7,-5)},scale=.9]
\node[left,scale=.95] at (-3.1,0) {$ \ds \sum_{e} \left[\cL;v\right]^{d e}_{d' c'} \left[\cL;v\right]^{b e }_{b' c'} %\sqrt{\frac{\d_\cL\,g_{c'}}{g_{e}}}
$};
       \draw[thick,color=BLUE, fill=white!90!BLUE] (-2,0) ellipse (0.75 and 1);
\pgfmathsetmacro{\Y}{0}
\pgfmathsetmacro{\X}{cos(0)}
\pgfmathsetmacro{\YY}{0.75*sin(180)}
\pgfmathsetmacro{\XX}{2+cos(180)}
\draw[BLUE,thick] plot[smooth, tension=1.5] coordinates{(-2+.6*\xA,-.32-\yA) (-2,-0.4) (-2-.6*\xA,-.32-\yA) };
\node[BLUE,circle,fill,scale=0.3] at (-2-.6*\xA,-.32-\yA) {} ;
\node[BLUE,circle,fill,scale=0.3] at (-2+.6*\xA,-.32-\yA) {} ;
\node[BLUE,above] at (-2,-0.4) {$\cL$};

\pgfmathsetmacro{\xA}{3/4 * cos(30)}
\pgfmathsetmacro{\yA}{sin(30)}
\pgfmathsetmacro{\xB}{-2 + 3/4 * cos(30)}
\pgfmathsetmacro{\yB}{sin(30)}
\draw[black,thick] (\xA,\yA) -- (\xB,\yB);
\draw[fill=black]   (\xB,\yB) circle (0.05);

\pgfmathsetmacro{\xAA}{3/4 * cos(-30)}
\pgfmathsetmacro{\yAA}{sin(-30)}
\pgfmathsetmacro{\xBB}{-2 + 3/4 * cos(-30)}
\pgfmathsetmacro{\yBB}{sin(-30)}
\draw[black,thick] (\xAA,\yAA) -- (\xBB,\yBB);
\draw[fill=black]   (\xBB,\yBB) circle (0.05);

\draw[color=white!70!gray, fill=white!90!gray, opacity=0.55] (0,1) arc (90:270: 0.75 and 1) -- (-2,-1) arc (270:90: 0.75 and 1) -- (0,1) -- cycle; 
\pgfmathsetmacro{\xxA}{3/4 * cos(140)}
\pgfmathsetmacro{\yyA}{sin(140)}
\pgfmathsetmacro{\xxB}{-2 + 3/4 * cos(140)}
\pgfmathsetmacro{\yyB}{sin(140)}

\pgfmathsetmacro{\xxAA}{3/4 * cos(-140)}
\pgfmathsetmacro{\yyAA}{sin(-140)}
\pgfmathsetmacro{\xxBB}{-2 + 3/4 * cos(-140)}
\pgfmathsetmacro{\yyBB}{sin(-140)}

        \draw[thick,color=gray, fill=white!90!gray, opacity=0.65] (0,0) node[black] {} ellipse (0.75 and 1);
        \draw[black,thick] (\xxAA+.008,\yyAA) -- (\xxBB,\yyBB);
        \draw[fill=black]   (\xxBB,\yyBB) circle (0.05);
        \draw[black,thick] (\xxA+.008,\yyA) -- (\xxB,\yyB);
        -\draw[fill=black]   (\xxB,\yyB) circle (0.05);
        \node[gray,above] at (-2,1) {$a$}; \node[gray,below] at (-2,-1) {$c'$}; 
        \node[gray,right] at (-1.5,-1) {$e$}; \node[gray,left] at (-2.5,-1) {$e$}; 
        \node[gray,right] at (-1.35,0.15) {$b$}; \node[gray,left] at (-2.65,0.15) {$d$}; 
        \draw[black,thick]  (\xAA,\yAA) arc (45:150:.78 and .7) ;
        \draw[black,thick]  (\xA,\yA) arc (-45:-150:.78 and .7) ;
         \draw[black,thick,double] (0,0.3) -- (0,-0.3);
        \node[black] at (0.3,0) {$\chi$};
    \end{scope}
\end{tikzpicture}
\eea
The equality between the two figures can be explicitly verified using the pentagon-like identity i.e.~the second formula in \eqref{eq:pentagonlikephistar}.

In Section \ref{sec: Bootstrap}, we will use these projectors to implement the S-matrix bootstrap with the fusion category symmetry. \footnote{The projectors and the Ward identities have other applications. For instance, they are needed to ensure that integrability survives even in the twisted sectors of categorical defects in integrable QFTs: more precisely, given a symmetry line $\cL$, one can construct mutually commuting transfer matrices $T_\cL(u)$ on the $\cL$-twisted Hilbert space if and only if the symmetry Ward identities are satisfied. We hope to report on this soon \cite{inprogress}.}

\section{S-matrix bootstrap with fusion category symmetry}\label{sec: Bootstrap}
The S-matrix bootstrap program aims at mapping out the space of possible scattering amplitudes imposing constraints coming from general principles such as Lorentz invariance, unitarity and causality.
Many different classes of theories have been studied with this approach, for a review see \cite{Kruczenski:2022lot}.
Although it is well-understood how to include the existence of global group symmetries in terms of irreducible representations and projectors, theories with generalized symmetries have been a completely uncharted territory for the S-matrix bootstrap. Here we start the systematic exploration of the S-matrix in such theories and study the space of scattering amplitudes for (1+1)d gapped theories with categorical symmetries.

As we saw in the previous section, the presence of categorical symmetries highly constrains the spectrum of the theory and --as we put forward in \cite{Copetti:2024rqj}-- has important implications for scattering amplitudes, giving modified crossing rules. 

In the following we review the consistency conditions we should impose on the amplitudes. First, we impose the global symmetry given by the fusion category $\mathcal C$ by projecting the amplitude into the different fusion channels. We then review what unitarity, (modified) crossing and analyticity imply for the amplitudes in different channels.  

\subsection{Review of S-matrix bootstrap in (1+1)d}\label{subsec:reviewB}
In this section we review how to bootstrap amplitudes in (1+1)d. The reader familiar with S-matrix bootstrap can safely skip this section and proceed to the next one. 

\begin{figure}[t]
\begin{subfigure}{0.49\textwidth}\centering
\includegraphics[width=.35\textwidth]{ 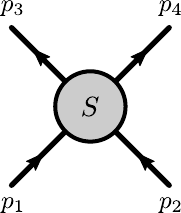}
\caption{}\label{fig:Smatrixa}
\end{subfigure}\hspace{0.3cm}
\begin{subfigure}{0.49\textwidth}\centering
\includegraphics[width=.8\textwidth]{ 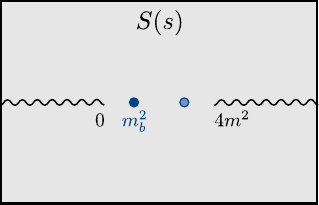}
\caption{}\label{fig:Smatrixb}
\end{subfigure}
\caption{(a) Two-body scattering of particles with the same mass $m$ giving the amplitude $S(s)$. (b) Analytic structure of the amplitude in the complex $s$ plane. Singularities of the amplitude $S(s)$ lie on the real axis, including two-particle branch cuts starting at $s=0,4m^2$. Possible bound states appear as poles in $0<s<4m^2$, here we show in blue one bound state of mass $m_b$ and its crossing symmetric image (lighter blue).
}
\end{figure}

We consider the two-body scattering of the lightest particle of mass $m$ in a gapped QFT, as shown in Figure~\ref{fig:Smatrixa}. Because of Lorentz invariance, this scattering amplitude should be a function of the three Mandelstam invariants $s=(p_1+p_2)^2$, $t=(p_1-p_3)^2$ and $u=(p_1-p_4)^2$ satisfying the usual relation $s+t+u=4m^2$. However, in 1+1d momentum conservation implies that $\{p_1,p_2\}=\{p_3,p_4\}$ so that we can set $u=0$ and hence we have a single independent kinematic variable which we choose to be the center of mass energy squared, $s$. This $2 \to 2$ amplitude $S(s)$ is highly constrained by the basic principles of analyticity, crossing and unitarity which we briefly explain below.

\paragraph{Analyticity} emerges form causality and states that the physical amplitude is the boundary value of a complex function $S(s)$ in the complex plane $s\in\mathbb C$. Moreover, this function is analytic except for possible singularities associated to on-shell processes like poles for bound states and branch cuts for multiparticle thresholds.\footnote{Since we are considering scattering of the lightest particle we exclude anomalous thresholds which would give extra singularities associated to Landau diagrams \cite{Landau:1959fi,Cutkosky:1960sp}. In contrast with higher dimensions, the latter singularities are poles in (1+1)d.} An example of the analytic structure for $S(s)$ in a theory with one bound state is given in Figure~\ref{fig:Smatrixb}.

\paragraph{Crossing symmetry} relates different scattering processes in which we exchange in and out particles, stemming from the interpretation of a particle as an antiparticle moving backward in time. In terms of the analytic function $S(s)$, it tells us that different kinematic channels are boundary values of the same analytic function. For instance, the s-channel $2\to2$ amplitude with $s\geq4m^2$ is related by analytic continuation to the t-channel one where $t=4m^2-s\geq4m^2$.\footnote{Crossing symmetry of $2\to2$ scattering of massive (local) particles has been proven in \cite{Bros:1965kbd}. As explained in \cite{Caron-Huot:2023ikn}, in more general cases crossing symmetry is highly non-trivial and relates scattering amplitudes to other type of asymptotic observables.} The explicit relation is then $S(s)=S(t=4m^2-s)$.

\paragraph{Unitarity} tells us that the full S-matrix $\mathbb{S}$ giving the map between in and out states $|p_1,\ldots,p_n\rangle_{in}=\mathbb S|p_1,\ldots,p_n\rangle_{out}$ is a unitary operator, i.e. $\mathbb S^\dagger \mathbb S=\unit$. This is indeed what is required from conservation of probabilities in quantum mechanics. For the $2\to2$ amplitude ${}_{out}\langle p_3,p_4|p_1,p_2 \rangle_{in}=(2\pi)^2 2\sqrt s\sqrt{s-4m^2}\delta^2(p_1+p_2-p_3-p_4) S(s)$ unitarity implies that $|S(s)|^2\leq1$ for physical values of the center of mass energy $s\geq4m^2$. Using real analyticity $S^*(s)=S(s^*)$ one can also write it as $S(s)S(s^*)\leq1$.

\paragraph{Bootstrap} Now that we know the conditions we want to impose, we can bound the space of consistent amplitudes. This can be done by choosing a family of functionals\footnote{Typical choices for functionals are effective quartic coupling given by the amplitude itself $S(s_*)$ for $0<s_*<4m^2$ (where the amplitude is real), scattering lengths, and cubic couplings given by the residue at bound state poles.} of the S-matrix $\mathcal F[S(s)]$ and maximizing them. In practice, we write down an ansatz for the amplitude which trivializes analyticity and crossing and perform the maximization imposing unitarity separately as pioneered in \cite{Paulos:2016but}. Since the analytic structure of the amplitude depends on the spectrum of the theory, one needs to first fix the latter. For example, for theories without bound states one can use Cauchy's theorem to write the dispersion relation\footnote{In case the amplitude does not fall off fast enough at infinity one can include subtractions.} 
\begin{equation}\label{eq:dispersion}
    S(s)=\frac{1}{2\pi i} \oint ds' \frac{S(s')}{s'-s}=S_\infty+\frac{1}{2\pi i}\int_{4m^2}^\infty ds'\[\frac{1}{s'-s}+\frac{1}{s'-4m^2+s}\] \, D(s')\,,
\end{equation}
where $D(s)$ is the discontinuity across the branch cut $D(s)\equiv {\rm Disc}\,S(s)$, which using real analyticity reduces to the imaginary part $D(s)=2i\Im S(s)$. Most of the time the maximization has to be done numerically, so that one writes a discretized ansatz for the amplitude and imposes unitarity in a grid of physical values of the energy. One way of making a discretized ansatz is to approximate $D(s)$ in \eqref{eq:dispersion} by a piece-wise linear function,
\begin{equation}
D(s)= \frac{(s-s_{j-1})D_j+(s_j-s)D_{j-1}}{s_j-s_{j-1}} \qquad (s_{j-1}\leq s\leq s_j)\,.
\end{equation}
Using the dispersion relation \eqref{eq:dispersion}, this provides an ansatz for the S-matrix, parametrized by discrete data $\{D_j\}$.
With this crossing symmetric, analytic ansatz, one can proceed to the optimization of a functional subject to unitarity (imposed at discrete points $s=s_j$). Explicit examples for the numerical implementation can be found in appendix~\ref{sec:appBoot}. This is known as the \textit{primal approach}, in which one explores the space of allowed amplitudes from the inside, by constructing explicit amplitudes that satisfy all axioms. The bounds we find for $\mathcal F$ in this way are however not rigorous and may vary as we increase the number of parameters in the ansatz.

A complimentary approach called \textit{dual} allows us to establish rigorous bounds. This approach was first proposed in \cite{Cordova:2019lot} and then developed in various contexts in \cite{Guerrieri:2020kcs,Kruczenski:2020ujw,He:2021eqn,EliasMiro:2021nul,Guerrieri:2021tak,Correia:2022dyp,Cordova:2023wjp,Guerrieri:2024ckc}. Here we explain the key points and give details on the derivation and examples in appendix~\ref{sec:appBoot}. The idea is to write a dual functional $\mathcal F_d$ satisfying the inequality $\mathcal F \leq \mathcal F_d$, so that we are approaching the optimal bound from above. The dual functional $\mathcal F_d$ is written in terms of dual variables $K(s)$ which act as Lagrange multipliers for the S-matrix constraints, so that the optimization problem satisfies $\max_{S} \mathcal F \leq \min_{K} \mathcal F_d$. The dual functional takes the form 
\begin{equation}
    \mathcal F_d = \frac{2}{\pi} \int\limits_{4m^2}^\infty ds | K(s) | \,,
\end{equation}
where $K(s)$ is analytic in the cut plane (except for possible poles if the primal functional is of the form $S(s_*)$), decays fast enough at large $s$ (at least like $s^{-3/2}$) and obeys an anticrossing condition $K(4m^2-s)=-K(s)$.\footnote{In the case with global symmetry where we have amplitudes in different channels satisfying $S_\chi(4m^2-s)= C_{\chi \chi'} S_{\chi'}(s)$, the dual variables $K_a(s)$  should obey anticrossing with the transpose of the crossing matrix $K_\chi(4m^2-s)=-C_{\chi' \chi} K_{\chi'}(s)$.} It is simple to implement the dual optimization numerically by writing an ansatz for $K(s)$ with the former properties and performing the integration by quadrature.

\paragraph{Global symmetry} So far we have focused on the scattering of identical particles. If we consider theories with a global symmetry where different particles are grouped into symmetry multiplets some modifications are needed. The first modification is that there will be more analytic functions (as many as irreducible representations the two external particles can form) related to each other by crossing. Take for instance the two-body scattering of particles transforming in the vector representation of $O(N)$. We can write this amplitude in terms of the allowed scattering channels, namely singlet, antisymmetric and symmetric representations
\begin{gather}
    S^{ij}_{lk}(s)=\sum\limits_\chi A_\chi(s) (P_\chi)^{ij}_{lk}\,,\\
   (P_\text{sing})^{ij}_{lk}=\frac{1}{N} \delta_{ij} \delta_{kl} \,,\quad (P_\text{anti})^{ij}_{lk}=\frac{\delta_{ik} \delta_{jl}-\delta_{il} \delta_{kj}}{2}  \,,\quad (P_\text{sym})^{ij}_{lk}=\frac{\delta_{ik} \delta_{jl}+\delta_{il} \delta_{kj}}{2}-\frac{1}{N} \delta_{ij} \delta_{kl}
\end{gather}
where $i,j,k,l=1,\ldots,N$. In this projector basis, unitarity simply reads $|A_\chi(s)|^2\leq1$ for each representation. Crossing symmetry $S^{ij}_{lk}(s)=S^{jk}_{li}(4m^2-s)$ then gives non-trivial relations between the different channel amplitudes 
\begin{equation}\label{eq:crossON}
    A_\chi(4m^2-s) = \sum\limits_{\chi'} C_{\chi\chi'} A_{\chi'}(s)\,,
\end{equation}
with $C_{\chi\chi'}$ the crossing matrix which for $O(N)$ reads
\begin{equation}
    C=\left(
\begin{array}{ccc}
\frac{1}{N}&-\frac{N}{2}+\frac{1}{2}\,\,&\frac{N}{2}+\frac{1}{2}-\frac{1}{N}\\
-\frac{1}{N}&\frac{1}{2}&\frac{1}{2}+\frac{1}{N}\\
\frac{1}{N}&\frac{1}{2}&\frac{1}{2}-\frac{1}{N}
\end{array}
\right)\,.
\end{equation}
In the following we explain how these S-matrix properties translate into the setting where the global symmetry is described by a fusion category.

%%%%%%%%%%%%%%%%%%%%%%%%%%%%%%%%%%%%%%%%%%%%%%%%%%%%%%%%%%%%%%%%%%%%
\subsection{Projector basis}
As we have seen in Section \ref{sec: symaction}, to study kink scattering in a massive (1+1)d QFT with categorical symmetry $\cC$, we need few ingredients. First, we have the module category $\calM$ over $\cC$, which encodes the information about possible boundary conditions $a$-$d$ we identify with the infrared vacua and physically tell us the pattern of spontaneous breaking of the symmetry $\cC$. Then, we have the kink $K_{a,b}^v$ interpolating between vacua which acts as a boundary changing operator to which we associate a simple line $v$ in the dual category $\cC_\calM^\ast$. Finally, to ensure kink scattering amplitudes are compatible with the symmetry $\cC$, we expand the amplitude in projectors $P_\chi$ which solve the corresponding Ward identities: 
\be\label{eq:Sproj2}
S^{ab}_{dc}(s) = \sum_\chi A_\chi(s) \left( P_\chi \right)^{a b}_{d c} \, ,
\ee
where $\chi \in \cC^*_{\calM}$ specifies the fusion channel $v \times v \supset \chi$ and $A_\chi(s)$ are the partial amplitude containing the dynamical information. The projectors can be explicitly written in terms of the boundary $F$-symbols for the dual category $\varphi^*$ as follows
\bea
\left( P_\chi \right)^{a b}_{d c} = \frac{\sqrt{\d_\chi}}{\d_v^2 \sqrt{g_b g_d}} 
\begin{tikzpicture}[baseline={(0,0)},scale=.95]
    \pgfmathsetmacro{\yA}{1*sin(45)}
\pgfmathsetmacro{\xA}{cos(45)}
 \draw[color=BLUE, thick, fill=white!90!gray] (0,0) ellipse (1 and 1); 
 \node[black,circle,fill,scale=0.3]  at (\xA,\yA) {};
 \node[black,circle,fill,scale=0.3]  at (-\xA,\yA)  {};
 \node[black,circle,fill,scale=0.3]  at (\xA,-\yA)  {};
 \node[black,circle,fill,scale=0.3]  at (-\xA,-\yA)  {};
 \node[black,right] at (-0.05,0) {$\chi$};
  \node[black] at (\xA-.1,\yA-.4) {$v$};
  \node[black] at (-\xA+.1,\yA-.4) {$v$};
  \node[black] at (\xA-.1,-\yA+.45) {$v$};
  \node[black] at (-\xA+.1,-\yA+.4) {$v$};
 \draw[black,thick] plot[smooth, tension=1.5] coordinates{(\xA,\yA) (0,0.35) (-\xA,\yA) }; 
 \draw[black,thick] plot[smooth, tension=1.5] coordinates{(\xA,-\yA) (0,-0.35) (-\xA,-\yA) }; 
 \draw[black,thick,double] (0,0.35) -- (0,-0.35);
 \node[gray,below] at (0,-1) {$c$}; \node[gray,above] at (0,1) {$a$}; \node[gray,right] at (1,0) {$b$}; \node[gray,left] at (-1,0) {$d$};
% \node[gray] at (1.25,1) {$\cT_{\calM}^{\cC^*_{\calM}}$};
\end{tikzpicture} =   \left( \varphi^* \right)_{v v \chi}^{b a d} \left(\varphi^*\right)_{v v \chi}^{d c b} 
=\sqrt{\d_a \d_c}\, \d_\chi 
    \begin{bmatrix}
     v & v & \chi\\
    d & b & a
    \end{bmatrix}_\ast
    \begin{bmatrix}
    v & v & \chi\\
    d & b & c
    \end{bmatrix}_\ast \,,
\eea
where in the last equation we used $\begin{bmatrix}     v_1 & v_2 & v_3\\ v_4 & v_5 & v_6 \end{bmatrix}_\ast$ to denote the tetrahedral symbols of the dual category  $\cC^*_{\calM}$. If we consider the case in which all symmetries are spontaneously broken, we have the regular module category for which  $\cC_\calM^\ast=\cC$ and we recover our previous result \cite{Copetti:2024rqj}
\begin{equation}\label{eq:Pdef}
    (P_\chi )^{ab}_{dc} =\frac{\sqrt{\d_\chi}}{\d_v^2 \sqrt{\d_b \d_d}}
\begin{tikzpicture}[baseline={(0,0)},scale=.95]
\draw[BLUE,thick] (0,0) circle (1);    
 \pgfmathsetmacro{\yA}{1*sin(45)}
\pgfmathsetmacro{\xA}{cos(45)}
  \node[BLUE] at (\xA-.1,\yA-.4) {$v$};
  \node[BLUE] at (-\xA+.1,\yA-.4) {$v$};
  \node[BLUE] at (\xA-.1,-\yA+.4) {$v$};
  \node[BLUE] at (-\xA+.1,-\yA+.4) {$v$};
  \draw[BLUE,thick] plot[smooth, tension=1.5] coordinates{(\xA,\yA) (0,0.35) (-\xA,\yA) }; 
 \draw[BLUE,thick] plot[smooth, tension=1.5] coordinates{(\xA,-\yA) (0,-0.35) (-\xA,-\yA) }; 
  \draw[BLUE,thick,double] (0,0.35) -- (0,-0.35);
  \node[BLUE,right] at (-0.05,0) {$\chi$};
 \node[BLUE,below] at (0,-1) {$c$}; \node[BLUE,above] at (0,1) {$a$}; \node[BLUE,right] at (1,0) {$b$}; \node[BLUE,left] at (-1,0) {$d$};
\end{tikzpicture}
    =\sqrt{\d_a \d_c}\, \d_\chi 
    \begin{bmatrix}
     v & v & \chi\\
    d & b & a
    \end{bmatrix}
    \begin{bmatrix}
    v & v & \chi\\
    d & b & c
    \end{bmatrix} \,.
\end{equation}

Let us now see how unitarity and (modified) crossing are implemented in this basis:
\paragraph{Unitarity} 
As explained in the previous section, unitarity of the full S-matrix $\mathbb S^\dagger\mathbb S=\unit$ implies the following inequality for the two-body scattering
\be
\sum_e S^{eb}_{dc} (s) S^{ab}_{de} (s^*) \leq \delta_{ac}\,,
%\quad \theta\in\mathbb R\,,
\ee
which in the projector basis \eqref{eq:Sproj2} is simply:
\be \label{eq:unitarity}
A_\chi(s) A_\chi(s^*) \leq 1  \quad  (s\geq4m^2)\, .
\ee

\paragraph{(Modified) crossing} As shown in our previous work \cite{Copetti:2024rqj} for the regular representation and in Section \ref{sssec:modifiedcross} for the generic case, the modified crossing equations for the S-matrix read\footnote{In \cite{Copetti:2024rqj} we used the rapidity $\theta$ related to $s$ as $s=4m^2\cosh^2(\theta/2)$, in this variable crossing maps $\theta$ to $i\pi-\theta$.}
\be \label{eq: modifiedcrossingsmat}
   S^{ab}_{dc}(s) = \sqrt{\frac{g_a g_c}{g_b g_d}}\, S^{bc}_{ad}(4m^2-s) \, ,
\ee
where $g_a$ is the relative Euler term defined in \eqref{eq:gdef}, which coincides with the quantum dimension $\d_a$ in the case of the regular representation. For the partial amplitudes $A_\chi(s)$ it amounts to the relation
\begin{equation}\label{eq:modcrossing}
     A_\chi(s) = \sum\limits_{\chi'} \d_{\chi'}
     \begin{bmatrix}
     v & v & \chi\\
    v & v & \chi'
    \end{bmatrix}_\ast A_{\chi'}(4m^2-s)\,.
\end{equation}
which can be obtained by comparing projectors in the $s-$ and $t-$ channel.  This gives us the analogue of the crossing matrix $C_{\chi\chi'}$ in \eqref{eq:crossON} in terms of the module category data.

\paragraph{Analyticity} The last main ingredient is analyticity: We assume the amplitudes $A_\chi(s)$ are analytic except for possible singularities related to stable particles (such as poles for bound states) and physical thresholds (branch cuts from multiparticle intermediate states). This means that each $A_\chi(s)$ is analytic in the physical sheet where we evaluate unitarity (see Figure~\ref{fig:Smatrixb}) except for possible bound state poles between $0$ and $4m^2$.\footnote{In the examples discussed here the external particles are also the lightest ones, so that there are no Landau singularities.}

\paragraph{Bootstrap}The bootstrap problem is the following: given a category, what is the space of possible fusion channel amplitudes $A_\chi(\theta)$ compatible with analyticity, unitarity \eqref{eq:unitarity} and (modified) crossing \eqref{eq:modcrossing}? 

We will explore the space of possible amplitudes by maximizing functionals of the form $\mathcal F=\sum_\chi n_\chi A_\chi(s_*)$, where we evaluate the partial amplitudes at a fixed energy $s_*$ and repeat the maximization for many coefficients $n_\chi$.\footnote{This is the \textit{normal} type of functionals considered in \cite{Cordova:2019lot}, which highlight the points at the boundary with more curvature. We also use the dual approach for the radial type where we fix a direction $A_\chi(s_*)=t\, n_\chi$ and maximize $t$.} In the following we bound the space of amplitudes for theories with $\mathcal A_n$ and Fibonacci categories. Since these spaces include integrable amplitudes, we first state the conditions integrable amplitudes with these categories should satisfy.

%\subsection
\paragraph{Yang-Baxter equations}
If the theory is integrable, the $2\rightarrow2$ amplitude obeys the Yang-Baxter equation, imposing the factorization of three-body scattering in different orders. Expanding the amplitudes in terms of projectors gives the following equation
\begin{equation}
    \sum\limits_{\eta_i} A_{\eta_1}(\theta_2) A_{\eta_2}(\theta_1+\theta_2) A_{\eta_3}(\theta_1) \quad \includegraphics[height=2.5cm,valign=c]{ 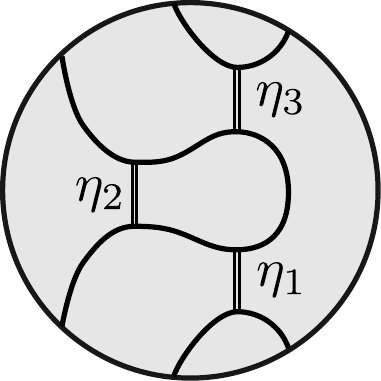}
    =\sum\limits_{\chi_i} A_{\chi_1}(\theta_1) A_{\chi_2}(\theta_1+\theta_2) A_{\chi_3}(\theta_2) \quad \includegraphics[height=2.5cm,valign=c]{ 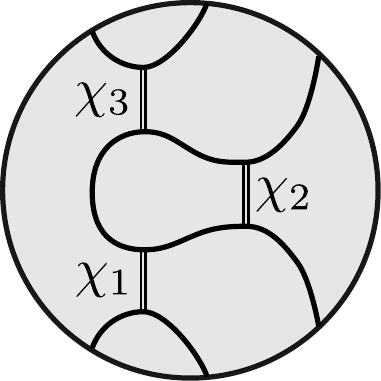}
\end{equation}
Through a series of F-moves, one can bring both sides of the equation to the following network of lines 
\begin{equation*}
    \includegraphics[height=2.5cm,valign=c]{ 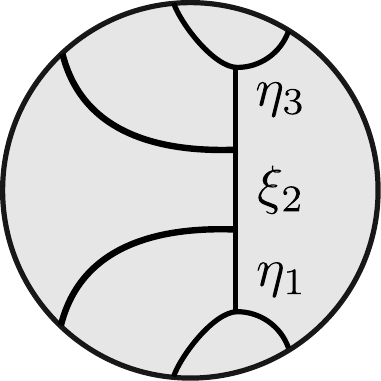}
\end{equation*}
and get the condition the partial amplitudes $A_\chi(\theta)$ should satisfy in case the theory is integrable:
\begin{equation}\label{eq:YBchi}
\begin{split}
    &\sum\limits_{\eta_2} \d_{\eta_2} 
    \begin{bmatrix}
     v & v & \eta_2\\
    \xi_2 & v & \eta_1
    \end{bmatrix}_\ast
    \begin{bmatrix}
     v & v & \eta_2\\
    \xi_2 & v & \eta_3
    \end{bmatrix}_\ast
    A_{\eta_1}(\theta_2) A_{\eta_2}(\theta_1+\theta_2) A_{\eta_3}(\theta_1) = \\
    & \sum\limits_{\chi_1,\chi_2,\chi_3} \d_{\chi_1} \d_{\chi_2} \d_{\chi_3} 
    \begin{bmatrix}
     v & v & \chi_2\\
    \xi_2 & v & \chi_1
    \end{bmatrix}_\ast
    \begin{bmatrix}
     v & v & \chi_2\\
    \xi_2 & v & \chi_3
    \end{bmatrix}_\ast
    \begin{bmatrix}
     v & v & \chi_1\\
    \xi_2 & v & \eta_1
    \end{bmatrix}_\ast
    \begin{bmatrix}
     v & v & \chi_3\\
    \xi_2 & v & \eta_3
    \end{bmatrix}_\ast
    A_{\chi_1}(\theta_1) A_{\chi_2}(\theta_1+\theta_2) A_{\chi_3}(\theta_2) \,.
\end{split}
\end{equation}
As we shall see, integrable amplitudes appear at special points of the allowed space of amplitudes.

\subsection{\texorpdfstring{$\mathcal A_n$ category}{An category}}
One of the simplest categories to consider is $\mathcal A_n$, with $n-1$ symmetry lines labeled by $a=0,1/2,\ldots$. We study the case in which we have $n-1$ vacua, corresponding to the regular representation of the module category discussed around \eqref{eq:Fregular}. This is the category present in the $\phi_{1,3}$ deformations of minimal models, described by integrable amplitudes that make their appearance at special features of our bounds. We focus on the scattering of kinks $K_{ab}$ interpolating between neighbouring vacua. Since they form a symmetry multiplet they all have the same mass $m$ \cite{Cordova:2024vsq}. 

The symmetry line associated to such kinks is $v=1/2$, which results in two possible fusion channels $\chi=0,1$. The modified crossing rules \eqref{eq:modcrossing} for the partial amplitudes $A_\chi(s)$ then read
\begin{equation}\label{eq:Anamps}
    \begin{pmatrix}
     A_0(s) \\
     A_1(s) 
    \end{pmatrix}
    =\frac{1}{\d_{1/2}}
    \begin{pmatrix}
     1 & \d_1 \\
    1 & -1
    \end{pmatrix}
    \begin{pmatrix}
     A_0(4m^2-s) \\
     A_1(4m^2-s) 
    \end{pmatrix}\,,\qquad \d_a=\frac{\sin \pi(2a+1)/n}{\sin\pi/n}\,.
\end{equation}
In the equation above we have evaluated the fact that in the regular representation the dual boundary $F$-symbols are the same as the bulk ones. Their explicit expressions can be found in appendix A of \cite{Copetti:2024rqj}. 

\begin{figure}[t]
\centering
\includegraphics[width=.7\textwidth]{ 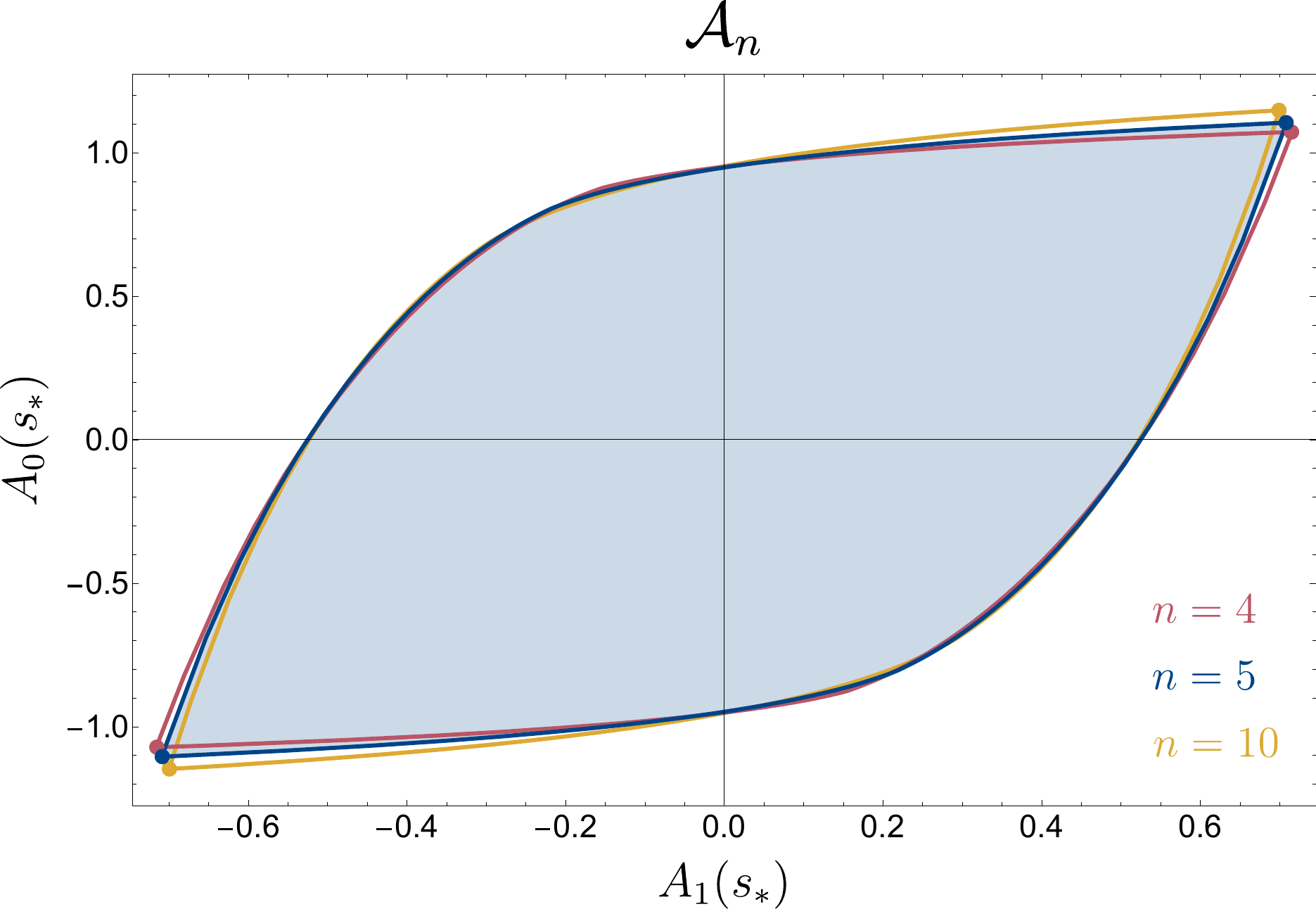}\vspace{0.2cm}
\caption{Space of allowed amplitudes $A_{\chi}(s)$ for $\mathcal A_n$ category, section at $s_*\approx3.41m^2$. Amplitudes consistent with analyticity, unitarity and modified crossing lie inside these bounds. The lines indicate the optimal bounds found with the dual/primal approaches explained in the main text, different colors show the dependence on $n$. The points at the vertices are integrable solutions of which the bottom left ones correspond to the $\mathcal M_n-\lambda\phi_{1,3}$ RG flows.
}
\label{fig:An}
\end{figure}

It is straightforward to bootstrap the space of amplitudes consistent with analyticity, unitarity and the modified crossing equations above. To fix the analytic properties, we consider theories where the kinks do not form bound states (in any fusion channel), so that the partial amplitudes $A_\chi(s)$ are analytic in the cut plane. To explore the space of allowed amplitudes one can maximize functionals of the form $\mathcal F=\sum_\chi n_\chi A_\chi(s_*)$ for many different $n_\chi$. The primal and corresponding dual bootstrap approaches described in \ref{subsec:reviewB} result in the optimal bounds shown in Figure~\ref{fig:An}.\footnote{Notice that the optimization procedure not only produces the bound but also gives the extremal amplitudes saturating those bounds, so that we can study generic features of these amplitudes.}

The first feature is that the lower and upper bounds are related by an overall change of sign $A_\chi(s)\rightarrow-A_\chi(s)$, since multiplying by an overall minus gives an equally consistent amplitude. A more interesting feature is that for each $n$ there are two clear vertices, which are actually the only two integrable points at the boundary of the allowed space. They are again related by an overall sign and are the Yang-Baxter solutions first proposed in \cite{Bernard:1990cw,Zamolodchikov:1991vh,Fendley:1993xa} whose modified crossing version we wrote in \cite{Copetti:2024rqj}. The solutions located on the lower vertices read
\begin{gather}
     A_0(\theta)=Z(\theta)\sinh\(\frac{i\pi+\theta}{n}\)\,, \quad A_1(\theta)=Z(\theta)\sinh\(\frac{i\pi-\theta}{n}\)\,, \label{eq:YBAn} \\
    Z(\theta)=\frac{1}{\sinh{\frac{\theta-i\pi}{n}}} \exp\left\{\frac{i}{2} \int\limits_{-\infty}^\infty \frac{dk}{k}\, \sin{k\theta}\, \frac{\sinh{\frac{k\pi}{2}(n-1)}}{\sinh{\frac{n k\pi}{2}\, \cosh{\frac{k\pi}{2}}}} \right\}\,,\nonumber
\end{gather}
where we used the rapidity variable $\theta$ satisfying $s=4m^2\cosh^2\frac{\theta}{2}$. Importantly, these amplitudes describe known theories, namely the $\phi_{1,3}$ deformations of unitary minimal models $\mathcal M_n$. The physical models for their minus sign counterparts are not known.
In a by now standard result in S-matrix bootstrap, the rest of the boundary amplitudes are not integrable but saturate two-particle unitarity.\footnote{In theories that are not integrable there should be particle production, but the optimization problem is blind to this fact since we are not including multi-particle data. The bounds are however rigorous and the extremal amplitudes are perfectly consistent at low energies.}
They exhibit a relatively simple analytic structure\footnote{To be more precise, there is a single tower of poles and zeros in the imaginary rapidity axis, with the ``fractal" structure discussed in \cite{Cordova:2019lot} and only pair of zeros (poles) at $\theta=i\pi(2n+1)+\alpha$ ($\theta=i\pi2n+\alpha$) for integer $n$ and real $\alpha$.} and crucially, an absence of oscillatory behaviour in $\theta$, in contrast with the results of other global group symmetries such as $\mathbb Z_2$, $\mathbb Z_4$ and $O(N)$ \cite{Homrich:2019cbt,Bercini:2019vme,Cordova:2018uop,Cordova:2019lot}. This is reminiscent of the fact that in the original proposal for the $\mathcal M_n-\phi_{1,3}$ amplitudes $S^{ab}_{dc}(\theta)$ there was an oscillating factor of the form $\(\frac{\d_a \d_c}{\d_b \d_d}\)^\frac{i\theta}{2\pi}$ which nicely disappears when using the correct crossing rules.

Although the only known physical model we have made contact with are the $\phi_{1,3}$ deformations of minimal models, let us stress that any QFT (integrable or not) with this $\mathcal A_n$ categorical symmetry and spectrum should have a two-particle amplitude inside these bounds.

%%%%%%%%%%%%%%%%%%%%%%%%%%%%%%%%%%%%%%%%%%%%%%%%%%%%%%%%%%
\subsection{Fibonacci category} \label{ssec: Fibboot}
\begin{figure}[th!]
\begin{subfigure}{0.99\textwidth}\centering
\includegraphics[width=.7\textwidth]{ 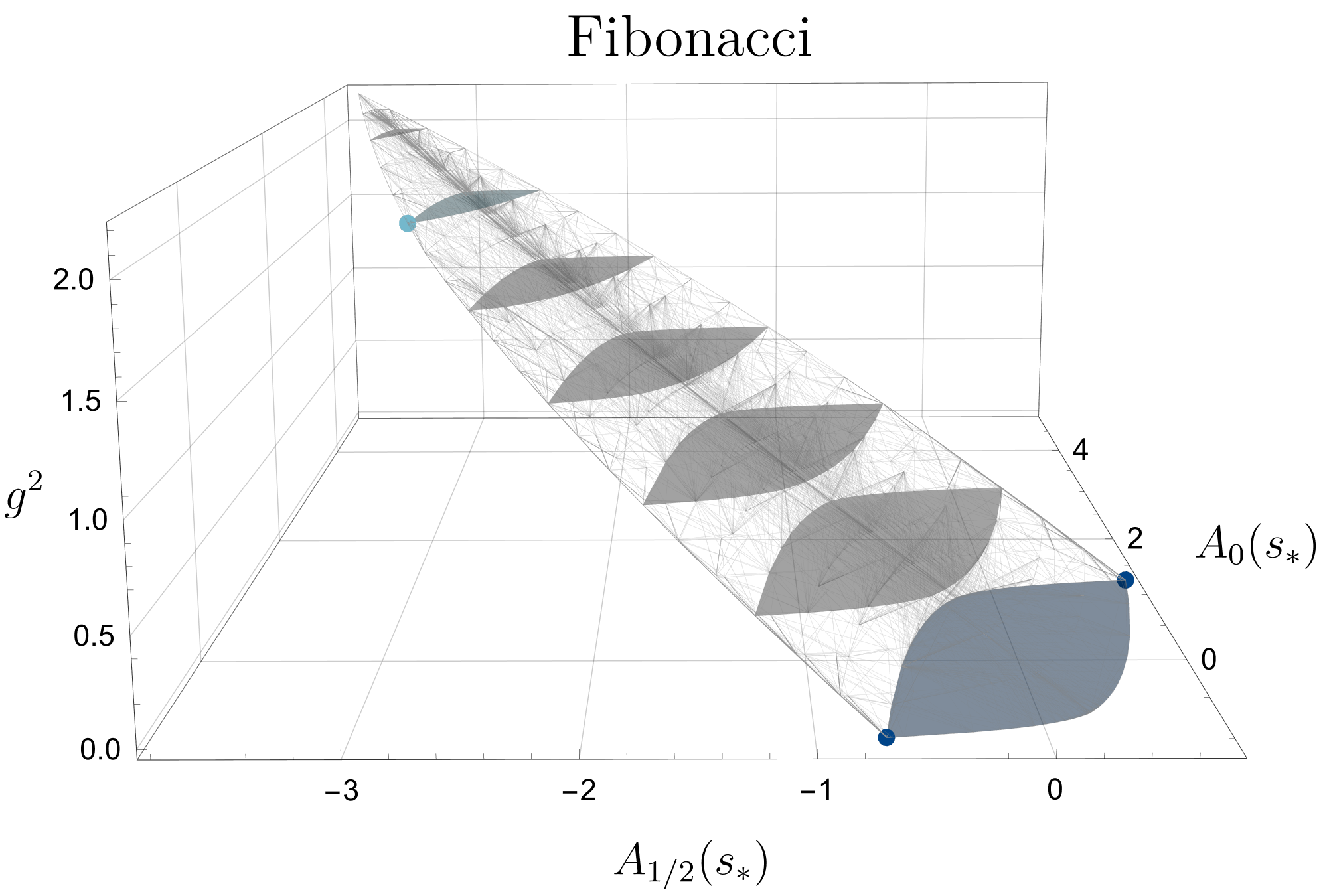}
\caption{}\label{fig:Fibo3d}
\vspace{.7cm}
\end{subfigure}\hspace{0.3cm}
\begin{subfigure}{0.99\textwidth}\centering
\includegraphics[width=.7\textwidth]{ 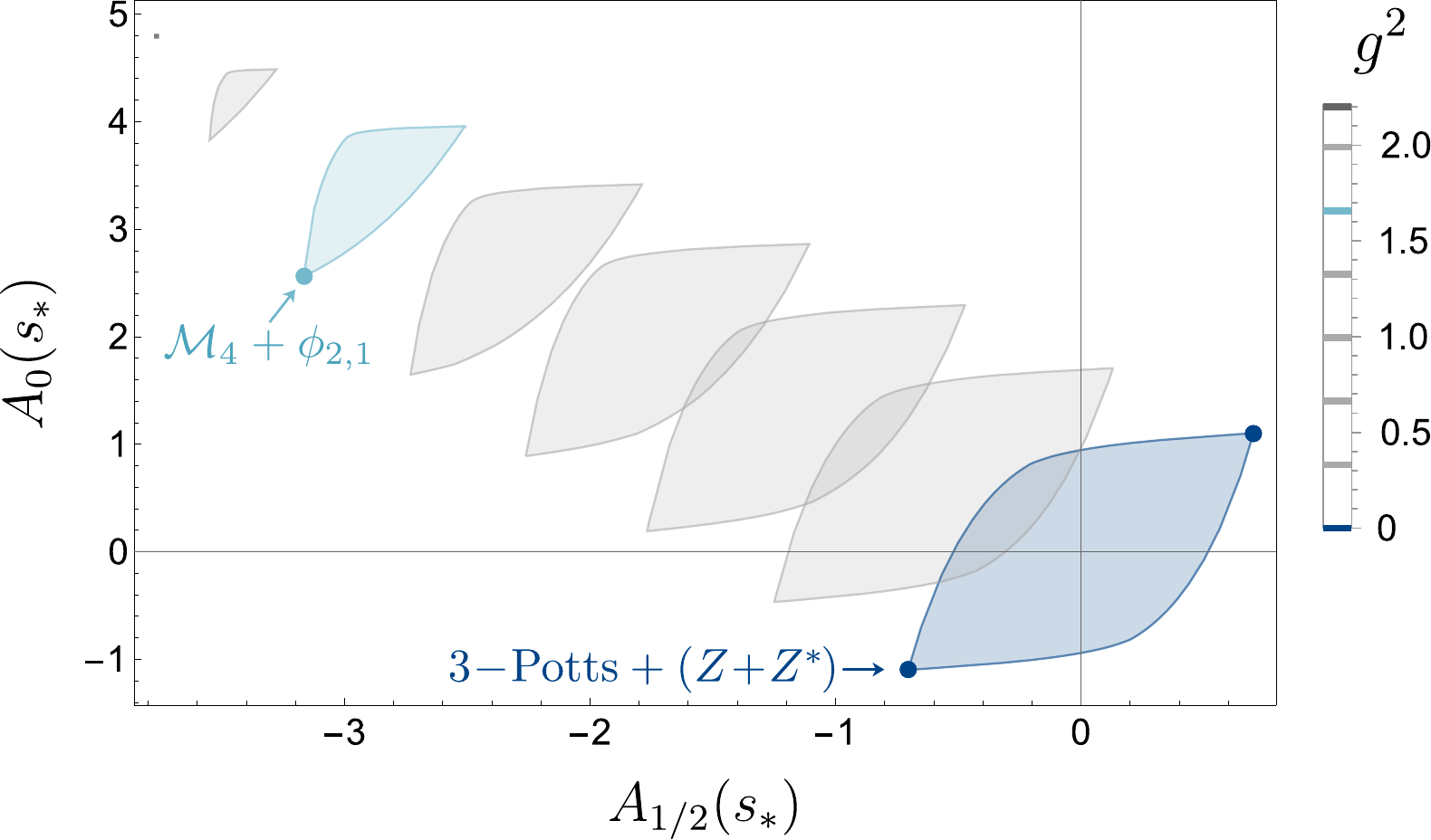}
\caption{}\label{fig:gsections}
\end{subfigure}
\caption{Space of allowed amplitudes $A_\chi(s)$ for Fibonacci category. (a) Optimal bounds for the two fusion channel amplitudes at $s_*\approx3.41m^2$ and the (squared) cubic coupling $g^2$. The dark blue section at the bottom has $g=0$ and matches the previous bounds found for $\mathcal A_5$ (see Figure~\ref{fig:An}. The three dots signal integrable solutions. (b) Projection into the $\{A_{1/2}(s_*),A_0(s_*)\}$ plane. The shaded regions are the fixed $g^2$ sections marked in the 3D plot above. The lower vertices in blue indicate the integrable amplitudes corresponding to known physical models: the subleading magnetic deformation of tricritical Ising (light blue) and the 3-state Potts deformed by the operator $Z+Z^*$ (dark blue). 
}
\label{fig:Fibo}
\end{figure}
The next category we study is Fibonacci with two elements $1,W$ and non trivial fusion rule $W^2=1+W$. Considering again the regular representation of the module category leaves us with two vacua $a=0,1/2$ which we identify with symmetry lines $1,W$. The kink is associated to the $v=1/2=W$ line, which gives the following vacua configurations as part of the same symmetry multiplet: $K_{1/2,0}$, $K_{0,1/2}$ and $K_{1/2,1/2}$. These are precisely the kink, anti-kink and breather discussed around \eqref{eq:WFibmultiplet}. Given the fusion rule above, there are two partial amplitudes obeying the modified crossing condition
\begin{equation}
    \begin{pmatrix}
     A_0(s) \\
     A_{1/2}(s) 
    \end{pmatrix}
    =\frac{1}{\d_{1/2}}
    \begin{pmatrix}
     1 & \d_{1/2} \\
    1 & -1
    \end{pmatrix}
    \begin{pmatrix}
     A_0(4m^2-s) \\
     A_{1/2}(4m^2-s) 
    \end{pmatrix}\,,\qquad \d_{1/2}=\frac{1+\sqrt{5}}{2}\,.
\end{equation}

Because of the fusion rule $v\times v\supset v$, allows us to have a cubic coupling between the scattered particles. Indeed, we can interpret the breather $K_{1/2,1/2}$ as a bound state of the kink and anti-kink. What this means for the amplitudes is that we can have a pole in the $v=1/2$ fusion channel $A_{1/2}(s)\sim\frac{g^2}{s-m^2}$. The bound state mass is fixed to be the same as the external particles since the states we are considering are all part of the same symmetry multiplet and therefore degenerate in mass, while the residue given by the (square of the) non-perturbative cubic coupling is a free parameter.\footnote{In general we are free to consider bound states in any fusion channel $\chi$ with any mass $m_b^{(\chi)}$. Here we choose the minimum spectrum allowed by the categorical symmetry. This is in contrast with the $\mathcal A_n$ example where the minimum spectrum is the one without bound states since the fusion rule $v\times v=0+1$ does not allow the scattered particles to be considered as bound states.} 

A natural space to bootstrap is then $\{A_0(s_*), A_{1/2}(s_*), g^2\}$. We do so with the same primal and dual methods as before, with the only difference that now the amplitudes have poles related to the bound state.\footnote{Note that while $A_0(s)$ has no s-channel pole, the crossing equations above produce a pole in the t-channel (i.e. at $s=4m^2-m^2=3m^2$) in both fusion channels.}  In practice, we fix the coupling to some value and use the normal functionals to produce a slice of the 3d plot and repeat for many couplings. The resulting bounds are presented in Figure~\ref{fig:Fibo}, where the allowed space has a cone/spear-like shape. 

Let us first discuss the $g^2=0$ section in dark blue. This space is actually the same what we found for $\mathcal A_5$ in Figure~\ref{fig:An}. The reason is quite simple: mathematically, the case without bound states brings us back to the same system of equations as we had before since $A_\chi(s)$ are free of poles and for $n=5$ the quantum dimensions in \eqref{eq:Anamps} are $\d^{\mathcal A_5}_{1/2}=\d^{\mathcal A_5}_1=\frac{1+\sqrt{5}}{2}=\d^\text{Fibo}_{1/2}$. Note however that the interpretation for the actual kink scattering is different: here we are dealing with theories which flow to a gapped phase with two vacua instead of four. The relation between the two is the $\mathbb Z_2$ orbifold alluded to in Section \ref{sssec: fibz2}. In practice, all one needs to do is identify the $\mathcal A_5$ labels as $0\leftrightarrow3/2$, $1/2\leftrightarrow1$. In this way we see for instance that the scattering process described by the amplitude $S^{0, 1/2}_{1/2,1}(s)$ in $\mathcal A_5$ becomes  $S^{0, 1/2}_{1/2,1/2}(s)$ in the orbifolded version which is indeed allowed in the Fibonacci category. However, the full symmetry preserved by this orbifold procedure is Fib$\times\mathbb Z_2$, so that we see an enhancement of the symmetry in this $g^2=0$ slice. This is only possible when $g=0$ since Fib$\times\mathbb Z_2$ does not admit the interpretation of the breather as a kink anti-kink bound state, see discussion around \eqref{eq:fusionW'}. The two dots at the vertices mark the integrable solutions obtained by setting $n=5$ in \eqref{eq:YBAn} for the lower vertex and multiplying by overall minus sign for the upper one. The lower vertex is the orbifold version of the $\mathcal{M}_5-\phi_{1,3}$ and describes another physical system: the 3-state Potts critical model deformed by $Z+Z^*$, where these fields are the two primaries with Kac labels $r=1,s=3$. Once more, the associated physical model for the upper vertex is unknown.  

Passing to the rest of the bounds, note there is a sharp tip at $g^2\approx2.19$ which can be understood from the fact that there is a unique amplitude with the maximum cubic coupling allowed by analyticity, modified crossing and unitarity. The $A_\chi(s)\rightarrow-A_\chi(s)$ symmetry in the $g^2=0$ bounds is lost for $g\neq0$, as is clear from Figure~\ref{fig:gsections}. This is because the residue at the bound state pole has a definite (positive) sign. There is a third integrable point at the boundary of the allowed region, marked with a lighter blue point. It corresponds to the $\phi_{2,1}$ deformation of the tricritical Ising model $\mathcal M_4$ proposed in \cite{Smirnov:1991uw} and revised in the context of non-invertible symmetries in our previous work \cite{Copetti:2024rqj}. The partial amplitudes in terms of the rapidity $\theta$ are
\begin{gather}
     A_0(\theta)=R(\theta)\, \sinh\(\frac{i\pi+\theta}{5/9}\)\,, \quad A_1(\theta)=R(\theta)\, \sinh\(\frac{i\pi-\theta}{5/9}\)\,, \label{eq:YBFibo} \\
     R(\theta)=-\sinh^{-1}\(\frac{i\pi+\theta}{5/9}\)\, f_{-2/5}\(9\theta/5\)\, f_{3/5}(9\theta/5)\, F_{-1/9}(\theta)\, F_{2/9}(\theta) \,,\nonumber 
\end{gather}
where $f_\alpha(\theta)=\frac{\sinh\(\frac{\theta+i\alpha\pi}{2}\)}{\sinh\(\frac{\theta-i\alpha\pi}{2}\)}$ and $F_\alpha(\theta)=-f_\alpha(\theta)f_\alpha(i\pi-\theta)$. Although not obvious from the plots, this point is also a cusp in the 3D shape, which nicely matches our expectation that physical theories take a special place in the allowed space of amplitudes. We sampled the rest of the extremal amplitudes and found a similar pattern as for $\mathcal A_n$. The amplitudes saturate unitarity, do not satisfy the Yang-Baxter equations \eqref{eq:YBchi} and display a simple analytic structure with no oscillating behaviour in energy. The points signaled in Figure~\ref{fig:gsections} are all the physical systems we can locate with certainty, but as emphasized in the previous section all models with this minimal spectrum and a global Fibonacci symmetry should be contained in these bounds.

%%%%%%%%%%%%%%%%%%%%%%%%%%%%%%%%%%%%%%%%%%%%%%%%%%%%%%%%%%%%%%%%%%%%%%%%%%%%%%%
\section{Conclusion and future directions}\label{sec: conclusions}
In this work we have demonstrated that non-invertible (categorical) symmetries can be efficiently implemented in an S-matrix bootstrap framework, opening the way to the study of non-invertible-symmetric scattering in 1+1 dimensions.

In the first part, we described kink multiplets and their scattering using the Symmetry Topological Field Theory (SymTFT), extending our previous results  \cite{Copetti:2024rqj} to cover general spontaneous breaking patterns of fusion categories and general kink multiplets. The SymTFT is a universal framework for studying categorical symmetries, which works also in higher dimensions. We thus expect that the approach in this paper, along with recent findings by one of the authors \cite{Copetti:2024onh}, paves the way for higher-dimensional generalizations.
Additionally, this formulation intuitively describes the fusion structure of kink multiplets, imposing nontrivial selection rules on bound states, as we have seen in Sections \ref{sssec: fibz2} and \ref{ssec: Fibboot}.

In the second part, we incorporated non-invertible symmetries into the S-matrix bootstrap by introducing a specialized basis of projectors. These projectors automatically incorporate both the Ward identity of non-invertible symmetries and the modified crossing rules. 
By charting the space of symmetric S-matrices, we showed that known integrable systems appear as cusps in the space of allowed partial-wave amplitudes. 
%Furthermore, we connected the vanishing of certain bootstrap parameters with the enhancement of non-invertible symmetry, providing more stringent constraints on the space of allowed resonances. 

This work serves as a proof-of-concept, illustrating the potential of combining categorical symmetries and the S-matrix bootstrap to impose stringent constraints on the space of consistent S-matrices. It naturally leaves many interesting open questions:
\begin{itemize}
\item {\it S-matrix bootstrap for Haagerup fusion category.} Interesting targets for the S-matrix bootstrap are theories with Haagerup fusion category symmetry. Haagerup fusion category is an ``exceptional" fusion category, initially constructed from a subfactor of von Neumann algebra \cite{haagerup1994principal,grossman2012quantum}. Unlike other fusion categories that are related to finite groups or quantum groups, its QFT realization remains unclear. Recently, evidence suggests that CFTs with Haagerup fusion category may arise from the continuum limit of lattice models \cite{Huang:2021nvb,Vanhove:2021zop}. However, the detailed properties of these CFTs remain unknown. Applying the S-matrix bootstrap to theories with Haagerup fusion category could illuminate the nature of such theories, offering insights into both the ultraviolet (UV) CFT and its relevant deformations. The techniques developed in this paper will be instrumental in pursuing this line of investigation. 
 Another important question is to construct integrable S-matrices with Haagerup fusion category by directly imposing the Yang-Baxter equation. Works in these directions are currently in progress.
\item {\it Form factor bootstrap for theories with non-invertible symmetries.} Incorporating information on the UV fixed point into the S-matrix bootstrap can be achieved by studying form factors of local operators \cite{Karateev:2019ymz,Correia:2022dyp,Cordova:2023wjp}. In the presence of non-invertible symmetries, crossing rules for form factors are likely modified. Deriving these modified crossing rules and performing a (numerical) form factor bootstrap are promising future directions. We anticipate that the SymTFT approach developed in this paper will be useful for this purpose.
\item {\it Modified crossing in $2+1$ dimensions.} An important open question is to generalize our results to higher dimensions. A natural next step in this direction would be to revisit modified crossing rules in Chern-Simons-matter theories, found in \cite{Mehta:2022lgq}, in light of our analysis; in particular from the SymTFT perspective. Note that a SymTFT approach for Chern-Simons-matter theories in (2+1)d has been proposed recently in \cite{Argurio:2024oym}.
    \item {\it 3+1 dimensions}. Finding concrete examples of modified crossing rules in 
$3+1$ dimensions would be extremely interesting. One promising avenue is the monopole-fermion scattering. Using the extension of the spinor-helicity formalism, recent work \cite{Csaki:2020inw} has shown that the standard crossing rule is violated in such processes. Additionally, studies suggest that non-invertible symmetries play a significant role in the monopole-fermion scattering \cite{vanBeest:2023dbu,vanBeest:2023mbs}. Studying these processes using the SymTFT framework could be a first step in this direction.
Alternatively, studying scattering processes involving extended objects, such as surfaces or domain walls, would be interesting, since non-invertible symmetries in higher dimensions often act on extended operators. SymTFT descriptions of extended operators can be achieved following recent results  \cite{Copetti:2024onh}. 

    \item {\it Soft dynamics and SymTFT}. As mentioned in our previous paper \cite{Copetti:2024rqj} and also in \cite{Mehta:2022lgq}, the examples analyzed here can be seen as toy models for the soft dynamics of gauge theories and gravity, where the complicated IR dynamics of soft gluons and gravitons is replaced with simpler TQFT dynamics. Our findings suggest that crossing rules in these theories (in $3+1$ dimensions) are also modified, with the modifications determined by the dynamics in the soft sector. To make further progress, it would be important to study asymptotic symmetries and soft physics using the SymTFT framework.
\end{itemize}

\paragraph{Acknowledgements} The authors thank Lorenzo Di Pietro  for discussions. CC wishes to thank L. Bhardwaj and S. Schafer-Nameki for sharing some of their results \cite{Bhardwaj:2024kvy} prior to publication and the aforementioned plus D. Pajer for collaboration on a related project \cite{CC:boundaries}. CC is supported by STFC grant ST/X000761/1.

\appendix

\section{Flows from minimal models and preserved symmetry}\label{app: flows}
Relevant flows from (unitary) minimal models have a long history. For diagonal minimal models $\cM_{n+1,n}$, the set of topological Verlinde lines $\cL_{r,s}$ is isomorphic to their set of primary operators $\phi_{r,s}$ \cite{Petkova:2000ip,Chang:2018iay}. The symmetry action on a primary operator is encoded in the modular S-matrix:
\bea
\begin{tikzpicture}[baseline={(0,-0.125)}]
    \draw[BLUE,thick] (0,0) circle (0.75); \node[above,BLUE] at (0,0.75) {$\cL_{r,s}$};
    \draw[fill=black] (0,0) node[above] {$\phi_{r',s'}$} circle (0.05);
\end{tikzpicture}
&= \frac{S_{r,s ; r' ,s'}}{S_{1,1;r',s'}} 
\begin{tikzpicture}
    \draw[fill=black] (0,0) node[above] {$\phi_{r',s'}$} circle (0.05);
\end{tikzpicture} \, , \\ 
S_{r,s;r',s'} &= \sqrt{\frac{8}{n(n+1)}} (-)^{1+ r s' + r' s} \sin\left( \frac{\pi n r r'}{n+1} \right) \, \sin\left(\frac{\pi(n+1) s s'}{n} \right) \, .
\eea
The symmetry line $\cL_{r,s}$ is preserved by the $\phi_{r',s'}$ deformation iff:
\be
\frac{S_{r,s ; r' ,s'}}{S_{1,1;r',s'}} = \frac{S_{r,s;1,1}}{S_{1,1;1,1}} \, .
\ee
This equation can be easily on a case-by-case basis. Some implications of the preserved symmetry are discussed e.g. in \cite{Chang:2018iay,Kikuchi:2021qxz,Kikuchi:2022gfi,Tanaka:2024igj}.

The most well known case is the universal $\phi_{1,3}$ deformation, which preserves:
\be
\cA_{n} \simeq \left\{ \cL_{r,1} , \, r = 1, ... , n -1\right\} \ .
\ee
We will label lines in the $\cA_n$ category by $a = 0, 1/2, ..., n/2-1$, with $r=2a + 1$. Their fusion rule is given by:
\be
a \times b = \sum_{\substack{c=|a-b| \\ c - a + b \, \text{mod}(1) =0}}^{\text{min}(a+b, n -2 + a + b )} \, c \, ,
\ee
the name derives from the fact that its objects can be thought of as the nodes of the $A_{n-1}$ Dynkin diagram:
\be
\begin{tikzpicture}
\draw (0,0) node[yshift=-0.55cm] {$0$} circle (0.25);    
\draw (1,0) node[yshift=-0.55cm] {$\frac{1}{2}$} circle (0.25);    
\draw (2,0) node[yshift=-0.55cm] {$1$} circle (0.25);    
\draw (4,0) node[yshift=-0.55cm] {$\frac{n}{2}-1$} circle (0.25);  
\node at (3,0) {$...$};
\draw (0.25,0) -- (0.75,0); \draw (1.25,0) -- (1.75,0); \draw (2.25,0) -- (2.75,0);  \draw (3.25,0) -- (3.75,0);
\end{tikzpicture}
\ee

In the paper we have also studied RG flows from the $c=4/5$ critical Potts model, which is the $\bZ_2$ orbifold of the $\cM_{5}$ diagonal unitary minimal model. 
In Potts language the $\phi_{1,3}$ deformation is mapped to the charge conjugation singlet $Z + Z^*$, see \cite{Chang:2018iay} (sec. 5.2.1), preserving the $\bZ_2$ charge conjugation symmetry as well as the Fibonacci line. 
Since the integrable $\phi_{1,3}$ flows have no kink bound states, it follows that neither do the kinks in the Potts model.

\section{\texorpdfstring{$\cA_5$ and $\cA_5/\bZ_2$ categories}{A5 and A5/Z2 categories}}
We now give some more details on the example considered in the main text: the orbifold of the $\cM_{5}$ minimal model.

The symmetry category of the $\cM_{n}$ unitary minimal model deformed by the relevant $\phi_{1,3}$ operator is the $\cA_{n}$ category, whose lines are labelled by nodes of the $A_{n-1}$ Dynkin diagrams with $F$-symbols:
\begin{gather}\label{eq:tetradef}
    \begin{bmatrix}
     a & b & c\\
    d & e & f
    \end{bmatrix}= 
    (-1)^p \begin{Bmatrix}
     a & b & c\\
    d & e & f
    \end{Bmatrix}_q \,,\quad q=e^{2\pi i/n}\,,\\[0.5em]
    \resizebox{0.6\hsize}{!}{$
    p=\frac{1}{2}\[3 (a + b + c + d + e + f)^2-(a+d)^2-(b+e)^2-(c+f)^2\]\,. \nonumber
    $}
\end{gather}
The $q$-deformed Wigner 6j-symbols are given in terms of the quantum dimensions $[a]=[a]_q=\d_a=\frac{\sin \[(2a+1)\pi/n\]}{\sin (\pi/n)}$ as follows
\begin{equation}
    \begin{split}
    &\begin{Bmatrix}
     a & b & c\\
    d & e & f
    \end{Bmatrix}_q = \Delta(a, b, c)\, \Delta(a, e, f)\, \Delta(d, b, f)\, \Delta(d, e, c) \times
    \\
    &\resizebox{.55\hsize}{!}{$\sum\limits_z
    %^{\min(a + b + d + e, a + c + d + f, b + c + e + f)}_{z=\max(a + b + c, c + d + e, b + d + f, a + e + f)}
    \frac{(-1)^z\[z+1\]!}{\[a+b+d+e-z\]!\[a+c+d+f-z\]!\[b+c+e+f-z\]!}\times
    $}\\
    &\resizebox{.55\hsize}{!}{$
    \frac{1}{\[-a-b-c+z\]!\[-c-d-e+z\]!\[-b-d-f+z\]!\[-a-e-f+z\]!}
    $}\,,
    \end{split} 
\end{equation}
where 
\begin{equation}
%\resizebox{\hsize}{!}{$
  \Delta(a,b,c)=
  \begin{cases}
      \sqrt{\frac{[a + b - c]! [a - b + c]! [-a + b + c]!}{[1 + a + b + c]!}} & \text{if}\ N_{abc}=1 \\
      0 & \text{otherwise}\,.
      \end{cases}
 % $}
\end{equation}
For $n=5$ we have four lines $1, \, \widetilde{W} , \, \widetilde{W}' , \, \widetilde{\eta}$ with the following identification:
\be
\begin{tikzpicture}
\draw (0,0) node[yshift=-0.55cm] {$1$} circle (0.25);    
\draw (1,0) node[yshift=-0.55cm] {$\widetilde{W}'$} circle (0.25);    
\draw (2,0) node[yshift=-0.55cm] {$\widetilde{W}$} circle (0.25);    
\draw (3,0) node[yshift=-0.55cm] {$\widetilde{\eta}$} circle (0.25);    
\draw (0.25,0) -- (0.75,0); \draw (1.25,0) -- (1.75,0); \draw (2.25,0) -- (2.75,0);
\end{tikzpicture}
\ee
The fusion algebra of this category is $\text{Fib} \times \bZ_2$, which is the only possibility for a self-dual fusion category of rank four \cite{dong2018non}. 

Orbifolding this theory by the $\bZ_2$ generated by $\widetilde{\eta}$ we obtain a new fusion category, $\cA_5/\bZ_2$, with the \emph{same} fusion ring but \emph{inequivalent} to $\cA_5$ at the level of $F$-symbols.\footnote{In order for them to be equivalent we must have that there exists a line $\cL$ such that $\cL \times \cL^\dagger = 1 + \widetilde{\eta}$ \cite{Diatlyk:2023fwf}. This is clearly not the case for our example.}

The interface between these two symmetries has two components --$1$, $W$-- and transforms in the regular representation under both Fib subcategories, generated by $\widetilde{W}$ and $W$, respectively.
Thus
\be
\varphi_{u v w}^{a b c} = (\varphi^*)^{a b c}_{\widetilde{u} \widetilde{v} \widetilde{w}} = \frac{1}{\d_w \d_b} \begin{bmatrix}
  v   & c & b  \\ a & u & w 
\end{bmatrix}_{\text{Fib}} \, , \ \ a,b,c\, ; u,v,w   \, \in \, (1,W)  \, ,
\ee
while both $\bZ_2$ symmetries leave the boundary condition invariant:
\be \label{eq: appetaeta}
\varphi^{a a a}_{\eta \eta 1} = (\varphi*)^{a a a }_{\widetilde{\eta} \widetilde{\eta} 1} = 1\, .
\ee
Instead of trying to determine the whole structure of the module category, we focus on the input needed for the S-matrix bootstrap, namely dual boundary $F$-symbols:
\be
(\varphi^*)^{abc}_{\widetilde{W}' \widetilde{W}' \widetilde{u}} \, , \ \ \ u = (1, W) \, .
\ee
In particular, we want to show that, in a judicious gauge:
\be
(\varphi^*)^{abc}_{\widetilde{W}' \widetilde{W}' \widetilde{u}} = (\varphi^*)^{abc}_{\widetilde{W} \widetilde{W} \widetilde{u}} \, ,
\ee
and thus correspond to those of the Fibonacci category. To do this we nucleate a $\widetilde{\eta}$ line between two $\widetilde{W}$ lines:
\be
\begin{tikzpicture}
\draw[color=white!90!gray, fill=white!90!gray] (0,0) -- (3,0) -- (3,2) -- (0,2);
\draw[gray] (0,0) -- (3,0);
\draw[black] (1,0) -- (1,2) node[above,black] {$\widetilde{W}$}; \draw[black] (2,0) -- (2,2) node[above,black] {$\widetilde{W}$};
\draw[black] (1.5,1) circle (0.2); \node at (1.5,1.5) {$\widetilde{\eta}$};
\draw[fill=black] (1,0) circle (0.05); \draw[fill=black] (2,0) circle (0.05);
\node[below,gray] at (0.5,-0.1) {$a$}; \node[below,gray] at (1.5,0) {$b$}; \node[below,gray] at (2.5,-0.1) {$c$};
\end{tikzpicture}
\ee
We can resolve this configuration either by shrinking the $\widetilde{\eta}$ loop and then joining the $\widetilde{W}$ lines or by the following schematic set of moves:
\be
\resizebox{0.75\width}{!}{$
\begin{tikzpicture}[baseline={(0,0.75)}]
    \draw[color=white!90!gray, fill=white!90!gray] (0,0) -- (3,0) -- (3,2) -- (0,2);
\draw[gray] (0,0) -- (3,0);
\draw[black] (1,0) -- (1,2); \draw[black] (2,0) -- (2,2);
\draw[black] (1.25,0) arc (180:0:0.25);
\end{tikzpicture}
\leadsto
\begin{tikzpicture}[baseline={(0,0.75)}]
     \draw[color=white!90!gray, fill=white!90!gray] (0,0) -- (3,0) -- (3,2) -- (0,2);
\draw[gray] (0,0) -- (3,0);
\draw[black] (1,0) -- (1,2); \draw[black] (2,0) -- (2,2);
\draw[black] (1,1) -- (2,1);
\end{tikzpicture}
\leadsto 
\begin{tikzpicture} [baseline={(0,0.75)}]
     \draw[color=white!90!gray, fill=white!90!gray] (0,0) -- (3,0) -- (3,2) -- (0,2);
\draw[gray] (0,0) -- (3,0);
\draw[black] (1,2) -- (1.25,1); \draw[black] (2,2) -- (1.75,1);; \draw[black] (1.25,1) -- (1.75,1); \draw[black] (1.25,1) -- (1.5,0.5) -- (1.75,1); \draw[black] (1.5,0.5) -- (1.5,0);
\end{tikzpicture}
\leadsto
\begin{tikzpicture} [baseline={(0,0.75)}]
  \draw[color=white!90!gray, fill=white!90!gray] (0,0) -- (3,0) -- (3,2) -- (0,2);
\draw[gray] (0,0) -- (3,0);
\draw[black] (1.5,0) -- (1.5,1) -- (1,2); \draw[black] (2,2) -- (1.5,1);
\end{tikzpicture}
$}
\ee
leading to the equations:
\be
(\varphi^*)^{abc}_{\widetilde{W} \widetilde{W} \widetilde{u}} = \left[ \d_{\widetilde{W}'} \begin{bmatrix}
\widetilde{W} & \widetilde{W} & \widetilde{u} \\
\widetilde{W}' & \widetilde{W}' & \widetilde{\eta} 
\end{bmatrix} (\varphi^*)^{abb}_{\widetilde{W} \widetilde{\eta} \widetilde{W}'} (\varphi^*)^{bbc}_{\widetilde{\eta} \widetilde{W} \widetilde{W'}}  \right] \, (\varphi^*)^{ab c}_{\widetilde{W} \widetilde{W} \widetilde{u}}
\ee
The combination $\d_{\widetilde{W}'} \begin{bmatrix}
\widetilde{W} & \widetilde{W} & \widetilde{W} \\
\widetilde{W}' & \widetilde{W}' & \widetilde{\eta} 
\end{bmatrix}$ is equal to $-1$ and can be removed by redefining the bulk $\widetilde{W}' \widetilde{W}' \widetilde{\eta}$ junction. With a bit more work it is also possible to show that $(\varphi^*)^{aab}_{\widetilde{\eta} \widetilde{W} \widetilde{W}' } = (\varphi^*)^{abb}_{\widetilde{W} \widetilde{\eta} \widetilde{W}'}$. Thus transforming the $\widetilde{W}' \, a \, b$ junction by a factor $\left[(\varphi^*)^{aab}_{\widetilde{\eta} \widetilde{W} \widetilde{W}' } \right]^\dagger $ gives the desired gauge.
Let us also discuss the $\bZ_2$ charge of the kink multiplets. This is described by the matrix:
\be
[\eta; v]_{a b}^{a b} \, ,
\ee
which is equal to $\pm 1$ from \eqref{eq: appetaeta}. Since $[\eta; \widetilde{\eta}]_{a a}^{a a} = -1$, as the two symmetries are dual of each other we find that $[\eta; \widetilde{W}]_{a b}^{a b} = - [\eta; \widetilde{W}']_{a b}^{ a b}$. The fusion rule $\widetilde{W}^2 = 1 + \widetilde{W}$ forces the charge of $\widetilde{W}$ to be 1, so:
\be
 [\eta; \widetilde{W}']_{a b}^{ a b} = -1 \, .
\ee
The action of the remaining lines $W$ and $W'$ can be determined from:
\be
\left[ u ; \widetilde{v} \right]_{a b}^{c d} = \sqrt{\d_a \d_b} \begin{bmatrix}
    u & a & b \\ v & d & c
\end{bmatrix}_{\text{Fib}} \, , \ \ \text{for} \ u,v,a,b,c,d \, \in \, (1,W) \, ,
\ee
and the knowledge of $\varphi^{a b c}_{\eta W W'}$, $(\varphi^*)^{a b c}_{\widetilde{\eta} \widetilde{W} \widetilde{W'}}$.
%%%%%%%%%%%%%%%%%%%%%%%%%%%%%%%%%%%%%%%%%%%%%%%%%%%%%%%%%%%%%%%%%%
\section{Normalization of junctions and boundary bubbles}\label{app:junctions}
We briefly discuss our choice of normalization that leads to the identity  \eqref{eq:bdybubbleremoval}. In fusion category (without boundaries), the identity \eqref{eq:orthocompleteness} follows from choosing specific normalizations for junction vector spaces \cite[Eq.~195]{Kitaev:2005hzj} (see also \cite{Barkeshli:2014cna}). This choice is isotropic, that is, it allows line junctions to be moved around freely while keeping endpoints fixed. For example:\footnote{This is correct provided we assume trivial Frobenius-Shur indicators for self-dual lines.}
\be
\begin{tikzpicture}[baseline={(0,0)}]
\draw[thick] (0,0) circle (1);
\draw[thick] plot[smooth, tension=1.5] coordinates{(225:1) (-0.25,0) (135:1)};
\draw[fill=black] (225:1) circle (0.05); \draw[fill=black] (135:1) circle (0.05); 
\node[left] at (180:1) {$a$}; \node[right] at (-0.25,0) {$b$}; \node[right] at (1,0) {$c$};
\end{tikzpicture}
=
\begin{tikzpicture}[baseline={(0,0)}]
\draw[thick] (0,0) circle (1);
\draw[thick] (0,-1) -- (0,1);
\draw[fill=black] (0,-1) circle (0.05); \draw[fill=black] (0,1) circle (0.05); 
\node[left] at (180:1) {$a$}; \node[left] at (0,0) {$b$}; \node[right] at (1,0) {$c$};
\end{tikzpicture}
=
\begin{tikzpicture}[baseline={(0,0)}]
\draw[thick] (0,0) circle (1);
\draw[thick] plot[smooth, tension=1.5] coordinates{(45:1) (0.25,0) (-45:1)};
\draw[fill=black] (45:1) circle (0.05); \draw[fill=black] (-45:1) circle (0.05); 
\node[left] at (180:1) {$a$}; \node[left] at (0.25,0) {$b$}; \node[right] at (1,0) {$c$};
\end{tikzpicture}
=\sqrt{\d_a \, \d_b \, \d_c} \, .
\ee
Note that, in \cite{Fuchs:2002cm} a different convention is used in which junction spaces are orthonormal: the orthonormal basis can always be constructed using the Gram-Schmidt orthogonalization. This is not isotropic, but it can be made so by rescaling vectors in $V_{ab}^c$ by $\left(\frac{\d_c}{\d_a \d_b}\right)^{-1/4}$.

It is natural to use an isotropic basis even in the presence of boundaries. As explained above, we can construct such a basis by first performing the Gram-Schmidt procedure to  obtain an orthonormal basis of boundary junctions $V_{a v}^b$, and rescaling them by a factor $\left( \frac{g_a}{g_b \, \d_v} \right)^{-1/4}$.
The normalization of the junction vertices then gives,
\be
\begin{tikzpicture}[baseline={(0,0)}]
\draw[thick, color=BLUE, fill=white!90!gray] (0,0) circle (1);
\draw[thick] plot[smooth, tension=1.5] coordinates{(225:1) (-0.25,0) (135:1)};
\draw[fill=black] (225:1) circle (0.05); \draw[fill=black] (135:1) circle (0.05); 
\node[left] at (180:1) {$a$}; \node[right] at (-0.25,0) {$v$}; \node[right] at (1,0) {$b$};
\end{tikzpicture}
=
\begin{tikzpicture}[baseline={(0,0)}]
\draw[thick, color=BLUE, fill=white!90!gray] (0,0) circle (1);
\draw[thick] (0,-1) -- (0,1);
\draw[fill=black] (0,-1) circle (0.05); \draw[fill=black] (0,1) circle (0.05); 
\node[left] at (180:1) {$a$}; \node[left] at (0,0) {$v$}; \node[right] at (1,0) {$b$};
\end{tikzpicture}
=
\begin{tikzpicture}[baseline={(0,0)}]
\draw[thick, color=BLUE, fill=white!90!gray] (0,0) circle (1);
\draw[thick] plot[smooth, tension=1.5] coordinates{(45:1) (0.25,0) (-45:1)};
\draw[fill=black] (45:1) circle (0.05); \draw[fill=black] (-45:1) circle (0.05); 
\node[left] at (180:1) {$a$}; \node[left] at (0.25,0) {$v$}; \node[right] at (1,0) {$b$};
\end{tikzpicture}
=\sqrt{\d_v \, g_a \, g_b} \, ,
\ee
which also implies \eqref{eq:bdybubbleremoval}. 

%%%%%%%%%%%%%%%%%%%%%%%%%%%%%%%%%%%%%%%%%%%%%%%%%%%%%%%%%%%%%%%%%%
\section{More on bootstrap and numerical implementation}\label{sec:appBoot}
In this appendix we give more details on how to optimize primal (dual) functionals $\mathcal F$ ($\mathcal F_d$). Starting with the primal approach, we choose a linear functional to maximize $\cF[S(s)]$. To solve this optimization problem numerically, we write an ansatz for $S(s)$ with the required analyticity and crossing symmetry and impose unitarity separately. To give a concrete example, suppose we want to bound the partial amplitudes at a given energy below threshold $A_\chi(s_*)$. For theories without bound states we would write an ansatz of the form
\begin{gather}
    A_\chi(s)= \sum\limits_{n=0}^{N_\text{max}} \[ a^{(n)}_{\chi}\, \rho^n_{s_0}(s) +C_{\chi\chi'} \, a^{(n)}_{\chi'} \, \rho^n_{s_0}(4m^2-s)\]\,,\\
    \rho_{s_0}(s)\equiv\frac{\sqrt{4m^2-s_0}-\sqrt{4m^2-s}}{\sqrt{4m^2-s_0}+\sqrt{4m^2-s}}\,,
\end{gather}
where $C_{\chi\chi'}$ is the crossing matrix, $s_0$ is a subtraction point we are free to choose and $a^{(n)}_{\chi}$ are the variables for the optimization problem. The ``$\rho$-variables" make sure the analytic properties of the amplitude are satisfied, as they put branch points at $s=0,4m^2$.\footnote{These were first used for S-matrix bootstrap in \cite{Paulos:2017fhb} but have their origin in similar variables used in conformal bootstrap introduced in \cite{Hogervorst:2013sma}.} To impose unitarity we choose a grid of physical values of $s_j\geq4m^2$ and demand that $|A_\chi(s_j)|^2\leq1$. A convenient choice for the unitarity grid can be written in terms of the zeros of Chebyshev polynomials \cite{Karateev:2019ymz}:
 \begin{equation}\label{eq:Ugrid}
    s_j=\frac{8m^2}{1+\cos\phi_j}\,,\quad \phi_j=\frac{\pi}{2}\left[1-\cos\left(\frac{2j-1}{N_\text{grid}}\pi\right)\right]\,,\quad j=1,\ldots,N_\text{grid}\,,
    \end{equation}
which conveniently puts more points close to threshold and infinity. 

To find bounds on $A_\chi(s_*)$ we can maximize various types of functionals\cite{Cordova:2019lot}; for instance the ``normal" type $\cF_\text{norm}=\sum_\chi n_\chi A_\chi(s_*)$ which highlight the points with more curvature or the ``radial" type where we maximize along a given direction $\cF_\text{rad}=t$ with the constraint $A_\chi(s_*)=t\, n_\chi$. In either case, we repeat the optimization procedure for many different $n_\chi$. For instance, to produce the bounds in Figure~\ref{fig:An} we used the normal functionals for 128 $n_\chi$'s uniformly distributed on a circle, with $N_\text{grid}=2N_\text{max}$ and $N_\text{max}=10$.\footnote{The low energy values $A_\chi(s_*)$ rapidly converge to the optimal result, however to get a refined picture of the analytic properties of the amplitudes in the $s$ plane and large energy behaviour we used $N_\text{max}=30$ at sampled points.} 

As explained in the main text, this approach explores the space of consistent amplitudes from the inside, by explicitly constructing amplitudes satisfying all axioms but has the bounds are not rigorous, since we might find a larger value for $\max\cF$ as we vary $N_\text{max}$ and $N_\text{grid}$. In contrast, the dual approach establishes rigorous upper bounds on $\cF$ through the minimization of a dual functional $\cF_d$ satisfying $\cF\leq\cF_d$. The dual problem is a standard way to solve convex optimization problems (see e.g. \cite{boyd2004convex} chapter 5) and was worked out in the context of the modern non-perturbative S-matrix bootstrap first in \cite{Cordova:2019lot} and in subsequent works \cite{Guerrieri:2020kcs,Kruczenski:2020ujw,He:2021eqn,EliasMiro:2021nul,Guerrieri:2021tak,Correia:2022dyp,Cordova:2023wjp,Guerrieri:2024ckc}. 

The idea is to introduce dual variables which act as Lagrange multipliers for the S-matrix constraints. This can be done by writing a Lagrangian $\mathscr L$ satisfying $\cF[A_\chi(s)]\leq \mathscr L$ with the following form
\begin{equation}
   \mathscr L= \cF[A_\chi(s)] + \frac{1}{2\pi i}\sum_\chi \oint ds\, K_\chi(s) A_\chi(s) + \frac{1}{\pi}\sum_\chi\int\limits_{4m^2}^\infty ds\, \lambda_\chi(s) \[1-|A_\chi(s)|^2\] \,,
\end{equation}
where  $K_\chi(s)$ and $\lambda_\chi(s)\geq0$ are respectively the Lagrange multipliers for analyticity+crossing and unitarity. Next we want to find constraints on the dual variables $K_\chi(s)$ and $\lambda_\chi(s)$ such that the primal variables are eliminated from the problem. In practice, we open up the contour for the first integral and ask that $K(s)$ transforms under crossing as $K_\chi(s)=-C_{\chi'\chi}K_{\chi'}(4m^2-s)$ and falls of at least like $s^{-3/2}$ at infinity so that we can group terms into a single integral over physical values of energy:
\begin{equation}
    0=\frac{1}{2\pi i}\sum_\chi \oint ds\, K_\chi(s) A_\chi(s) = \frac{2}{\pi} \sum_\chi \int\limits_{4m^2}^\infty ds\, \Im\[K_\chi(s) A_\chi(s)\] +\text{possible residues}\,.
\end{equation}
The possible residues might come from $A_\chi(s)$ or $K_\chi(s)$, depending on the problem at hand. The point is that we can fix the analytic properties of $K(s)$ so that we cancel the primal functional $\cF$. For the example we had earlier with no bound states we would ask for a pole $K(s)\sim  \frac{\tilde n_\chi/2}{s-s_*}$ (and its crossing image) where the residue satisfies $\tilde n_\chi=n_\chi$ for the normal functional $\cF_\text{norm}$ and $\sum_\chi\tilde n_\chi n_\chi=1$ for the radial one $\cF_\text{rad}$. In this way we are left with
\begin{equation}
    \mathscr L= \frac{1}{\pi} \sum_\chi \int\limits_{4m^2}^\infty ds\,\left\{2 \Im\[K_\chi(s) A_\chi(s)\] + \lambda_\chi(s) \[1-|A_\chi(s)|^2\] \right\} \,,
\end{equation}
which we can now extremize over $A_\chi(s)$. The result of the maximization sets $S(s)=i K^*(s)/\lambda(s)$ and gives
\begin{equation}
   \underset{A_\chi}{\max}\, \mathscr L = \frac{1}{\pi} \sum_\chi\int\limits_{4m^2}^\infty ds\,\left\{\frac{|K_\chi(s)|^2}{\lambda_\chi(s)} +\lambda_\chi(s)\right\}\,.
\end{equation}
The expression above already gives a dual functional that depends only on the Lagrange multipliers. Since we are interested in the best bound possible, we minimize over the dual variables. This can be done trivially for $\lambda(s)$ which results in the dual functional in the main text\footnote{In the case where there are more constraints in the primal problem, such as fixing the value of the amplitude at a given $s$, we would need to consider more dual variables/Lagrange multipliers and minimize over them as well which might result in extra terms in the dual functional.}
\begin{equation} \label{eq:Fdual}
 \cF_d \equiv \frac{2}{\pi} \sum_\chi \int\limits_{4m^2}^\infty ds\,|K_\chi(s)|=\underset{\lambda}{\min} \, \underset{A_\chi}{\max}\, \mathscr L\,,
\end{equation}
which satisfies $\underset{A_\chi}{\max} \,\cF \leq \underset{K_\chi}{\min} \,\cF_d$. Let us stress that by minimizing $\cF_d$ we are putting a rigorous upper bound on the original optimization problem. Also, if we find the same result with the primal and dual problems then we can be sure we have the optimal bound.\footnote{The difference between $\underset{K_\chi}{\min} \,\cF_d$ and $\underset{A_\chi}{\max} \,\cF$ is known as duality gap. For convex optimization problems such as the ones we are studying it is enough to find a dual functional satisfying all conditions for the duality gap to close (see Slater's criterion in \cite{boyd2004convex}).}

To perform the dual optimization numerically we follow the same logic: we first write an ansatz for $K_\chi(s)$ compatible with analyticity in the cut plane and fall-off at infinity, evaluate the integral by quadrature and finally minimize. 

To give a concrete example, let us go back to the case without bound states and set our primal functional to be $\cF_\text{rad}=t$ with $A\chi(s_*)=t n_\chi$.\footnote{This functional has the advantage that one can read off directly the values for $A_\chi(s_*)$ after the dual optimization.} We write first the following ansatz
\begin{equation}
    K_\chi(s)= \frac{\sqrt 2}{\sqrt{s}\sqrt{4-s}}\sum\limits_{n=0}^{M_\text{max}} \[ \frac{1}{s-s_*}\,b^{(n)}_{\chi}\, \rho^n_{s_0}(s) -\frac{1}{4m^2-s-s_*}\, C_{\chi'\chi} \, b^{(n)}_{\chi'} \, \rho^n_{s_0}(4m^2-s)\]\,.
\end{equation}
To integrate \eqref{eq:Fdual} numerically we can use a polynomial approximation. The idea is to transform the integral into a weighted sum of the integrand evaluated at some points in the integration domain:
\begin{equation}\label{eq:numint}
    \int\limits_{-1}^1 dx f(x)\approx \int\limits_{-1}^1 dx \sum_i f(x_i) P_i(x) = \sum_i f(x_i) w_i\,, \quad P_i(x)=\prod_{j\neq i}^n\frac{x-x_j}{x_i-x_j}\,, \; w_i=\int\limits_{-1}^1 dx P_i(x)\,,
\end{equation}
where the weights $w_i$ are evaluated using $\int_{-1}^1 dx \,x^n=\frac{1+(-1)^n}{n+1}$. A suitable change of variables like $x=\frac{2m^2-\sqrt{s-4m^2}}{2m^2+\sqrt{s-4m^2}}$ brings the last equation to the form of \eqref{eq:Fdual} and the grid points can be for instance \eqref{eq:Ugrid}. The final step is to minimize this discretized integral subject to the constraint $2\sum_\chi \,\text{Res}_{s=s_*} \[K_\chi(s)\] \,n_\chi=1$, which can be done in e.g. Mathematica. With this method (setting $M_\text{max}=5$ and $N_\text{grid}=30$) we reproduced the bounds presented in the main text so that we know they are optimal. 

%%%%%%%%%%%%%%%%%%%%%%%%%%%%%%%%%%%%%%%%%%%%%%%%%%%
 \bibliographystyle{JHEP}
 \small
 \baselineskip=.75\baselineskip
 \bibliography{intebib}

\end{document}